\documentclass[12pt,a4]{article}

\usepackage{a4wide}
\usepackage{xspace}
\usepackage{amsmath}
\usepackage{amssymb}
\usepackage{booktabs}
\usepackage{graphicx}
\usepackage{url}
\usepackage[breaklinks=true]{hyperref}
\usepackage{color}
\definecolor{comment}{rgb}{0,0.3,0}
\definecolor{identifier}{rgb}{0.0,0,0.3}
\usepackage{listings}
\global\emergencystretch = .9\hsize

\newcommand{\fastjet}{\texttt{FastJet}\xspace}

\newcommand{\ttt}[1]{{\small\texttt{#1}}}

\newcommand{\order}[1]{{\cal O}\left(#1\right)}
\newcommand{\ie}{{\it i.e.}\ }

\newcommand{\ee}{e^+e^-}
\newcommand{\Dzero}{D\O\xspace}
\newcommand{\GeV}{\,\text{GeV}}

\newcommand{\PseudoJet}{\ttt{PseudoJet}\xspace}
\newcommand{\PJ}{\ttt{PseudoJet}\xspace}
\newcommand{\ClusterSequence}{\ttt{ClusterSequence}\xspace}
\newcommand{\CS}{\ttt{ClusterSequence}\xspace}

\newcommand{\throws}{{\it throws}}

\title{\sf FastJet user manual%
  \\ \large (for version
  3.0.1
  )
}
\author{Matteo Cacciari,$^{1,2}$ Gavin P. Salam$^{3,4,1}$ and Gregory Soyez$^{5}$\\[10pt]
  \normalsize
  $^1$LPTHE, UPMC Univ.~Paris 6 and CNRS UMR 7589, Paris, France\\
  \normalsize
  $^2$Universit\'e Paris Diderot, Paris, France\\
  \normalsize
  $^3$CERN, Physics Department, Theory Unit, Geneva, Switzerland\\
  \normalsize
  $^4$Department of Physics, Princeton University, Princeton, NJ 08544,USA\\
  \normalsize
  $^5$Institut de Physique Th\'eorique, CEA Saclay, France
}

\date{}

\begin{document}

\maketitle

\vspace{-10cm}
\begin{flushright}
  CERN-PH-TH/2011-297
\end{flushright}
\vspace{9cm}

\begin{abstract}

  \fastjet is a \ttt{C++} package that provides a broad range of jet
  finding and analysis tools.
  It includes efficient native implementations of all widely used $2\to 1$
  sequential recombination jet algorithms for $pp$ and $e^+e^-$
  collisions, as well as access to 3rd party jet algorithms through a
  plugin mechanism, including all currently used cone algorithms.
  \fastjet also provides means to facilitate the manipulation of jet
  substructure, including some common boosted heavy-object taggers, as
  well as tools for estimation of pileup and underlying-event noise
  levels, determination of jet areas and subtraction or suppression
  of noise in jets.

\end{abstract}

\newpage
\tableofcontents
\newpage

\section{Introduction}

Jets are the collimated sprays of hadrons that result from the
fragmentation of a high-energy quark or gluon.
They tend to be visually obvious structures when one looks at an
experimental event display, and by measuring their energy and
direction one can approach the idea of the original ``parton'' that
produced them.
Consequently jets are both an intuitive and quantitatively essential
part of collider experiments, used in a vast array of analyses, from
new physics searches to studies of Quantum Chromodynamics (QCD).
For any tool to be so widely used, its behaviour must be well defined
and reproducible: it is not sufficient that one be able to visually
identify jets, but rather one should have rules that project a set of
particles onto a set of jets.
Such a set of rules is referred to as a jet algorithm.
Usually a jet algorithm involves one or more parameters that govern
its detailed behaviour.
The combination of a jet algorithm and its parameters is known as a
jet definition.
Suitable jet definitions can be applied to particles,
calorimeter towers, or even to the partonic events of perturbative QCD
calculations, with the feature that the jets resulting from these
different kinds of input are not just physically close to the
concept of partons, but can be meaningfully be compared to each other.

Jet finding dates back to seminal work by Sterman and
Weinberg~\cite{StermanWeinberg} and several reviews have been written
describing the various kinds of jet finders, their different uses and
properties, and even the history of the
field, for example~\cite{Moretti:1998qx,RunII-jet-physics,Ellis:2007ib,Salam:2009jx,Ali:2010tw}.

It is possible to classify most jet algorithms into one of two broad
classes: sequential recombination algorithms and cone algorithms.

Sequential recombination algorithms usually identify the pair of
particles that are closest in some distance measure, recombine them,
and then repeat the procedure over and again, until some stopping
criterion is reached. 
The distance measure is usually related to the structure of
divergences in perturbative QCD.
The various sequential recombination algorithms differ mainly in their
particular choices of distance measure and stopping criterion.

Cone algorithms put together particles within specific conical angular
regions, notably such that the momentum sum of the particles contained
in a given cone coincides with the cone axis (a ``stable cone'').
Because QCD radiation and hadronisation leaves the direction of a
parton's energy flow essentially unchanged, the stable cones are
physically close in direction and energy to the original partons.
Differences between various cone algorithms are essentially to do with
the strategy taken to search for the stable cones (e.g.\ whether
iterative or exhaustive) and the procedure used to deal with cases
where the same particle is found in multiple stable cones (e.g.\
split--merge procedures).

One of the aims of the \fastjet C++ library is to provide
straightforward, efficient implementations for the majority of widely
used sequential-recombination algorithms, both for hadron-hadron and
$e^+e^-$ colliders, and easy access also to cone-type jet algorithms.
It is distributed under the terms of version~2 of the GNU General Public License
(GPL)~\cite{GPLv2}.

To help introduce the terminology used throughout \fastjet and this
manual, let us consider the longitudinally-invariant $k_t$ algorithm
for hadron colliders~\cite{ktexcl,ktincl}.
This was the first jet algorithm to be implemented in
\fastjet~\cite{fastjet} and its 
structure, together with that of other sequential recombination
algorithms, has played a key role in the design of \fastjet's interface.
The $k_t$ algorithm involves a (symmetric) distance measure, $d_{ij}$,
between all pairs of particles $i$ and $j$,
\begin{equation}
  \label{eq:dij-illustr}
  d_{ij} = d_{ji} = \min(p_{ti}^2, p_{tj}^2) \frac{\Delta R_{ij}^2}{R^2}\,,
\end{equation}
where $p_{ti}$ is the transverse momentum of particle $i$ with respect
to the beam ($z$) direction and $\Delta R_{ij}^2 = (y_i - y_j)^2 +
(\phi_i - \phi_j)^2$, with $y_i = \frac12 \ln \frac{E_i + p_{zi}}{E_i
  - p_{zi}}$ and $\phi_i$ respectively $i$'s rapidity and azimuth.
The $k_t$ algorithm also involves a distance measure between every
particle $i$ and the beam
\begin{equation}
  \label{eq:diB-illustr}
  d_{iB} = p_{ti}^2\,.
\end{equation}
$R$ in eq.~(\ref{eq:dij-illustr}), usually called the jet radius, is a
parameter of the algorithm that determines its angular reach.
In the original, so-called ``exclusive'' formulation of the $k_t$
algorithm~\cite{ktexcl} (generally used with $R=1$), one identifies
the smallest of the $d_{ij}$ and $d_{iB}$. 
If it is a $d_{ij}$, one replaces $i$ and $j$ with a single new object
whose momentum is $p_i + p_j$ --- often this object is called a
``pseudojet'', since it is neither a particle, nor yet a full
jet.\footnote{In \fastjet we actually will use \texttt{PseudoJet} to denote any
  generic object with 4-momentum.}
If instead the smallest distance is a $d_{iB}$, then one removes $i$
from the list of particles/pseudojets and declares it to be part of
the ``beam'' jet.
One repeats this procedure until the smallest $d_{ij}$ or $d_{iB}$ is
above some threshold $d_{\text{cut}}$; all particles/pseudojets that
are left are then that event's (non-beam) jets.

In the ``inclusive'' formulation of the $k_t$ algorithm~\cite{ktincl},
the $d_{ij}$ and $d_{iB}$ distances are the same as above.
The only difference is that when a $d_{iB}$ is smallest, then $i$ is
removed from the list of particles/pseudojets and added to the list of
final ``inclusive'' jets (this is instead of being incorporated into a
beam jet).
There is no $d_{\text{cut}}$ threshold and the clustering continues
until no particles/pseudojets remain.
Of the final jets, generally only those above some transverse momentum
are actually used.\footnote{This transverse momentum cut has some
  similarity to $d_{\text{cut}}$ in the exclusive case, since in the
  exclusive case pseudojets with $p_t < \sqrt{d_{\text{cut}}}$ become
  part of the beam jets, i.e.\ are discarded.}
Because the distance measures are the same in the inclusive and
exclusive algorithms, the clustering sequence is common to both
formulations (at least up to $d_{\text{cut}}$), a property that will
be reflected in \fastjet's common interface to both formulations.

Having seen these descriptions, the reader may wonder why a special
library is needed for sequential-recombination jet finding. 
Indeed, the $k_t$ algorithm can be easily implemented in just a few
dozen lines of code.
The difficulty that arises, however, is that at hadron colliders,
clustering is often performed with several hundreds or even thousands
of particles.
Given $N$ particles, there are $N(N-1)/2$ $d_{ij}$ distances to
calculate, and since one must identify the smallest of these
$\order{N^2}$ distances at each of $\order{N}$ iterations of the
algorithm, original implementations of the $k_t$
algorithm~\cite{KtClus,KtJet} involved $\order{N^3}$ operations to
perform the clustering.
In practice this translates to about 1\,s for $N=1000$. Given that
events with pileup can have multiplicities significantly in excess of
1000 and that it can be necessary to cluster hundreds of millions of
events, $N^3$ timing quickly becomes prohibitive, all the more so in
time-critical contexts such as online triggers.
To alleviate this problem, \fastjet makes use of the
observation~\cite{fastjet} that the smallest pairwise distance remains
the same if one uses the following alternative (non-symmetric)
$d_{ij}$ distance measure:
\begin{equation}
  \label{eq:dij-illustr-asym}
  d_{ij} = p_{ti}^2 \frac{\Delta R_{ij}^2}{R^2}\,,\qquad
  d_{ji} = p_{tj}^2 \frac{\Delta R_{ij}^2}{R^2}
\end{equation}
For a given $i$, the smallest of the $d_{ij}$ is simply found by
choosing the $j$ that minimises the $\Delta R_{ij}$, i.e.\ by
identifying $i$'s geometrical nearest neighbour on the $y-\phi$
cylinder.
Geometry adds many constraints to closest pair and nearest neighbour
type problems, e.g.\ if $i$ is geometrically close to $k$ and $j$ is
geometrically close to $k$, then $i$ and $j$ are also geometrically
close; such a property is not true for the $d_{ij}$. The factorisation
of the problem into momentum and geometrical parts makes it possible
to calculate and search for minima among a much smaller set of
distances.
This is sufficiently powerful that with the help of the external
Computational Geometry Algorithms Library (CGAL)~\cite{CGAL}
(specifically, its Delaunay triangulation modules), \fastjet achieves
expected $N\ln N$ timing for many sequential recombination algorithms.
This $N\ln N$ strategy is supplemented in \fastjet with several other
implementations, also partially based on geometry, which help optimise
clustering speed up to moderately large multiplicities, $N \lesssim
30000$.
The timing for $N=1000$ is thus reduced to a few milliseconds.
The same techniques apply to a range of sequential recombination
algorithms, described in section~\ref{sec:native-algs}.

At the time of writing, sequential recombination jet algorithms are
the main kind of algorithm in use at CERN's Large Hadron Collider
(LHC), notably the anti-$k_t$ algorithm~\cite{antikt}, which simply
replaces $p_{t}^2$ with $p_{t}^{-2}$ in
eqs.~(\ref{eq:dij-illustr},\ref{eq:diB-illustr}). Sequential
recombination algorithms were also widely used at HERA and LEP.
However at Fermilab's Tevatron, and in much preparatory LHC work, cone
algorithms were used for nearly all studies.
For theoretical and phenomenological comparisons with these results,
it is therefore useful to have straightforward access also to cone
algorithm codes.
The main challenge that would be faced by someone wishing to write
their own implementation of a given cone algorithm comes from the
large number of details that enter into a full specification of such
algorithms, e.g.\ the precise manner in which stable cones are found,
or in which the split--merge step is carried out.
The complexity is such that in many cases the written descriptions
that exist of specific cone algorithms are simply insufficient to
allow a user to correctly implement them.
Fortunately, in most cases, the authors of cone algorithms have
provided public implementations and these serve as a reference for the
algorithm.
While each tends to involve a different interface, a different
4-momentum type, etc.,
\fastjet has a ``plugin'' mechanism, which makes it possible
to provide a uniform interface to these different third party jet
algorithms.
Many plugins (and the corresponding third party code) are distributed
with \fastjet. 
Together with the natively-implemented sequential-recombination
algorithms, they ensure easy access to all jet algorithms used at
colliders in the past decade (section~\ref{sec:plugins}).
Our distribution of this codebase is complemented with some limited
curatorial activity, e.g.\ solving bugs that become apparent when
updating compiler versions, providing a common build infrastructure,
etc.

In the past few years, research into jets has evolved significantly
beyond the question of just ``finding'' jets.
This has been spurred by two characteristics of CERN's LHC
experimental programme.
The first is that the LHC starts to probe momentum scales that are far
above the the electroweak scale, $M_{EW}$, e.g.\ in the search for new
particles or the study of high-energy $WW$ scattering.
However, even in events with transverse momenta $\gg M_{EW}$, there can
simultaneously be hadronic physics occurring on the electroweak scale
(e.g.\ hadronic $W$ decays).
Jet finding then becomes a multi-scale problem, one manifestation of
which is that hadronic decays of W's, Z's and top quarks may be so
collimated that they are entirely contained within a single jet.
The study of this kind of problem has led to the development of a wide
array of jet substructure tools for identifying ``boosted'' object
decays, as reviewed in~\cite{Abdesselam:2010pt}.
As was the situation with cone algorithms a few years ago, there is
considerable fragmentation among these different tools, with some
public code available from a range of different sources, but
interfaces that differ from one tool to the next.
Furthermore, the facilities provided with version 2 of \fastjet did not
always easily accommodate tools to manipulate and transform jets.
Version 3 of \fastjet aims to improve this situation, providing 
implementations of the most common jet substructure tools\footnote{To
  some extent there is overlap here with SpartyJet~\cite{SpartyJet},
  however we believe there are benefits to being able to easily carry
  jet structure manipulations natively within the framework of
  \fastjet.} and a framework to help guide third party authors who wish
to write further such tools using a standard interface
(section~\ref{sec:transformers}).
In the near future we also envisage the creation of a \fastjet
``contrib'' space, to provide a common location for access to these
new tools as they are developed.

The second characteristic of the LHC that motivates facilities beyond
simple jet finding in \fastjet is the need to use jets in high-noise
environments.
This is the case for proton-proton ($pp$) collisions, where in
addition to the $pp$ collision of interest there are many additional
soft ``pileup'' $pp$ collisions, which contaminate jets with a high
density of low-momentum particles.
A similar problem of ``background contamination'' arises also for
heavy-ion collisions (also at RHIC) where the underlying event in the
nuclear collision can generate over a TeV of transverse momentum per
unit rapidity, part of which contaminates any hard jets that are
present.
One way of correcting for this involves the use of jet ``areas'',
which provide a measure of a given jet's susceptibility to soft
contamination.
Jet areas can be determined for example by examining the clustering of
a large number of additional, infinitesimally soft ``ghost''
particles~\cite{CSSAreas}.
Together with a determination of the level of pileup or
underlying-event noise in a specific event, one can then perform
event-by-event and 
jet-by-jet subtraction of the
contamination~\cite{cs,Cacciari:2010te}. 
\fastjet allows jet clustering to be performed in such a way that the
jet areas are determined at the same time as the jets are identified,
simply by providing an ``area definition'' in addition to the jet
definition (section~\ref{sec:areas}).
Furthermore it provides the tools needed to estimate the density of
noise contamination in an event and to subtract the appropriate amount
of noise from each jet (section~\ref{sec:BackgroundEstimator}).
The interface here shares a number of characteristics with the
substructure tools, some of which also serve to remove noise
contamination. Both the substructure and pileup removal make use
also of a ``selectors'' framework for specifying and combining simple
cuts (section~\ref{sect:selectors}).

While \fastjet provides a broad range of facilities, usage for basic
jet finding is straightforward.
To illustrate this, a quick-start guide is provided in
section~\ref{sec:quick-start}, while the core classes
(\ttt{PseudoJet}, \ttt{JetDefinition} and \ttt{ClusterSequence}) are
described in section~\ref{sec:core-classes}.
For more advanced usage, one of the design considerations in \fastjet
has been to favour user extensibility, for example through plugins,
selectors, tools, etc. This is one of the topics covered in the
appendices.
Further information is also available from the extensive 
``doxygen'' documentation, available online at
\url{http://fastjet.fr}.

\section{Quick-start guide}
\label{sec:quick-start}

For the impatient, the \fastjet package can be set up and run as follows.

\begin{itemize}
\item Download the code and the unpack it
\begin{verbatim}
 curl -O http://fastjet.fr/repo/fastjet-X.Y.Z.tar.gz 
 tar zxvf fastjet-X.Y.Z.tar.gz
 cd fastjet-X.Y.Z/
\end{verbatim}
replacing \ttt{X.Y.Z} with the appropriate version number. On some
systems you may need to replace ``\texttt{curl -O}'' with
``\texttt{wget}''. 

\item Compile and install (choose your own preferred prefix), and when 
you're done go back to the original directory
\begin{verbatim}
 ./configure --prefix=`pwd`/../fastjet-install
 make
 make check
 make install
 cd ..
\end{verbatim}
If you copy and paste the above lines from one very widespread PDF
viewer, you should note that the first line contains \emph{back-quotes}
not forward quotes but that your PDF viewer may nevertheless paste
forward quotes, causing problems down the line (the issue arises again
below). 

\item
Now paste the following piece of code into a file called \tt{short-example.cc}
\begin{lstlisting}
#include "fastjet/ClusterSequence.hh"
#include <iostream>
using namespace fastjet;
using namespace std;

int main () {
  vector<PseudoJet> particles;
  // an event with three particles:   px    py  pz      E
  particles.push_back( PseudoJet(   99.0,  0.1,  0, 100.0) ); 
  particles.push_back( PseudoJet(    4.0, -0.1,  0,   5.0) ); 
  particles.push_back( PseudoJet(  -99.0,    0,  0,  99.0) );

  // choose a jet definition
  double R = 0.7;
  JetDefinition jet_def(antikt_algorithm, R);

  // run the clustering, extract the jets
  ClusterSequence cs(particles, jet_def);
  vector<PseudoJet> jets = sorted_by_pt(cs.inclusive_jets());

  // print out some info
  cout << "Clustered with " << jet_def.description() << endl;

  // print the jets
  cout <<   "        pt y phi" << endl;
  for (unsigned i = 0; i < jets.size(); i++) {
    cout << "jet " << i << ": "<< jets[i].perp() << " " 
                   << jets[i].rap() << " " << jets[i].phi() << endl;
    vector<PseudoJet> constituents = jets[i].constituents();
    for (unsigned j = 0; j < constituents.size(); j++) {
      cout << "    constituent " << j << "'s pt: "<< constituents[j].perp() << endl;
    }
  }
}
\end{lstlisting}

\item \rm Then compile and run it with
\begin{verbatim}
 g++ short-example.cc -o short-example \
     `fastjet-install/bin/fastjet-config --cxxflags --libs --plugins`
 ./short-example
\end{verbatim}
(watch out, once again, for the back-quotes if you cut and paste from the PDF).
\end{itemize}
\noindent
The output will consist of a banner, followed by the lines
\begin{verbatim}
Clustering with Longitudinally invariant anti-kt algorithm with R = 0.7 
and E scheme recombination
        pt y phi
jet 0: 103 0 0
    constituent 0's pt: 99.0001
    constituent 1's pt: 4.00125
jet 1: 99 0 3.14159
    constituent 0's pt: 99
\end{verbatim}

More evolved example programs, illustrating many of the capabilities of \fastjet,
are available in the \ttt{example/} subdirectory of the
\fastjet distribution.

\section{Core classes}
\label{sec:core-classes}

All classes are contained in the \ttt{fastjet} namespace. For brevity this namespace
will usually not be explicitly written below, with the possible exception of the first
appearance of a \fastjet class, and code excerpts will
assume that a ``\ttt{using namespace fastjet;}'' statement is present in the user
code.
For basic
usage, the user is exposed to three main classes:
\begin{lstlisting}
  class fastjet::PseudoJet;
  class fastjet::JetDefinition;
  class fastjet::ClusterSequence;
\end{lstlisting}
\ttt{PseudoJet} provides a jet object with a four-momentum and some
internal indices to situate it in the context of a jet-clustering
sequence. 
The class \ttt{JetDefinition} contains a specification of how
jet clustering is to be performed. 
\ttt{ClusterSequence} is the class that carries out
jet-clustering and provides access to the final jets.

\subsection{\tt fastjet::PseudoJet}
\label{sec:PseudoJet}

All jets, as well as input particles to the clustering (optionally)
are \ttt{PseudoJet} objects.  They can be created using one of the
following constructors
\begin{lstlisting}
  PseudoJet (double px, double py, double pz, double  E);
  template<class T> PseudoJet (const T & some_lorentz_vector);
\end{lstlisting}
where the second form allows the initialisation to be obtained from
any class \ttt{T} that allows subscripting to return the components of
the momentum (running from $0\ldots3$ in the order $p_x,p_y,p_z,E$).
The default constructor for a \PJ sets the momentum components to
zero.

The \ttt{PseudoJet} class includes the following member functions for
accessing the components
{
\begin{lstlisting}
  double E()        const ; // returns the energy component
  double e()        const ; // returns the energy component
  double px()       const ; // returns the x momentum component
  double py()       const ; // returns the y momentum component
  double pz()       const ; // returns the z momentum component
  double phi()      const ; // returns the azimuthal angle in range $0\ldots2\pi$
  double phi_std()  const ; // returns the azimuthal angle in range $-\pi\ldots\pi$
  double rap()      const ; // returns the rapidity
  double rapidity() const ; // returns the rapidity
  double pseudorapidity() const ; // returns the pseudo-rapidity
  double eta()      const ; // returns the pseudo-rapidity
  double pt2()      const ; // returns the squared transverse momentum
  double pt()       const ; // returns the transverse momentum
  double perp2()    const ; // returns the squared transverse momentum
  double perp()     const ; // returns the transverse momentum
  double m2()       const ; // returns squared invariant mass
  double m()        const ; // returns invariant mass ($-\sqrt{-m^2}$ if $m^2 < 0$)
  double mt2()      const ; // returns the squared transverse mass = $k_t^2+m^2$
  double mt()       const ; // returns the transverse mass
  double mperp2()   const ; // returns the squared transverse mass = $k_t^2+m^2$
  double mperp()    const ; // returns the transverse mass
  double operator[] (int i) const; // returns component i
  double operator() (int i) const; // returns component i

  /// return a valarray containing the four-momentum (components 0-2
  /// are 3-momentum, component 3 is energy).
  valarray<double> four_mom() const;
\end{lstlisting}} 
\noindent The reader may have observed that in some cases more than
one name can be used to access the same quantity. This is intended to
reflect the diversity of common usages within the
community.\footnote{The \texttt{pt()}, \texttt{pt2()}, \texttt{mt()},
  \texttt{mt2()} names are available only from version 3.0.1 onwards.}

There are two ways of associating user information with a
\ttt{PseudoJet}.
The simpler method is through an integer called the user index
\begin{lstlisting}
  /// set the user_index, intended to allow the user to label the object (default is -1)
  void set_user_index(const int index);
  /// return the user_index
  int user_index() const ;
\end{lstlisting}
A more powerful method, new in \fastjet 3, involves passing a pointer to a derived class
of \ttt{PseudoJet::UserInfoBase}. The two essential calls are
\begin{lstlisting}
  /// set the pointer to user information (the PseudoJet will then own it)
  void set_user_info(UserInfoBase * user_info);
  /// retrieve a reference to a dynamic cast of type L of the user info
  template<class L> const L & user_info() const;
\end{lstlisting}
Further details are to be found in appendix~\ref{app:user-info} and in
\ttt{example/09-user\_info.cc}.

A \verb:PseudoJet: can be reset with
\begin{lstlisting}
  /// Reset the 4-momentum according to the supplied components, put the user
  /// and history indices and user info back to their default values (-1, unset) 
  inline void reset(double px, double py, double pz, double E);
  /// Reset just the 4-momentum according to the supplied components,
  /// all other information is left unchanged
  inline void reset_momentum(double px, double py, double pz, double E);
\end{lstlisting}
and similarly taking as argument a templated
\verb:some_lorentz_vector: or a \verb:PseudoJet: (in the latter case,
or when \verb:some_lorentz_vector: is of a type derived from
\verb:PseudoJet:, \ttt{reset} also copies the user and internal indices and
user-info).

Additionally, the \ttt{+}, \ttt{-}, \ttt{*} and \ttt{/} operators are
defined, with \ttt{+}, \ttt{-} acting on pairs of \ttt{PseudoJet}s and
\ttt{*}, \ttt{/} acting on a \ttt{PseudoJet} and a \ttt{double}
coefficient.
The analogous \ttt{+=}, etc., operators, are also defined.\footnote{
  The \texttt{+}, \texttt{-} operators return a \texttt{PseudoJet}
  with default user information; the \texttt{*} and \texttt{/} operators
  return a \texttt{PseudoJet} with the same user information as the
  original \texttt{PseudoJet}; the 
  \texttt{+=}, \texttt{-=}, etc., operators all preserve the user
  information of the \texttt{PseudoJet} on the left-hand side of the operator.  }

There are also equality testing operators: \ttt{(jet1 == jet2)}
returns true if the two jets have identical 4-momenta, structural
information and user information;
the \ttt{(jet == 0.0)} test returns true if all the components of the
4-momentum are zero.
The \ttt{!=} operator works analogously.

Finally, we also provide routines for taking an unsorted vector of
\ttt{PseudoJet}s and returning a sorted vector,
\begin{lstlisting}
  /// return a vector of jets sorted into decreasing transverse momentum
  vector<PseudoJet> sorted_by_pt(const vector<PseudoJet> & jets);
  
  /// return a vector of jets sorted into increasing rapidity
  vector<PseudoJet> sorted_by_rapidity(const vector<PseudoJet> & jets);
  
  /// return a vector of jets sorted into decreasing energy
  vector<PseudoJet> sorted_by_E(const vector<PseudoJet> & jets);
\end{lstlisting}
These will typically be used on the jets returned by
\ttt{ClusterSequence}.

A number of further \PJ member functions provide access to information
on a jet's structure. They are documented below in
sections~\ref{sec:constituents} and \ref{sec:composite-jet}.

\subsection{\tt fastjet::JetDefinition}
\label{sec:JetDefinition}

The class \ttt{JetDefinition} contains a full specification
of how to carry out the clustering. According to the Les Houches convention 
detailed in~\cite{Buttar:2008jx}, a `jet definition' should include the 
jet algorithm name, its parameters (often the radius $R$) and the 
recombination scheme.
Its constructor is
\begin{lstlisting}
  JetDefinition(fastjet::JetAlgorithm jet_algorithm,
                double R,
                fastjet::RecombinationScheme recomb_scheme = E_scheme,
                fastjet::Strategy strategy = Best);
\end{lstlisting}
The jet algorithm is one of the entries of the \ttt{JetAlgorithm}
\ttt{enum}:\footnote{As of v2.3, the \ttt{JetAlgorithm} name replaces the old \ttt{JetFinder}
one, in keeping with the Les Houches convention. Backward compatibility is
assured at the user level by a typedef and a doubling of the methods' names.
Backward compatibility (with versions $<$ 2.3) is however broken for 
user-written derived classes of \ttt{ClusterSequence}, as the protected 
variables \ttt{\_default\_jet\_finder} and  \ttt{\_jet\_finder} have been 
replaced by \ttt{\_default\_jet\_algorithm} and \ttt{\_jet\_algorithm}.}
\begin{lstlisting}
  enum JetAlgorithm {kt_algorithm, cambridge_algorithm,
                     antikt_algorithm, genkt_algorithm,
                     ee_kt_algorithm, ee_genkt_algorithm, ...};
\end{lstlisting}
Each algorithm is described in detail in section~\ref{sec:native-algs}.
The $\ldots$ represent additional values that are present for 
internal or testing purposes. They include
\ttt{plugin\_algorithm}, automatically set when plugins are used
(section~\ref{sec:plugins}) and  \ttt{undefined\_jet\_algorithm}, 
which is the value set in \ttt{JetDefinition}'s default constructor.

The parameter \texttt{R} specifies the value of $R$ that appears in
eq.~(\ref{eq:dij-illustr}) and in the various definitions of
section~\ref{sec:native-algs}.
For one algorithm, \verb:ee_kt_algorithm:, there is no $R$ parameter,
so the constructor is to be called without the \verb:R: argument.
%
For the generalised $k_t$ algorithm and its $e^+e^-$ version, one
requires $R$ and (immediately after $R$) an extra
parameter $p$.
%
Details are to be found in sections~\ref{sec:genkt}--\ref{sec:kt-ee-alg}.
If the user calls a constructor with the incorrect number of arguments
for the requested jet algorithm, a \ttt{fastjet::Error()} exception
will be thrown with an explanatory message.

The recombination scheme is set by an \ttt{enum} of type
\ttt{RecombinationScheme}, and it is related to the choice of how to
recombine the 4-momenta of \ttt{PseudoJet}s during the clustering procedure.
The default in \fastjet is the $E$-scheme, where the four components of
two 4-vectors are simply added.
This scheme is used when no explicit choice is made in the
constructor. 
Further recombination schemes are described below in
section~\ref{sec:recomb_schemes}.

The strategy selects the algorithmic strategy to use while clustering
and is an \ttt{enum} of type \ttt{Strategy}. The default option of
\ttt{Best} automatically determines and selects a strategy that
should be close to optimal in speed for each event, based on its
multiplicity. 
A discussion of the main available strategies together with their
performance is given in appendix~\ref{app:strategies}.

A textual description of the jet definition can be obtained by a call
to the member function \ttt{std::string description()}.

\subsection{\tt fastjet::ClusterSequence}
\label{sec:ClusterSequence}

To run the jet clustering, create a \ttt{ClusterSequence}
object
through the following constructor
\begin{lstlisting}
  template<class L> ClusterSequence(const std::vector<L> & input_particles,
                                    const JetDefinition & jet_def); 
\end{lstlisting}
where \ttt{input\_particles} is the vector of initial particles of any
type (\ttt{PseudoJet}, \ttt{HepLorentzVector}, etc.) that can be
used to initialise a \ttt{PseudoJet} and \ttt{jet\_def} contains
the full specification of the clustering (see Section
\ref{sec:JetDefinition}).

\subsubsection{Accessing inclusive jets}

Inclusive jets correspond to all objects that have undergone a
``beam'' clustering (i.e. $d_{iB}$ recombination) in the description
following Eq.~(\ref{eq:diB-illustr}).
For nearly all hadron-collider algorithms, the ``inclusive'' jets
above some given transverse momentum cut are the ones usually just
referred to as the ``jets''.

To access inclusive jets, the following member function should be used
\begin{lstlisting}
  /// return a vector of all jets (in the sense of the inclusive
  /// algorithm) with pt >= ptmin. 
  vector<PseudoJet> inclusive_jets (const double & ptmin = 0.0) const;
\end{lstlisting}
where \ttt{ptmin} may be omitted, then implicitly taking the value
zero.
Note that the order in which the inclusive jets are provided depends
on the jet algorithm. To obtain a specific ordering, such as
decreasing $p_t$, the user should perform a sort themselves, e.g.\
with the \ttt{sorted\_by\_pt(...)} function, described in
section~\ref{sec:PseudoJet}. 

With a zero transverse momentum cut, the number of jets found in the
event is not an infrared safe quantity (adding a single soft particle
can lead to one extra soft jet).
However it can still be useful to talk of all the objects returned by
\ttt{inclusive\_jets()} as being ``jets'', e.g. in the context of the
estimation underlying-event and pileup densities, cf.\
section~\ref{sec:BackgroundEstimator}. 

\subsubsection{Accessing exclusive jets}
There are two ways of accessing exclusive jets,
one where one specifies
$d_{cut}$, the other where one specifies that the clustering is taken
to be stopped once it reaches the specified number of jets.
\begin{lstlisting}
  /// return a vector of all jets (in the sense of the exclusive algorithm) that would 
  /// be obtained when running the algorithm with the given dcut.
  vector<PseudoJet> exclusive_jets (const double & dcut) const;

  /// return a vector of all jets when the event is clustered (in the exclusive sense) 
  /// to exactly njets. Throws an error if the event has fewer than njets particles.
  vector<PseudoJet> exclusive_jets (const int & njets) const;

  /// return a vector of all jets when the event is clustered (in the exclusive sense) 
  /// to exactly njets. If the event has fewer than njets particles, it returns all
  /// available particles.
  vector<PseudoJet> exclusive_jets_up_to (const int & njets) const;
\end{lstlisting}
The user may also wish just to obtain information about the number of
jets in the exclusive algorithm:
\begin{lstlisting}
  /// return the number of jets (in the sense of the exclusive algorithm) that would 
  /// be obtained when running the algorithm with the given dcut.
  int n_exclusive_jets (const double & dcut) const;
\end{lstlisting}
Another common query is to establish the $d_{\min}$ value associated
with merging from $n+1$ to $n$ jets. Two member functions are
available for determining this:
\begin{lstlisting}
  /// return the dmin corresponding to the recombination that went from n+1 to n jets 
  /// (sometimes known as $d_{n,n+1}$).
  double exclusive_dmerge (const int & n) const;

  /// return the maximum of the dmin encountered during all recombinations up to the one 
  /// that led to an n-jet final state; identical to exclusive_dmerge, except in cases 
  /// where the dmin do not increase monotonically.
  double exclusive_dmerge_max (const int & n) const;
\end{lstlisting}
The first returns the $d_{\min}$  in going from $n+1$ to $n$ jets. 
Occasionally however the $d_{\min}$ value does not increase
monotonically during successive mergings and using a $d_{cut}$ smaller
than the return value from \ttt{exclusive\_dmerge} does not guarantee
an event with more than $n$ jets.
For this reason the second function \ttt{exclusive\_dmerge\_max} is
provided --- using a $d_{cut}$ below its return value is guaranteed to
provide a final state with more than $n$ jets, while using a larger
value will return a final state with $n$ or fewer jets.

For $\ee$ collisions, where it is usual to refer to $y_{ij} =
d_{ij}/Q^2$ ($Q$ is the total (visible) energy) \fastjet provides the
following methods:
\begin{lstlisting}
  double exclusive_ymerge (int njets);
  double exclusive_ymerge_max (int njets);
  int n_exclusive_jets_ycut (double ycut);
  std::vector<PseudoJet> exclusive_jets_ycut (double ycut);
\end{lstlisting}
which are relevant for use with the $\ee$ $k_t$ algorithm and with the
Jade plugin (section~\ref{sec:ee-jade}).

\subsubsection{Other functionality}

\paragraph{Unclustered particles.}  
Some jet algorithms (e.g.\ a number of the plugins in
section~\ref{sec:plugins}) have the property that not all particles
necessarily participate in the clustering.
In other cases, particles may take part in the clustering, but not end
up in any final inclusive jet.
Two member functions are provided to obtain the list of these
particles.
One is
\begin{lstlisting}
  vector<PseudoJet> unclustered = clust_seq.unclustered_particles();
\end{lstlisting}
which returns the list of particles that never took part in the clustering.
The other additionally returns objects that are the result of
clustering but that never made it into a inclusive jet (i.e.\ into a
``beam'' recombination):
\begin{lstlisting}
  vector<PseudoJet> childless = clust_seq.childless_pseudojets();
\end{lstlisting}
A practical example where this is relevant is with plugins that
perform pruning~\cite{Ellis:2009su}, since they include a condition for
vetoing certain recombinations.%
\footnote{To obtain the list of all initial particles that never end up in any
  inclusive jet, one should simply concatenate the vectors of
  constituents of all the childless \texttt{PseudoJet}s.}

\paragraph{Copying and transforming a ClusterSequence.}
A standard copy constructor is available for {\CS}s. 
Additionally it is possible to copy the clustering history of a \CS
while modifying the momenta of the \ttt{PseudoJet}s at all (initial,
intermediate, final) stages of the clustering, with the \CS member
function
\begin{lstlisting}
  void transfer_from_sequence(const ClusterSequence & original_cs, 
                              const FunctionOfPseudoJet<PseudoJet> * action_on_jets = 0);
\end{lstlisting}
\ttt{FunctionOfPseudoJet<PseudoJet>} is an abstract base class whose
interface provides a \ttt{PseudoJet operator()(const PseudoJet \& jet)}
function, i.e.\ a function of a \PJ that returns a \PJ (cf.\ 
appendix~\ref{app:function-of-pj}).
As the clustering history is copied to the target \CS, each \PJ in the
target \CS is set to the result of
\ttt{action\_on\_jet(original\_pseudojet)}. 
One use case for this is if one wishes to obtain a Lorentz-boosted
copy of a \CS, which can be achieved as follows
\begin{lstlisting}
  #include "fastjet/tools/Boost.hh"
  // ...
  ClusterSequence original_cs(...);
  ClusterSequence boosted_cs;
  Boost boost(boost_4momentum);
  boosted_cs.transfer_from_sequence(cs, &boost);
\end{lstlisting}

\subsection{Recombination schemes}
\label{sec:recomb_schemes}

When merging particles (i.e. \ttt{PseudoJet}s) during 
the clustering procedure, one must
specify how to combine the momenta. The simplest procedure
($E$-scheme) simply adds the four-vectors. 
This has been advocated as a standard in~\cite{RunII-jet-physics}, was
the main scheme in use during Run~II of the Tevatron, is currently
used by the LHC experiments, and is the default option in \fastjet.
Other choices are listed in
table~\ref{tab:RecombSchemes}, and are described below.

\begin{table}
  \centering
  \begin{tabular}{|l|}\hline
    \ttt{E\_scheme}     \\\hline
    \ttt{pt\_scheme}    \\\hline
    \ttt{pt2\_scheme}   \\\hline
    \ttt{Et\_scheme}    \\\hline
    \ttt{Et2\_scheme}   \\\hline
    \ttt{BIpt\_scheme}  \\\hline
    \ttt{BIpt2\_scheme} \\\hline
  \end{tabular}
  \caption{Members of the \ttt{RecombinationScheme} enum; the last two
    refer to boost-invariant version of the $p_t$ and $p_t^2$ schemes
    (as defined in section~\ref{sec:recomb_schemes}).}
  \label{tab:RecombSchemes}
\end{table}

\paragraph{Other schemes for $\boldsymbol{pp}$ collisions.} Other
schemes provided by earlier $k_t$-clustering implementations
\cite{KtClus,KtJet} are the $p_t$, $p_t^2$, $E_t$ and $E_t^2$ schemes. They
all incorporate a `preprocessing' stage to make the initial momenta
massless (rescaling the energy to be equal to the 3-momentum for the
$p_t$ and $p_t^2$ schemes, rescaling to the 3-momentum to be equal to
the energy in the $E_t$ and $E_t^2$ schemes). Then for all schemes the
recombination $p_r$ of $p_i$ and $p_j$ is a massless 4-vector
satisfying
\begin{subequations}
  \begin{align}
    p_{t,r} &= p_{t,i} + p_{t,j}\,,\\
    \phi_r &= (w_i \phi_i + w_j \phi_j)/(w_i + w_j)\,,\\
    y_r &= (w_i y_i + w_j y_j)/(w_i + w_j)\,,
  \end{align}
\end{subequations}
where $w_i$ is $p_{t,i}$ for the $p_t$ and $E_t$ schemes, and is
$p_{t,i}^2$ for the $p_t^2$ and $E_t^2$ schemes. 

Note that for massive particles the schemes defined in the previous
paragraph are not invariant under longitudinal boosts. 
As a workaround for this issue, we propose boost-invariant $p_t$ and
$p_t^2$ schemes, which are identical to the normal $p_t$ and $p_t^2$
schemes, except that they omit the preprocessing stage. They are
available as \ttt{BIpt\_scheme} and \ttt{BIpt2\_scheme}.

\paragraph{Other schemes for $\boldsymbol{e^+e^-}$ collisions.}
On request, we may in the future provide dedicated schemes for $\ee$
collisions.

\paragraph{User-defined schemes.} The user may define their own,
custom recombination schemes, as described in Appendix~\ref{sec:recombiner}.

\subsection{Constituents and substructure of jets}
\label{sec:constituents}

For any \PseudoJet that results from a clustering, the user can
obtain information about its constituents, internal substructure,
etc., through  member functions of the \PseudoJet
class.\footnote{This is a new development in version 3 of
  \fastjet. In earlier versions, access to information about a jet's
  contents had to be made through its \CS. Those forms of access,
  though not documented here, will be retained throughout the 3.X
  series.}

\paragraph{Jet constituents.}
The constituents of a given \PseudoJet \verb|jet| can be
obtained through
\begin{lstlisting}
  vector<PseudoJet> constituents = jet.constituents();
\end{lstlisting}
Note that if the user wishes to identify these constituents with the
original particles provided to \ttt{ClusterSequence}, she or
he should have set a unique index for each of the original particles
with the \ttt{PseudoJet::set\_user\_index} function.
Alternatively more detailed information can also be set through the
\verb|user_info| facilities of \PseudoJet, as discussed in
appendix~\ref{app:user-info}.

\paragraph{Subjet properties.} To obtain the set of subjets at a specific
$d_{\mathrm{cut}}$ scale inside a given jet, one may use the following
\PseudoJet member function:
\begin{lstlisting}
  /// Returns a vector of all subjets of the current jet (in the sense of the exclusive 
  /// algorithm) that would be obtained when running the algorithm with the given dcut
  std::vector<PseudoJet> exclusive_subjets (const double & dcut) const;
\end{lstlisting}
If $m$ jets are found, this takes a time $\order{m \ln m}$ (owing to
the internal use of a priority queue). Alternatively one may obtain
the jet's constituents, cluster them separately and then carry out an
\ttt{exclusive\_jets} analysis on the resulting \ttt{ClusterSequence}.
The results should be identical. This second method is mandatory if
one wishes to identify subjets with an algorithm that differs from the
one used to find the original jets.

In analogy with the \ttt{exclusive\_jets(...)} functions of \CS, \PJ
also has \ttt{exclusive\_subjets(int nsub)} and
\ttt{exclusive\_subjets\_up\_to(int nsub)} functions.

One can also make use of the following methods, which allow one to
follow the merging sequence (and walk back through it):
\begin{lstlisting}
  /// If the jet has parents in the clustering, returns true and sets parent1 and parent2 
  /// equal to them. If it has no parents returns false and sets parent1 and parent2 to 0
  bool has_parents(PseudoJet & parent1, PseudoJet & parent2) const;

  /// If the jet has a child then returns true and sets the child jet otherwise returns 
  /// false and sets the child to 0
  bool has_child(PseudoJet & child) const;

  /// If this jet has a child (and so a partner), returns true and sets the partner, 
  /// otherwise returns false and sets the partner to 0
  bool has_partner(PseudoJet & partner) const;
\end{lstlisting}

\paragraph{Accessibility of structural information.}
If any of the above functions are used with a \PseudoJet that is not
associated with a \ClusterSequence, an error will be thrown. 
Since the information about a jet's constituents is actually stored in
the \CS and not in the jet itself, these methods will also throw an
error if the \CS associated with the jet has gone out of scope, been
deleted, or in any other way become invalid.
One can establish the status of a \PseudoJet's associated cluster
sequence with the following enquiry functions:
\begin{lstlisting}
  // returns true if this PseudoJet has an associated (and valid) ClusterSequence.
  bool has_valid_cluster_sequence() const;

  // returns a (const) pointer to the parent ClusterSequence (throws if inexistent 
  // or no longer valid) 
  const ClusterSequence* validated_cluster_sequence() const; 
\end{lstlisting}
There are also \ttt{has\_associated\_cluster\_sequence()} and
\ttt{associated\_cluster\_sequence()} member functions. The first
returns true even when the cluster sequence is not valid, and the
second returns a null pointer in that case.
Further information is to be found in
appendix~\ref{app:structure_table}.

There are contexts in which, within some function, one might create a
\ClusterSequence, obtain a jet from it and then return that jet from
the function. For the user to be able to access the information about
the jet's internal structure, the \ClusterSequence must not have gone
out of scope and/or been deleted.
To aid with potential memory management issues in this case, as long
the \ClusterSequence was created via a \ttt{new} operation,
then one can tell the \ClusterSequence that it should be automatically
deleted after the last external object (jet, etc.)  associated with it
has disappeared. 
The call to do this is
\ttt{ClusterSequence::delete\_self\_when\_unused()}. There must be at
least one external object already associated with the \ClusterSequence
at the time of the call (e.g. a jet, subjet or jet constituent).
Note that \ClusterSequence tends to be a large object, so this should be
used with care.

\subsection{Composite jets, general considerations on jet structure}
\label{sec:composite-jet}

There are a number of cases where it is useful to be able to take two
separate jets and create a single object that is the sum of the two,
not just from the point of view of its 4-momentum, but also as
concerns its structure.
For example, in a search for a dijet resonance, some user code may
identify two jets, \ttt{jet1} and \ttt{jet2}, that are thought to come
from a resonance decay and then wish to return a single object that
combines both \ttt{jet1} and \ttt{jet2}.
This can be accomplished with the function \ttt{join}:
\begin{lstlisting}
  PseudoJet resonance = join(jet1,jet2); 
\end{lstlisting}
The 4-momenta are added,\footnote{This corresponds to $E$-scheme
  recombination. If the user wishes to have the jets joined with a
  different recombination scheme he/she can pass a
  \texttt{JetDefinition::Recombiner} (cf.\ section~\ref{sec:recombiner})
  as the last argument to \texttt{join(...)}.} and in addition the
\ttt{resonance} remembers that it came from \ttt{jet1} and
\ttt{jet2}. So, for example, a call to \ttt{resonance.constituents()}
will return the constituents of both \ttt{jet1} and \ttt{jet2}.
It is possible to \ttt{join} 1, 2, 3 or 4 jets or a \ttt{vector} of
jets.
If the jets being joined had areas (section~\ref{sec:areas}) then the
joined jet will also have an area.

For a jet formed with \ttt{join}, one can find out which pieces it has
been composed from with the function
\begin{lstlisting}
  vector<PseudoJet> pieces = resonance.pieces();
\end{lstlisting}
In the above example, this would simply return a vector of size 2
containing \ttt{jet1} and \ttt{jet2}.
The \ttt{pieces()} function also works for jets that come from a \CS,
returning two pieces if the jet has parents, zero otherwise.

\paragraph{Enquiries as to available structural information.}

Whether or not a given jet has constituents, recursive substructure or
pieces depends on how it was formed. 
Generally a user will know how a given jet was formed, and so know if
it makes sense, say, to call \ttt{pieces()}.  
However if a jet is returned from some third-party code, it may not
always be clear what kinds of structural information it has. 
Accordingly a number of enquiry functions are available:
\begin{lstlisting}
  bool has_structure();         // true if the jet has some kind of structural info
  bool has_constituents();      // true if the jet has constituents 
  bool has_exclusive_subjets(); // true if there is cluster-sequence style subjet info
  bool has_pieces();            // true if the jet can be broken up into pieces
  bool has_area();              // true if the jet has jet-area information
  string description();         // returns a textual description of the type
                                // of structural info associated with the jet
\end{lstlisting}
Asking (say) for the \ttt{pieces()} of a jet for which
\ttt{has\_pieces()} returns false will cause an error to be thrown.
The structural information available for different kinds of jets is
summarised in appendix~\ref{app:structure_table}.

\subsection{Version information}
\label{sec:version-information}

Information on the version of \fastjet that is being run can be
obtained by making a call to 
\begin{lstlisting}
  std::string fastjet_version_string();
\end{lstlisting}
(defined in \ttt{fastjet/JetDefinition.hh}).  
In line with recommendations for other programs in high-energy
physics, the user should include this information in publications and
plots so as to facilitate reproducibility of the
jet-finding.\footnote{We devote significant effort to ensuring that
  all versions of the \fastjet\ program give identical, correct
  clustering results, and that any other changes from one version to
  the next are clearly indicated.
  However, as per the terms of the GNU General Public License (v2),
  under which \fastjet\ is released, we are not able to provide
  a warranty that \fastjet\ is free of bugs that might affect your use
  of the program.
  Accordingly it is important for physics publications to include a
  mention of the \fastjet version number used, in order to help trace
  the impact of any bugs that might be discovered in the future.  }
We recommend also that the main elements of the
\ttt{jet\_def.description()} be provided, together with citations to
the original article that defines the algorithm, as well as to this
manual and, optionally, the original \fastjet paper~\cite{fastjet}.

\section{\fastjet native jet algorithms}
\label{sec:native-algs}

\subsection[Longitudinally invariant $k_t$ jet algorithm]{Longitudinally invariant $\boldsymbol{k_t}$ jet algorithm}
The longitudinally invariant $k_t$ jet algorithm \cite{ktexcl,ktincl}
comes in inclusive and exclusive variants.
The inclusive variant (corresponding to \cite{ktincl}, modulo small
changes of notation) is formulated as follows:
\begin{itemize}
\item[1.] For each pair of particles $i$, $j$ work out the $k_t$
  distance\footnote{In the soft, small angle limit for $i$, the $k_t$
    distance is the (squared) transverse momentum of $i$ relative to $j$.}
  \begin{equation}
    \label{eq:dij}
    d_{ij} = \min(p_{ti}^2,{p_{tj}^2}) \, \Delta R_{ij}^2 / R^2
  \end{equation}
  with $\Delta R_{ij}^2 = (y_i-y_j)^2 + (\phi_i-\phi_j)^2$,
  where $p_{ti}$, $y_i$ and $\phi_i$ are the transverse momentum (with
  respect to the beam direction),
  rapidity and azimuth of particle $i$. $R$ is a jet-radius
  parameter usually taken of order $1$. For each parton $i$ also work
  out the beam distance $d_{iB} = p_{ti}^2$.
\item[2.] Find the minimum $d_{\min}$ of all the $d_{ij},d_{iB}$. If
  $d_{\min}$ is a $d_{ij}$ merge particles $i$ and $j$ into a single
  particle, summing their four-momenta (this is $E$-scheme
  recombination); if it is a $d_{iB}$ then declare particle $i$ to be
  a final jet and remove it from the list.
\item[3.] Repeat from step 1 until no particles are left.
\end{itemize}
The exclusive variant of the longitudinally invariant $k_t$ jet
algorithm \cite{ktexcl} is similar, except that (a) when a $d_{iB}$ is
the smallest value, that particle is considered to become part of the
beam jet (i.e.\ is discarded) and (b) clustering is stopped when all
$d_{ij}$ and $d_{iB}$ are above some $d_{cut}$. In the exclusive mode
$R$ is commonly set to $1$.

The inclusive and exclusive variants are both obtained through 
\begin{lstlisting}
   JetDefinition jet_def(kt_algorithm, R);
   ClusterSequence cs(particles, jet_def);
\end{lstlisting}
The clustering sequence is identical in the inclusive and exclusive
cases and the jets can then be obtained as follows:
\begin{lstlisting}
   vector<PseudoJet> inclusive_kt_jets = cs.inclusive_jets();
   vector<PseudoJet> exclusive_kt_jets = cs.exclusive_jets(dcut);
\end{lstlisting}

\subsection{Cambridge/Aachen jet algorithm}

Currently the $pp$ Cambridge/Aachen (C/A) jet
algorithm~\cite{CamOrig,CamWobisch} is provided only in an inclusive
version~\cite{CamWobisch}, whose formulation is identical to that of
the $k_t$ jet algorithm, except as regards the distance measures,
which are:
\begin{subequations}
  \label{eq:dij_cam}
  \begin{align}
    d_{ij} &= \Delta R_{ij}^2 / R^2\,,\\
    d_{iB} &= 1\,.
  \end{align}
\end{subequations}
To use this algorithm, define 
\begin{lstlisting}
   JetDefinition jet_def(cambridge_algorithm, R);
\end{lstlisting}
and then extract inclusive jets from the cluster sequence.

Attempting to extract exclusive jets from the Cambridge/Aachen
algorithm with a
$d_{cut}$ parameter simply provides the set of jets obtained up to the
point where all $d_{ij},d_{iB} > d_{cut}$. Having clustered with some
given $R$, this can actually be an effective way of viewing the event
at a smaller radius, $R_{eff} = \sqrt{d_{cut}} R$, thus allowing a
single event to be viewed at a continuous range of $R_{eff}$ within a
single clustering.

We note that the true exclusive formulation of the Cambridge
algorithm~\cite{CamOrig} (in $e^+e^-$) instead makes use an auxiliary ($k_t$)
distance measure and `freezes' pseudojets whose recombination would
involve too large a value of the auxiliary distance measure. Details
are given in section~\ref{sec:ee-cam}.

\subsection[Anti-$k_t$ jet algorithm]{Anti-$\boldsymbol{k_t}$ jet algorithm}
This algorithm, introduced and studied in~\cite{antikt}, is defined
exactly like the standard $k_t$ algorithm, except for the distance
measures which are now given by
\begin{subequations}
    \label{eq:dij_antikt}
  \begin{eqnarray}
    &&d_{ij} = \min(1/p_{ti}^2,1/{p_{tj}^2}) \, \Delta R_{ij}^2 / R^2 \, , \\
    &&d_{iB} = 1/p_{ti}^2 \, .
  \end{eqnarray}
\end{subequations}
While it is a sequential recombination algorithm like $k_t$ and
Cambridge/Aachen, the anti-$k_t$ algorithm behaves in some sense like a
`perfect' cone algorithm, in that its hard jets are exactly
circular on the $y$-$\phi$ cylinder~\cite{antikt}.
To use this algorithm, define 
\begin{lstlisting}
   JetDefinition jet_def(antikt_algorithm, R);
\end{lstlisting}
and then extract inclusive jets from the cluster sequence.
We advise against the use of exclusive jets in the context of the
anti-$k_t$ algorithm, because of the lack of physically meaningful
hierarchy in the clustering sequence.

\subsection[Generalised-$k_t$ jet algorithm]{Generalised $\boldsymbol{k_t}$ jet algorithm}
\label{sec:genkt}

The ``generalised $k_t$'' algorithm again follows the same procedure,
but depends on an additional continuous parameter $p$, and has the
following distance measure:
\begin{subequations}
  \label{eq:dij_genkt}
  \begin{eqnarray}
    &&d_{ij} = \min(p_{ti}^{2p},p_{tj}^{2p}) \, \Delta R_{ij}^2 / R^2 \, , \\
    &&d_{iB} = p_{ti}^{2p} \, .
  \end{eqnarray}
\end{subequations}
For specific values of $p$, it reduces to one or other of the
algorithms list above, $k_t$ ($p=1$), Cambridge/Aachen ($p=0$) and
anti-$k_t$ ($p=-1$). 
To use this algorithm, define 
\begin{lstlisting}
  JetDefinition jet_def(genkt_algorithm, R, p);
\end{lstlisting}
and then extract inclusive jets from the cluster sequence (or, for
$p\ge 0$, also the exclusive jets).

\subsection[Generalised $k_t$ algorithm for $e^+e^-$ collisions]
{Generalised $\boldsymbol{k_t}$ algorithm for $\boldsymbol{e^+e^-}$ collisions}
\label{sec:ee-algs}

\fastjet also provides native implementations of clustering algorithms
in spherical coordinates (specifically for $e^+e^-$ collisions) along
the lines of the original $k_t$ algorithms~\cite{eekt}, but extended
following  the generalised $pp$ algorithm of~\cite{antikt} and
section \ref{sec:genkt}. We define the two following distances:
\begin{subequations}
  \label{eq:dij_eegenkt}
\begin{align}
  \label{eq:dij_eegenkt_ij}
  d_{ij} &= \min(E_i^{2p}, E_j^{2p}) \frac{(1- \cos
    \theta_{ij})}{(1-\cos R)}\,,\\
  d_{iB} &= E_i^{2p}\,,
\end{align}
\end{subequations}
for a general value of $p$ and $R$. At a given stage of the clustering
sequence, if a $d_{ij}$ is smallest then $i$ and $j$ are recombined,
while if a $d_{iB}$ is smallest then $i$ is called an ``inclusive
jet''. 

For values of $R \le \pi$ in eq.~(\ref{eq:dij_eegenkt}), the
generalised $e^+e^-$ $k_t$ algorithm behaves in analogy with the $pp$
algorithms: when an object is at an angle $\theta_{iX} > R$ from all
other objects $X$ then it forms an inclusive jet.
With the choice $p=-1$ this provides a simple, infrared and collinear
safe way of obtaining a cone-like algorithm for $e^+e^-$ collisions,
since hard well-separated jets have a circular profile on the $3$D
sphere, with opening half-angle $R$.
To use this form of the algorithm, define 
\begin{lstlisting}
   JetDefinition jet_def(ee_genkt_algorithm, R, p);
\end{lstlisting}
and then extract inclusive jets from the cluster sequence.

For values of $R > \pi$, \fastjet replaces the factor $(1-\cos R)$ in
the denominator of eq.~(\ref{eq:dij_eegenkt_ij}) with $(3+\cos
R)$. With this choice (as long as $R < 3\pi$), the only time a $d_{iB}$ will
be relevant is when there is just a single particle in the event.
The \ttt{inclusive\_jets(...)} will then always return a single jet
consisting of all the particles in the event. 
In such a context it is only the \ttt{exclusive\_jets(...)} call that
provides non-trivial information.

\subsection[$k_t$ algorithm for $e^+e^-$ collisions]
{$\boldsymbol{k_t}$ algorithm for $\boldsymbol{e^+e^-}$ collisions}
\label{sec:kt-ee-alg}

The $e^+e^-$ $k_t$ algorithm~\cite{eekt}, often referred to also as
the Durham algorithm,  has a single distance:
\begin{align}
  \label{eq:dij_eekt}
  d_{ij} &= 2 \min(E_i^{2}, E_j^{2}) (1- \cos \theta_{ij})\,.
\end{align}
Note the difference in normalisation between the $d_{ij}$ in
eqs.~(\ref{eq:dij_eegenkt}) and (\ref{eq:dij_eekt}), and the fact that in neither
case have we normalised to the total energy $Q$ in the event, contrary
to the convention adopted originally in~\cite{eekt} (where the
distance measure was called $y_{ij}$).
To use the $e^+e^-$ $k_t$  algorithm, define 
\begin{lstlisting}
   JetDefinition jet_def(ee_kt_algorithm);
\end{lstlisting}
and then extract exclusive jets from the cluster sequence.

Note that the \ttt{ee\_genkt\_algorithm} with $ \pi < R < 3\pi$ and
$p=1$ gives a clustering sequence that is identical to that of the
\ttt{ee\_kt\_algorithm}.
The normalisation of the $d_{ij}$'s will however be different.

\section{Plugin jet algorithms}
\label{sec:plugins}

It can be useful to have a common interface for a range of jet
algorithms beyond the native ($k_t$, anti-$k_t$ and Cambridge/Aachen)
algorithms, notably for the many cone algorithms that are in
existence. 
It can also be useful to be able to use \fastjet features such as
area-measurement tools for these other jet algorithms.
In order to facilitate this, the
\fastjet package provides a \emph{plugin} facility, allowing almost
any other jet algorithm\footnote{Except those for which one particle
  may be assigned to more than one jet, e.g.\ algorithms such as
  ARCLUS~\cite{Lonnblad:1992qd}, which performs $3\to2$
  clustering.} to be used within the same framework.

Generic plugin use is described in the next subsection.
The plugins distributed with \fastjet are described afterwards in
sections~\ref{sec:siscone-plugin}--\ref{sec:other-ee-plugins}.
They are all accessible within the \ttt{fastjet} namespace and their code
is to be found in \fastjet's \ttt{plugins/} directory.
New user-defined plugins can also be implemented, as described 
in section~\ref{sec:new-plugin}. 
Some third-party plugins are linked to from the tools page at
\url{http://fastjet.fr/}$\,$.

Not all plugins are enabled by default in \fastjet. At configuration
time \verb:./configure --help: will indicate which ones get enabled
by default. To enable all plugins, run \verb|configure| with the
argument \verb|--enable-allplugins|, while to enable all but PxCone
(which requires a Fortran compiler, and can introduce link-time issues) use
\verb|--enable-allcxxplugins|.

\subsection{Generic plugin use}
\label{sec:generic-plugin-use}

Plugins are classes derived from the abstract base class
\ttt{fastjet::JetDefinition::Plugin}.
A \ttt{JetDefinition} can be constructed by providing a pointer to a
\ttt{JetDefinition::Plugin}; the resulting \ttt{JetDefinition} object
can then be used identically to the normal \ttt{JetDefinition} objects
used in the previous sections.
We illustrate this with an example based on the SISCone plugin:
\begin{lstlisting}
  #include "fastjet/SISConePlugin.hh"

  // allocate a new plugin for SISCone (for other plugins, simply
  // replace the appropriate parameters and plugin name)
  double cone_radius = 0.7;
  double overlap_threshold = 0.75;
  JetDefinition::Plugin * plugin = new SISConePlugin(cone_radius, overlap_threshold);

  // create a jet definition based on the plugin
  JetDefinition jet_def(plugin);

  // run the jet algorithm and extract the jets
  ClusterSequence clust_seq(particles, jet_def);
  vector<PseudoJet> inclusive_jets = clust_seq.inclusive_jets();

  // then analyse the jets as for native fastjet algorithms
  ...

  // only when you will no longer be using the jet definition, or
  // ClusterSequence objects that involve it, may you delete the plugin
  delete plugin;
\end{lstlisting}
In cases such as this where the plugin has been created with a
\ttt{new} statement and the user does not wish to manage the deletion
of the corresponding memory when the \ttt{JetDefinition} (and any
copies) using the plugin goes out of scope, then the user may wish to
call the \ttt{JetDefinition}'s \ttt{delete\_plugin\_when\_unused()}
function, which tells the \ttt{JetDefinition} to acquire ownership of
the pointer to the plugin and delete it when it is no longer needed.

\subsection{SISCone Plugin}
\label{sec:siscone-plugin}

SISCone~\cite{SISCone} is an implementation of a stable-cone jet
algorithm with a split--merge step (SC-SM in the notation
of~\cite{Salam:2009jx}).
As with most modern cone algorithms, it is divided into two parts:
first it searches for stable cones; then, because a particle can
appear in more than one stable cone, a `split--merge' procedure is
applied, which ensures that no particle ends up in more than one jet.
The stable cones are identified using an $\order{N^2 \ln N}$ seedless
approach. This (and some care in the the `split--merge' procedure)
ensures that the jets it produces are insensitive to additional soft
particles and collinear splittings, i.e.\ the algorithm is infrared
and collinear safe.

The plugin library and include files are to be be found in the
\verb:plugins/SISCone: directory, and the main header file is
\verb:SISConePlugin.hh:. The \verb:SISConePlugin: class has a
constructor with the following structure
\begin{lstlisting}
  SISConePlugin (double cone_radius,
                 double overlap_threshold,
                 int    n_pass_max = 0,
                 double protojet_ptmin = 0.0, 
                 bool   caching = false,
                 SISConePlugin::SplitMergeScale   
                                split_merge_scale = SISConePlugin::SM_pttilde);
\end{lstlisting}
A cone centred at $y_c,\phi_c$ is stable if the sum of momenta of all
particles $i$ satisfying $\Delta y_{ic}^2 + \Delta \phi_{ic}^2 <
\verb:cone_radius:^2$ has rapidity $y_c,\phi_c$.
The \verb:overlap_threshold: is the fraction of overlapping momentum
above which two protojets are merged in a Tevatron Run~II type
\cite{RunII-jet-physics} split--merge procedure.
The
radius and overlap parameters are a common feature of most modern cone
algorithms. Because some event particles are not to be found in any
stable cone \cite{EHT}, SISCone can carry out multiple stable-cone
search passes (as advocated in \cite{TeV4LHC}): in each pass one
searches for stable cones using just the subset of particles not
present in any stable cone in earlier passes. Up to \verb:n_pass_max:
passes are carried out, and the algorithm automatically stops at the
highest pass that gives no new stable cones. The default of
$\verb:n_pass_max: = 0$ is equivalent to setting it to $\infty$.

Concern had at some point been expressed that an excessive number of
stable cones might complicate cone jets in events with high
noise~\cite{RunII-jet-physics}, and in particular lead to large
``monster'' jets.
The \verb:protojet_ptmin: parameter allows one to use only protojets
with $p_t \ge \verb:protojet_ptmin:$ in the split--merge phase (all
others are thrown away), so as to limit this issue.
A large value of the split--merge overlap threshold, e.g.\ $0.75$,
also helps to disfavour the production of these monster jets through
repeated merge operations.

In many cases SISCone's most time-consuming step is the search for
stable cones. If one has multiple \verb:SISConePlugin:-based jet
definitions, each with \verb:caching=true:, a check will be carried
out whether the previously clustered event had the same set of
particles and the same cone radius and number of passes. If it did,
the stable cones are simply reused from the previous event, rather
than being recalculated, and only the split--merge step is repeated,
often leading to considerable speed gains.

A final comment concerns the \verb:split_merge_scale:
parameter. This controls both the scale used for ordering the
protojets during the split--merge step during the split--merge step,
and also the scale used to measure the degree of overlap between
protojets. While various options have been implemented, 
\begin{lstlisting}
  enum SplitMergeScale {SM_pt, SM_Et, SM_mt, SM_pttilde};
\end{lstlisting}
we recommend using only the last of them $\tilde p_t = \sum_{i \in
  \mathrm{jet}}|p_{t,i}|$, which is also the default scale. The other
scales are included only for historical comparison purposes: $p_t$
(used in several other codes) is IR unsafe for events whose hadronic
component conserves momentum, $E_t$ (advocated in
\cite{RunII-jet-physics}) is not boost-invariant, and $m_t = \sqrt{m^2
  + p_t^2}$ is IR unsafe for events whose hadronic component conserves
momentum and stems from the decay of two identical particles.

An example of the use of the SISCone plugin was given in
section~\ref{sec:generic-plugin-use}. 
As can be seen there, SISCone jets are to be obtained by requesting
inclusive jets from the cluster sequence.
Note that it makes no sense to ask for exclusive jets from a
SISCone based \verb:ClusterSequence:.

%
%
%
%
%

Some cone algorithms provide information beyond that simply about the
jets.
Such additional information, where available, can be accessed with the
help of the \ttt{ClusterSequence::extras()} function.
In the case of SISCone, one can access that information as follows:
\begin{lstlisting}
  const fastjet::SISConeExtras * extras = 
            dynamic_cast<const fastjet::SISConeExtras *>(clust_seq.extras());
\end{lstlisting}
To determine the pass at which a given jet was found, one then
uses\footnote{ In versions of \fastjet prior to 3.0.1, a jet's user
  index indicated the pass at which it had been found. The
  value was however incorrectly set for single-particle
  jets. The current choice is to leave the user index unchanged from
  its default.}
\begin{lstlisting}
  int pass = extras->pass(jet);
\end{lstlisting}
One may also obtain a list of the positions of original stable
cones as follows:
\begin{lstlisting}
  const vector<PseudoJet> & stable_cones = extras->stable_cones();
\end{lstlisting}
The stable cones are represented as \ttt{PseudoJet}s, for which only the
rapidity and azimuth are meaningful. The \verb:user_index(): indicates
the pass at which a given stable cone was found.

SISCone uses $E$-scheme recombination internally and also for
constructing the final jets from the list of constituents. 
For the latter task, the user may instead instruct SISCone to use the
jet-definition's own recombiner, with the command
\begin{lstlisting}
  plugin->set_use_jet_def_recombiner(true);
\end{lstlisting}
For this to work, \ttt{plugin} must explicitly be of type
\ttt{SISConePlugin *} rather than \ttt{JetDefinition::Plugin *}. 

Since SISCone is infrared safe, it may meaningfully be used also with
the \verb:ClusterSequenceArea: class. Note however that in that
case ones loses the cone-specific information from the jets, because
of the way \fastjet filters out the information relating to ghosts in
the clustering. If the user needs both areas and cone-specific
information, she/he may use the
\verb:ClusterSequenceActiveAreaExplicitGhosts: class (for usage
information, see the corresponding \verb:.hh: file). 

A final remark concerns speed and memory requirements: as mentioned
above, SISCone takes $\order{N^2 \ln N}$ time to find jets, and the
memory use is $\order{N^2}$; taking $N=10^3$ as a reference point, it
runs in a few tenths of a second, making it about 100 times slower
than native \fastjet algorithms.
These are `expected' results, i.e.\ valid for a suitably random set of
particles.\footnote{%
  In area determinations, the ghost particles are not entirely random,
  but distributed close to a grid pattern, all with similar transverse
  momenta.
  Run times and memory usage are then, in practice, somewhat larger
  than for a normal QCD event with the same number of particles.  
  We therefore recommend running with not too small a
  \texttt{ghost\_area} (e.g.\ $\sim 0.05$) and using
  \texttt{grid\_scatter=1} (cf.\ section~\ref{sec:areas}), which helps
  to reduce the number of stable cones (and correspondingly, the time
  and memory usage of the subsequent split--merge step). 
  An alternative, which has been found to be acceptable in many
  situations, is to use a passive area, since this is relatively fast
  to calculate with SISCone.
}

Note that the underlying implementation of SISCone is optimised for
large $N$.
An alternative implementation that is faster for $N \lesssim 10$ was
presented in~\cite{Weinzierl:2011jx}. 

\subsection{Other plugins for hadron colliders }
\label{sec:other-pp-plugins}

Most of the algorithms listed below are cone algorithms. 
They are usually either infrared (IR) or collinear unsafe.
The details are indicated for each algorithm following the
notation of Ref.~\cite{Salam:2009jx}: IR$_{n+1}$ means 
that the hard jets may be modified if, to an ensemble of $n$ hard
particles in a common neighbourhood, one adds a single soft particle;
Coll$_{n+1}$ means that for $n$ hard particles in a common
neighbourhood, the collinear splitting of one of them may modify the
hard jets.
The \fastjet authors (and numerous theory-experiment accords) advise
against the use of IR and/or collinear unsafe jet algorithms.  Interfaces to
these algorithms have been provided mainly for legacy comparison
purposes.

Except where stated, the usual way to access jets from these plugins
is through \ttt{ClusterSequence::inclusive\_jets()}.

\subsubsection{CDF Midpoint} 
This is one of the two cone algorithms used by CDF in Run~II of the Tevatron, based
on \cite{RunII-jet-physics}. It is a midpoint-type iterative cone with
a split--merge step.
\begin{lstlisting}
  #include "fastjet/CDFCones.hh"
  // ...
  CDFMidPointPlugin(double R, 
                    double overlap_threshold,
                    double seed_threshold = 1.0, 
                    double cone_area_fraction = 1.0);
\end{lstlisting}
The overlap threshold ($f$) used by CDF is usually $0.5$, the seed
threshold is $1$\,GeV. 
With a cone area fraction $\alpha < 1$, the search for stable
cones is performed with a radius that is $R \times
\sqrt{\alpha}$, i.e.\ it becomes the searchcone
algorithm of \cite{EHT}.
CDF has used both $\alpha = 0.25$ and $\alpha = 1.0$.
It is our understanding that the particular choice of $\alpha$ is not
always clearly documented in the corresponding publications.

Further control over the plugin can be obtained by consulting the
header file.

The original underlying code for this algorithm was provided on a
webpage belonging to Joey Huston~\cite{CDFCones} (with minor
modifications to ensure reasonable run times with optimising compilers
for 32-bit Intel processors --- these modifications do not affect the
final jets).

This algorithm is IR$_{3+1}$ unsafe in the limit of zero seed
threshold~\cite{SISCone} (with $\alpha \neq 1$ it
becomes IR$_{2+1}$ unsafe~\cite{TeV4LHC}).
With a non-zero seed threshold (and no preclustering into calorimeter
towers) it is collinear unsafe.
It is to be
deprecated for new experimental or theoretical analyses.

\subsubsection{CDF JetClu}

JetClu is the other cone algorithm used by CDF during Run~II, as well
as their main algorithm during Run~I~\cite{Abe:1991ui}. 
It is an iterative cone with split-merge and optional ``ratcheting''
if \ttt{iratch == 1} (particles that appear in one iteration of a cone
are retained in future iterations).
It can be obtained as follows:
\begin{lstlisting}
  #include "fastjet/CDFCones.hh"
  // ...
  CDFJetCluPlugin (double   cone_radius, 
		   double   overlap_threshold, 
		   double   seed_threshold = 1.0,
		   int      iratch = 1);
\end{lstlisting}
The overlap threshold is usually set to 0.75 in CDF analyses.
Further control over the plugin can be obtained by consulting the
header file.

The original underlying code for this algorithm was provided on a
webpage belonging to Joey Huston~\cite{CDFCones}.

This algorithm is IR$_{2+1}$ unsafe (for zero seed threshold; some
IR unsafety persists with non-zero seed threshold). 
It is to be
deprecated for new experimental or theoretical analyses.
Note also that the underlying implementation groups particles together
into calorimeter towers, with CDF-type geometry, before running the
jet algorithm. 

\subsubsection{\Dzero Run I cone}

This is the main algorithm used by \Dzero in Run~I of the
Tevatron~\cite{Abbott:1997fc}. It is an iterative cone algorithm
with a split-merge step. It comes in two versions
\begin{lstlisting}
  #include "fastjet/D0RunIpre96ConePlugin.hh"
  // ...
  D0RunIpre96ConePlugin (double R, 
                         double min_jet_Et, 
                         double split_ratio = 0.5);
\end{lstlisting}
and
\begin{lstlisting}
  #include "fastjet/D0RunIConePlugin.hh"
  // ...
  D0RunIConePlugin (double R, 
                    double min_jet_Et, 
                    double split_ratio = 0.5);
\end{lstlisting}
corresponding to versions of the algorithm used respectively before
and after 1996.
They differ only in the recombination scheme used to determine the jet
momenta once each jet's constituents have been identified.
In the pre-1996 version, a hybrid between an $E$-like scheme and an
$E_t$ scheme recombination is used, while in the post-1996 version it
is just the $E_t$ scheme~\cite{Abbott:1997fc}.

The algorithm places a cut on the minimum $E_t$ of the cones during
iteration (related to \verb|min_jet_Et|).
The \verb|split_ratio| is the same as the overlap threshold in other
split-merge based algorithms (\Dzero usually use 0.5).
It is the \fastjet authors' understanding that the value 
used for \verb|min_jet_Et| was 8\,GeV, corresponding to a cut of
$4\GeV$ on cones.
The publication that describes this algorithm~\cite{Abbott:1997fc}
mentions the use of a $1\GeV$ seed threshold applied to preclustered
calorimeter towers in order to obtain the seeds for the stable cone
finding. 
Such a threshold and the pre-clustering appear not to be included in the code
distributed with \fastjet.

The underlying code for this algorithm was provided by
Lars Sonnenschein.

Note: this algorithm is IR$_{2+1}$ unsafe. It is recommended that it
be used only for the purpose of comparison with Run~I data from
\Dzero.
It is to be deprecated for new experimental or theoretical analyse

\subsubsection{\Dzero Run II cone}
This is the main algorithm used by \Dzero in Run~II of the
Tevatron. It is a midpoint type iterative cone with split-merge step.
The \Dzero collaboration usually refers to
Ref.~\cite{RunII-jet-physics} when introducing the algorithm in its
articles.
That generic Tevatron document does not reflect all details of the
actual \Dzero algorithm, and for a complementary description
the reader is referred to Ref.~\cite{arXiv:1110.3771}.
\begin{lstlisting}
  #include "fastjet/D0RunIIConePlugin.hh"
  // ...
  D0RunIIConePlugin (double R, 
                     double min_jet_Et, 
                     double split_ratio = 0.5);
\end{lstlisting}
The parameters have the same meaning as in the \Dzero Run~I
cone. 
There is a cut on the minimum $E_t$ of the cones during iteration,
which by default is half of \verb|min_jet_Et|.
It is the \fastjet authors' understanding that two values have been
used for \verb|min_jet_Et|, 8\,GeV (in earlier publications) and
6\,GeV (in more recent publications).

As concerns seed thresholds and preclustering, \Dzero describes them
as being applied to calorimeter towers in data and in Monte Carlo studies
that include detector simulation~\cite{arXiv:1110.3771}.
However, for NLO calculations and Monte Carlo studies based on stable
particles, no seed threshold is applied.
The code distributed with \fastjet does not allow for seed thresholds.

The underlying code for this algorithm was provided by
Lars Sonnenschein.

Note: this algorithm is IR$_{3+1}$ unsafe (IR$_{2+1}$ for jets with
energy too close to \verb:min_jet_Et:). It is to be deprecated for new
experimental or theoretical analyses.

\subsubsection{ATLAS iterative cone}
This iterative cone algorithm, with a split-merge step, was used by
ATLAS during the preparation for the LHC.
\begin{lstlisting}
  #include "fastjet/AtlasConePlugin.hh"
  // ...
  ATLASConePlugin (double R, 
                   double seedPt = 2.0, 
                   double f = 0.5);
\end{lstlisting}
$f$ is the overlap threshold

The underlying code for this algorithm was extracted from an early
version of SpartyJet \cite{SpartyJet} (which itself was distributed
under the GPL license).
Since version 3.0 of \fastjet it is a slightly modified version that
we distribute, where an internal \ttt{sort} function has been replaced
with a \ttt{stable\_sort}, to ensure reproducibility of results across
compilers and architectures (results are unchanged when the results of
the sort are unambiguous).

Note: this algorithm is IR$_{2+1}$ unsafe (in the limit of zero seed
threshold). It is to be deprecated for new experimental or theoretical
analyses.

\subsubsection{CMS iterative cone}
This iterative cone algorithm with progressive removal was used by CMS
during the preparation for the LHC.
\begin{lstlisting}
  #include "fastjet/CMSIterativeConePlugin.hh"
  // ...
  CMSIterativeConePlugin (double ConeRadius, double SeedThreshold=0.0);
\end{lstlisting}

The underlying code for this algorithm was extracted from the CMSSW
web site, with certain small service routines having been rewritten by
the \fastjet authors. 
Permission to redistribute the resulting code with \fastjet has been
granted by CMS under the terms of the GPL license.
The code was validated by clustering 1000 events with the original
version of the CMS software and comparing the output to the clustering
performed with the \fastjet plugin.
The jet contents were identical in all cases. However the jet momenta
differed at a relative precision level of $10^{-7}$, related to the
use of single-precision arithmetic at some internal stage of the CMS
software (while the \fastjet version is in double precision).

Note: this algorithm is Coll$_{3+1}$ unsafe~\cite{antikt}. It is to be
deprecated for new experimental or theoretical analyses.

\subsubsection{PxCone}

The PxCone algorithm is an iterative cone with midpoints and a
split-drop procedure:
\begin{lstlisting}
  #include "fastjet/PxConePlugin.hh"
  // ...
  PxConePlugin (double  cone_radius      , 
		double  min_jet_energy = 5.0  , 
		double  overlap_threshold = 0.5,
                bool    E_scheme_jets = false); 
\end{lstlisting}
It includes a threshold on the minimum transverse energy for a cone
(jet) to be included in the split-drop stage.
If \verb|E_scheme_jets| is true then the plugin applies an $E$-scheme
recombination to provide the momenta of the final jets (by default an
$E_t$ type recombination scheme is used).

The base code for this plugin is written in Fortran and, on some
systems, problems have been reported at the link stage due to mixing
Fortran and C++.
The Fortran code has been modified by the \fastjet authors to provide
the same jets regardless of the order of the input particles.
This involved a small modification of the midpoint procedure, which
can have a minor effect on the final jets and should make the algorithm
correspond to the description of \cite{Seymour:2006vv}.

The underlying code for this algorithm was written by Luis del Pozo
and Michael Seymour with input also from David Ward~\cite{PxCone}
and they have granted permission for their code to be distributed with
\fastjet under the terms of the GPL license.

This algorithm is IR$_{3+1}$ unsafe. It is to be deprecated for
new experimental or theoretical analyses.

\subsubsection{TrackJet}
This algorithm has been used at the Tevatron for identifying jets from
charged-particle tracks in underlying-event studies~\cite{Affolder:2001xt}.
\begin{lstlisting}
  #include "fastjet/TrackJetPlugin.hh"
  // ...
  TrackJetPlugin (double radius, 
		  RecombinationScheme jet_recombination_scheme=pt_scheme, 
		  RecombinationScheme track_recombination_scheme=pt_scheme);
\end{lstlisting}
Two recombination schemes are involved: the first one indicates how
momenta are recombined to provide the final jets (once particle-jet
assignments are known), the second one indicates how momenta are
combined in the procedure that constructs the jets.

The underlying code for this algorithm was written by the \fastjet
authors, based on code extracts from the (GPL) Rivet implementation, written
by Andy Buckley with input from Manuel Bahr and Rick Field.
Since version 3.0 of \fastjet it is a slightly modified version that
we distribute, where an internal \ttt{sort} function has been replaced
with a \ttt{stable\_sort}, to ensure reproducibility of results across
compilers and architectures (results are unchanged when the results of
the sort are unambiguous, which is the usual case).

Note: this algorithm is believed to be Coll$_{3+1}$ unsafe. It is to
be deprecated for new experimental or theoretical analyses.

\subsubsection{GridJet}
GridJet allows you to define a grid and then cluster particles such
that all particles in a common grid cell combine to form a jet.
Its main interest is in providing fast clustering for high
multiplicities (the clustering time scales linearly with the number of
particles).
The jets that it forms are not always physically meaningful: for
example, a genuine physical jet may lie at the corner of 4 grid cells
and so be split up somewhat arbitrarily into 4 pieces.
It is therefore not intended to be used for standard jet finding.
However for some purposes (such as background estimation) this
drawback is offset by the greater uniformity of the area of the jets.
Its interface is as follows
\begin{lstlisting}
  #include "fastjet/GridJetPlugin.hh"
  // ...
  GridJetPlugin (double ymax, double requested_grid_spacing);
\end{lstlisting}
creating a grid that covers $|y|<$\ttt{ymax} with a grid spacing close
to the \ttt{requested\_grid\_spacing}: the spacings chosen in $\phi$
and $y$ are those that are closest to the requested spacing while also
giving an integer number of grid cells that fit exactly into the
rapidity and $0<\phi <2\pi$ ranges.

Note that for background estimation purposes the
\ttt{GridMedianBackgroundEstimator} is much faster than using the
\ttt{GridJetPlugin} with ghosts and a
\ttt{JetMedianBackgroundEstimator}.

The underlying code for this algorithm was written by the \fastjet
authors.

\subsection{Plugins for $e^+e^-$ collisions}
\label{sec:other-ee-plugins}

\subsubsection{Cambridge algorithm}
\label{sec:ee-cam}
The original $\ee$ Cambridge~\cite{CamOrig} algorithm is provided as a plugin:
\begin{lstlisting}
  #include "fastjet/EECambridgePlugin.hh"
  // ...
  EECambridgePlugin (double ycut);
\end{lstlisting}
This algorithms performs sequential recombination of the pair of
particles that is closest in angle, except when $y_{ij} =
\frac{2\min(E_i^2,E_j^2)}{Q^2}(1-\cos\theta) > y_{cut}$, in which case
the less energetic of $i$ and $j$ is labelled a jet, and the other
member of the pair remains free to cluster.

To access the jets, the user should use the \verb|inclusive_jets()|,
\ie as they would for the majority of the $pp$ algorithms.

The underlying code for this algorithm was written by the \fastjet
authors.

\subsubsection{Jade algorithm}
\label{sec:ee-jade}
The JADE algorithm~\cite{Bartel:1986ua,Bethke:1988zc}, a sequential
recombination algorithm with distance measure $d_{ij} = 2E_i E_j
(1-\cos\theta)$, is available through
\begin{lstlisting}
  #include "fastjet/JadePlugin.hh"
  // ...
  JadePlugin ();
\end{lstlisting}
To access the jets at a given $y_{cut} = d_{cut}/Q^2$, the user 
should call \ttt{ClusterSequence::exclusive\_jets\_ycut(double ycut)}.

Note: the JADE algorithm has been used with various recombination
schemes. The current plugin will use whatever recombination scheme the
user specifies with for the jet definition. The default $E$-scheme is
what was used in the original JADE publication \cite{Bartel:1986ua}.
To modify the recombination scheme, the user may first construct the
jet definition and then use either of 
\begin{lstlisting}
  void JetDefinition::set_recombination_scheme(RecombinationScheme recomb_scheme);
  void JetDefinition::set_recombiner(const Recombiner * recomb)
\end{lstlisting}
(cf.~sections~\ref{sec:recomb_schemes}, \ref{sec:recombiner}) to modify the
recombination scheme.

The underlying code for this algorithm was written by the \fastjet
authors.

\subsubsection{Spherical SISCone algorithm}
\label{sec:spherical-siscone}

The spherical SISCone algorithm is an extension~\cite{SpheriSISCone}
to spherical coordinates of the hadron-collider SISCone
algorithm~\cite{SISCone}.
\begin{lstlisting}
  #include "fastjet/SISConeSphericalPlugin.hh"
  // ...
  SISConeSphericalPlugin(double R, 
                         double overlap\_threshold
			 double protojet_Emin = 0.0, 
			 bool   caching = false,
			 SISConeSphericalPlugin::SplitMergeScale  
                           split_merge_scale = SISConeSphericalPlugin::SM_Etilde,
			 double split_merge_stopping_scale = 0.0);
\end{lstlisting}
The functionality follows directly that of \ttt{SISConePlugin}.

Note that the underlying implementation of spherical SISCone is
optimised for large $N$. 
An alternative implementation that is faster for $N \lesssim 10$ was
presented in~\cite{Weinzierl:2011jx}. That reference also contains a
nice description of the algorithm.

\section{Selectors}
\label{sect:selectors}

Analyses often place constraints (cuts) on jets' transverse momenta,
rapidity, maybe consider only some $N$ hardest jets, etc.
There are situations in which it is convenient to be able to define
a basic set of jet cuts in one part of a program and then have it used
elsewhere.
To allow for this, we provide a \ttt{fastjet::Selector} class, available
through 
\begin{lstlisting}
  #include "fastjet/Selector.hh"
\end{lstlisting}

\subsection{Essential usage}

As an example of how \ttt{Selector}s are used, suppose that we  have a vector of jets, \ttt{jets},
and wish to select those that have rapidities $|y|<2.5$ and transverse
momenta above $20\GeV$. We might write the following:
\begin{lstlisting}
  Selector select_rapidity = SelectorAbsRapMax(2.5);
  Selector select_pt       = SelectorPtMin(20.0);
  Selector select_both     = select_pt && select_rapidity;
  
  vector<PseudoJet> selected_jets = select_both(jets);
\end{lstlisting}
Here, \ttt{Selector} is a class, while \ttt{SelectorAbsRapMax} and
\ttt{SelectorPtMin} are functions that return an instance of the
\ttt{Selector} class containing the internal information needed to
carry out the selection.
\ttt{Selector::operator(const vector<PseudoJet> \& jets)} takes the
jets given as input and returns a vector containing those that pass
the selection cuts. The logical operations \ttt{\&\&}, \ttt{||} and
\ttt{!} enable different selectors to be combined.

Nearly all selectors, like those above, apply jet by jet (the function
\ttt{Selector::applies\_jet\_by\_jet()} returns \ttt{true}). 
For these, one can
query whether a single jet passes the selection with the help of the
function \ttt{bool Selector::pass(const PseudoJet \&)}.

There are also selectors that only make sense applied to an ensemble
of jets. 
This is the case specifically for \ttt{SelectorNHardest(unsigned
  int n)}, which, acting on an ensemble of jets, selects the $n$ jets with
largest transverse momenta. If there are fewer than $n$ jets, then all
jets pass.

When a selector is applied to an ensemble of jets one can also use
\begin{lstlisting}
  Selector::sift(vector<PseudoJet> & jets, 
                 vector<PseudoJet> & jets_that_pass, 
		 vector<PseudoJet> & jets_that_fail)		 
\end{lstlisting}
to obtain the vectors of \ttt{PseudoJet}s that pass or fail the selection.

For selectors that apply jet-by-jet, the selectors on either
side of the logical operators \ttt{\&\&} and \ttt{||} naturally
commute.
For operators that act only on the ensemble of jets the behaviour needs
specifying. 
The design choice that we have made is that
\begin{lstlisting}
  SelectorNHardest(2)    && SelectorAbsRapMax(2.5)
  SelectorAbsRapMax(2.5) && SelectorNHardest(2)
\end{lstlisting}
give identical results: in logical combinations of selectors, each
constituent selector is applied independently to the ensemble of jets,
and then a decision whether a jet passes is determined from the
corresponding logical combination of each of the selectors'
results. Thus, here only jets that are among the 2 hardest of the
whole ensemble and that have $|y|<2.5$ will be selected.
If one wishes to first apply a rapidity cut, and {\sl then} find the 2
hardest among those jets that pass the rapidity cut, then one should
instead use the \ttt{*} operator:
\begin{lstlisting}
  SelectorNHardest(2)  *  SelectorAbsRapMax(2.5)
\end{lstlisting}
In this combined selector, the right-hand selector is applied first,
and then the left-hand selector is applied to the results of the
right-hand selection.

A complementary selector can also be defined using the negation operator. For
instance
\begin{lstlisting}
  Selector sel_allbut2hardest = !SelectorNHardest(2);
\end{lstlisting}
Note that, if directly applying (as opposed to first defining) a similar 
negated selector to a collection
of jets, one should write
\begin{lstlisting}
  vector<PseudoJet> allbut2hardest = (!SelectorNHardest(2))(jets);
\end{lstlisting}
with the brackets around the selector definition being now necessary due to
\ttt{()} having higher precedence in C++ than Boolean operators.

A user can obtain a string describing a given \ttt{Selector}'s action by
calling its \ttt{description()} member function. 
This behaves sensibly also for compound selectors.

New selectors can be implemented as described in
section~\ref{sec:new-selectors}.

\subsubsection{Other information about selectors}

Selectors contain a certain amount of additional information that can
provide useful hints to the functions using them.

One such piece of information is a selector's rapidity extent,
accessible through a \ttt{get\_rapidity\_extent(rapmin,rapmax)} call,
which is useful in the context of background estimation
(section~\ref{sec:BackgroundEstimator}).
If it is not sensible to talk about a rapidity extent for a given
selector (e.g.\ for \ttt{SelectorPtMin}) the rapidity limits that are
returned are the largest (negative and positive) numbers that can be
represented as doubles.
The function \ttt{is\_geometric()} returns true if the selector places
constraints only on rapidity and azimuth.

Selectors may also have areas associated with them (in analogy with
jet areas, section~\ref{sec:areas}).
The \ttt{has\_finite\_area()} member function returns true if a selector has a
meaningful finite area. The \ttt{area()} function returns this area.
In some cases the area may be computed using ghosts (by default with
ghosts of area $0.01$; the user can specify a different ghost area as
an argument to the \ttt{area} function).

\subsection{Available selectors}

\subsubsection{Absolute kinematical cuts}

A number of selectors have been implemented following the naming
convention \ttt{Selector}{\it\{Var\}\{LimitType\}}. 
The {\it\{Var\}} indicates which variable is being cut on, and can be
one of 
\begin{lstlisting}
       pt, Et, E, Mass, Rap, AbsRap, Eta, AbsEta
\end{lstlisting}
The {\it\{LimitType\}} indicates whether one places a lower-limit on
the variable, an upper limit or a range, corresponding to the choices
\begin{lstlisting}
       Min, Max, Range
\end{lstlisting}
A couple of examples are
\begin{lstlisting}
  SelectorPtMin(25.0)        // Selects $p_t>25$ (units are user's default for momenta)
  SelectorRapRange(1.9,4.9)  // Selects $1.9<y<4.9$
\end{lstlisting}
Following a similar naming convention, there are also
\ttt{SelectorPhiRange(}$\phi_{\min},\phi_{\max}$\ttt{)} and
\ttt{SelectorRapPhiRange(}$y_{\min},y_{\max},\phi_{\min},\phi_{\max}$\ttt{)}.

\subsubsection{Relative kinematical cuts}

Some selectors take a \emph{reference jet}. 
They have been developed because it is can be useful for a selector to
make its decision based on information about some other jet. 
For example one might wish to select all jets within some distance of
a given reference jet; or all jets whose transverse momentum is at
least some fraction of a reference jet's.
That reference jet may change from event to event, or even from one
invocation of the Selector to the next, even though the Selector is
fundamentally performing the same underlying type of action.

The available selectors of this kind are:
\begin{lstlisting}
  SelectorCircle($R$)                   // a circle of radius R around the reference jet
  SelectorDoughnut($R_{in}$, $R_{out}$)          // a doughnut between $R_{in}$ and $R_{out}$
  SelectorStrip(half_width)           // a rapidity strip 2*half_width large 
  SelectorRectangle(half_rap_width, half_phi_width) // a rectangle in rapidity and phi
  SelectorPtFractionMin($f$)            // $p_t$ larger than $f p_t^{ref}$
\end{lstlisting}
One example of selectors taking a reference jet is the following. 
First, one constructs the selector, 
\begin{lstlisting}
  Selector sel = SelectorCircle(1.0);
\end{lstlisting}
Then if one is interested in the subset of \ttt{jets} near
\ttt{jet1}, and then those near \ttt{jet2}, one performs the following
operations:
\begin{lstlisting}
  sel.set_reference(jet1);
  vector<PseudoJet> jets_near_jet1 = sel(jets);

  sel.set_reference(jet2);
  vector<PseudoJet> jets_near_jet2 = sel(jets);
\end{lstlisting}
If one uses a selector that takes a reference without the reference having been
actually set, an exception will be thrown.
If one sets a reference for a compound selector, the reference is
automatically set for all components that take a reference.
One can verify whether a given selector takes a reference by calling
the member function
\begin{lstlisting}
  bool Selector::takes_reference() const;
\end{lstlisting}
Attempting to set a reference for a Selector that returns \ttt{false}
here will cause an exception to be thrown.

\subsubsection{Other selectors}

The following selectors are also available:

\begin{lstlisting}
  SelectorNHardest($n$)      //  selects the $n$ hardest jets
  SelectorIsPureGhost()    //  selects jets that are made exclusively of ghost particles
  SelectorIsZero()         //  selects jets with zero momentum
  SelectorIdentity()       //  selects everything. Included for completeness
\end{lstlisting}

\section{Jet areas}
\label{sec:areas}

Jet areas provide a measure of the surface in the $y$-$\phi$ plane
over which a jet extends, or, equivalently, a measure of a jet's
susceptibility to soft contamination.

Since a jet is made up of only a finite number of particles, one needs
a specific definition in order to make its area an unambiguous
concept. Three definitions of area have been proposed
in~\cite{CSSAreas} and implemented in \fastjet:
\begin{itemize}
\item {\it Active} areas add a uniform background of extremely soft massless
  `ghost' particles to the event and allow them to participate in the
  clustering. The area of a given jet is proportional to the number of
  ghosts it contains. 
  Because the ghosts are extremely soft (and sensible jet algorithms
  are infrared safe), the presence of the ghosts does not affect the
  set of user particles that end up in a given jet.
  Active areas give a measure of a jet's sensitivity to diffuse
  background noise.

\item {\it Passive} areas are defined as follows: one adds a single randomly
  placed ghost at a time to the event. One examines which jet (if any)
  the ghost ends up in. One repeats the procedure many times and the
  passive area of a jet is then proportional to the probability of it
  containing the ghost.
  Passive areas give a measure of a jet's sensitivity to point-like
  background noise.

\item The {\it Voronoi} area of a jet is the sum of the Voronoi areas of its
  constituent particles. The Voronoi area of a particle is obtained by
  determining the Voronoi diagram for the event as a whole, and
  intersecting the Voronoi cell of the particle with a circle of
  radius $R$ centred on the particle. Note that for the $k_t$
  algorithm (but not in general for other algorithms) the Voronoi area
  of a jet coincides with its passive area.
\end{itemize}
In the limit of very densely populated events, all area definitions
lead to the same jet-area results~\cite{CSSAreas}.\footnote{This can
  be useful when one area is particularly expensive to calculate: for
  example active areas for SISCone tend to be memory and CPU
  intensive; however, for dense events, they can be adequately
  replaced with passive areas, which, for SISCone, are computationally
  more straightforward.}

The area of a jet can be calculated either as a scalar, or as a 4-vector.
Essentially the scalar case corresponds to counting the number of
ghosts in the jet; the $4$-vector case corresponds to summing their
4-vectors, normalised such that for a narrow jet, the transverse
component of the $4$-vector is equal to the scalar area.

To access jet areas, the user is exposed to two main classes:
\begin{lstlisting}
  class fastjet::AreaDefinition;
  class fastjet::ClusterSequenceArea;
\end{lstlisting}
with input particles, a jet definition and an area definition being
supplied to a \ttt{ClusterSequenceArea} in order to obtain jets with
area information.
Typical usage would be as follows:
\begin{lstlisting}
  #include "fastjet/ClusterSequenceArea.hh"
  // ...
  double ghost_maxrap = 5.0; // e.g. if particles go up to y=5
  AreaDefinition area_def(active_area, GhostedAreaSpec(ghost_maxrap));
  ClusterSequenceArea clust_seq(input_particles, jet_def, area_def);
  vector<PseudoJet> jets = sorted_by_pt(clust_seq.inclusive_jets());
  double area_hardest_jet = jets[0].area();
\end{lstlisting}
Details are to be found below and an example program is given as
\ttt{example/06-area.cc}.

When jet areas are to be used to establish the level of a diffuse noise
that might be present in the event (e.g.\ from underlying event particles
or pileup), and maybe subtract it from jets, further classes such as
\ttt{fastjet::JetMedianBackgroundEstimator} and
\ttt{fastjet::Subtractor} are useful. This topic is discussed in
Section \ref{sec:BackgroundEstimator} and an example program is given
in \ttt{example/07-subtraction.cc}.

\subsection{\tt AreaDefinition}

Area definitions are contained in the \ttt{AreaDefinition}
class. Its two main constructors are:
\begin{lstlisting}
  AreaDefinition(fastjet::AreaType area_type, 
                 fastjet::GhostedAreaSpec ghost_spec);
\end{lstlisting}
for the various active and passive areas (which all involve ghosts)
and
\begin{lstlisting}
  AreaDefinition(fastjet::VoronoiAreaSpec voronoi_spec);
\end{lstlisting}
for the Voronoi area. A default constructor exists, and provides an
active area with a \ttt{ghost\_spec} that is acceptable for a majority
of area measurements with clustering algorithms and typical Tevatron
and LHC rapidity coverage.

Information about the current \ttt{AreaDefinition} can be retrieved
with the help of \ttt{description()}, \ttt{area\_type()},
\ttt{ghost\_spec()} and \ttt{voronoi\_spec()} member functions.

\subsubsection{Ghosted Areas (active and passive)}
\label{sec:ghosted-areas}

There are two variants each of the active and passive areas, as
defined by the \ttt{AreaType} \ttt{enum}:
\begin{lstlisting}
  enum AreaType{ [...],
                 active_area, 
                 active_area_explicit_ghosts,
                 one_ghost_passive_area, 
                 passive_area, 
                 [...]};
\end{lstlisting}
The two active variants give identical results for the areas of hard
jets.
The second one explicitly includes the ghosts when the user requests
the constituents of a jet and also leads to the presence of ``pure
ghost'' jets.
The first of the passive variants
explicitly runs through the procedure mentioned above, \ie it clusters
the events with one ghost at a time, and repeats this for very many
ghosts. This can be quite slow, so we also provide the
\ttt{passive\_area} option, which makes use of information specific to
the jet algorithm in order to speed up the passive-area
determination.\footnote{This ability is provided for $k_t$,
  Cambridge/Aachen, anti-$k_t$ and the SISCone plugin. In the case of
  $k_t$ it is actually a Voronoi area that is used, since this can be
  shown to be equivalent to the passive area~\cite{CSSAreas} (though
  some approximations are made for 4-vector areas). 
  For other algorithms it defaults back to the
  \texttt{one\_ghost\_passive\_area} approach.}

In order to carry out a clustering with a ghosted area determination,
the user should also create an object that specifies how to distribute
the ghosts.\footnote{Or accept a default --- which uses the default
  values listed in the explicit constructor and $\texttt{ghost\_maxrap} =
  6$} %
This is done via the class \ttt{GhostedAreaSpec} whose
constructor is
\begin{lstlisting}
  GhostedAreaSpec(double ghost_maxrap, 
                  int    repeat        = 1,   double ghost_area    = 0.01, 
                  double grid_scatter  = 1.0, double pt_scatter    = 0.1,
                  double mean_ghost_pt = 1e-100);
\end{lstlisting}
The ghosts are distributed on a uniform grid in $y$ and $\phi$, with
small random fluctuations to avoid clustering degeneracies.

\begin{figure}
  \centering
  \includegraphics[width=0.48\textwidth]{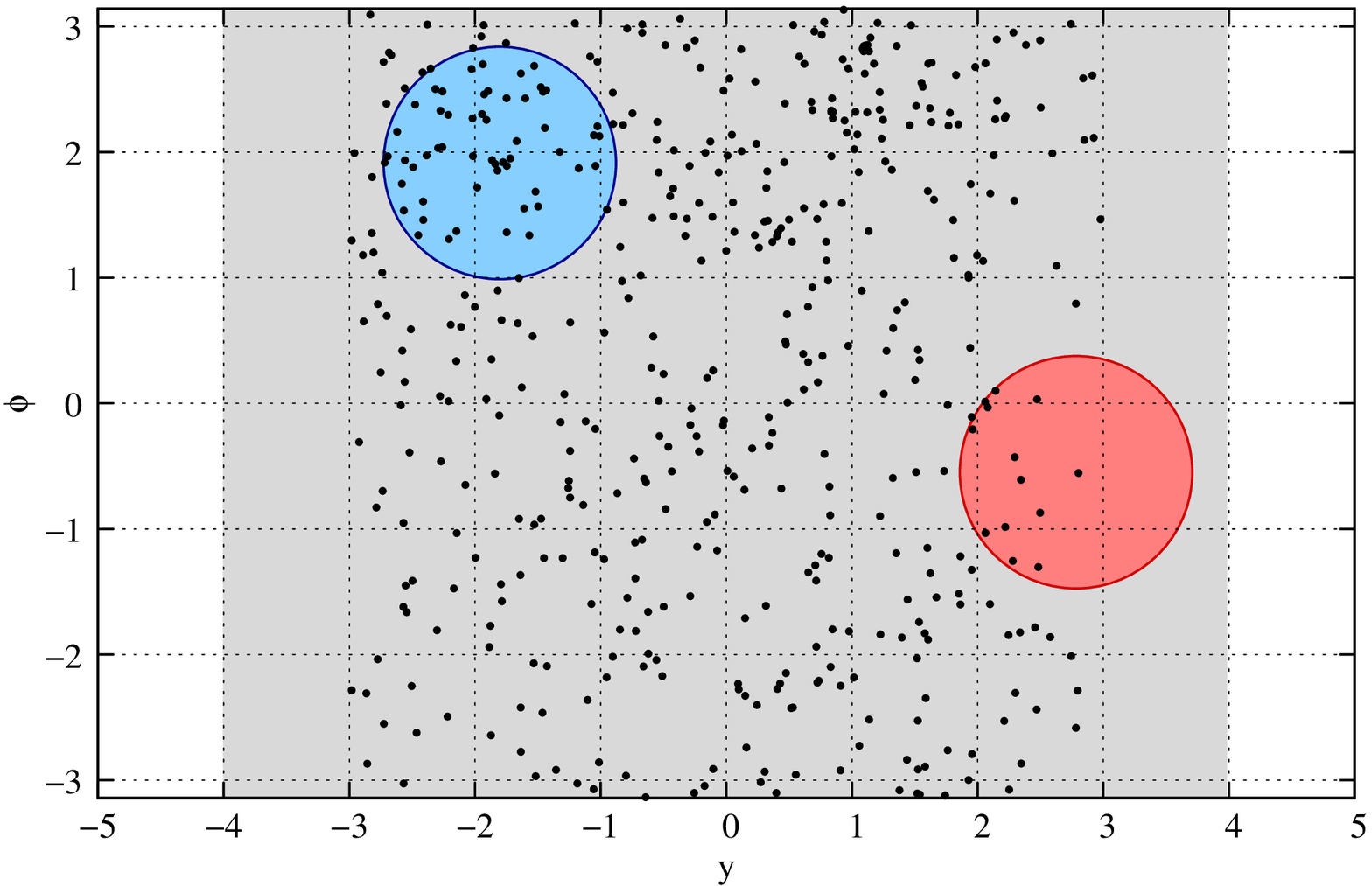}\hfill
  \includegraphics[width=0.48\textwidth]{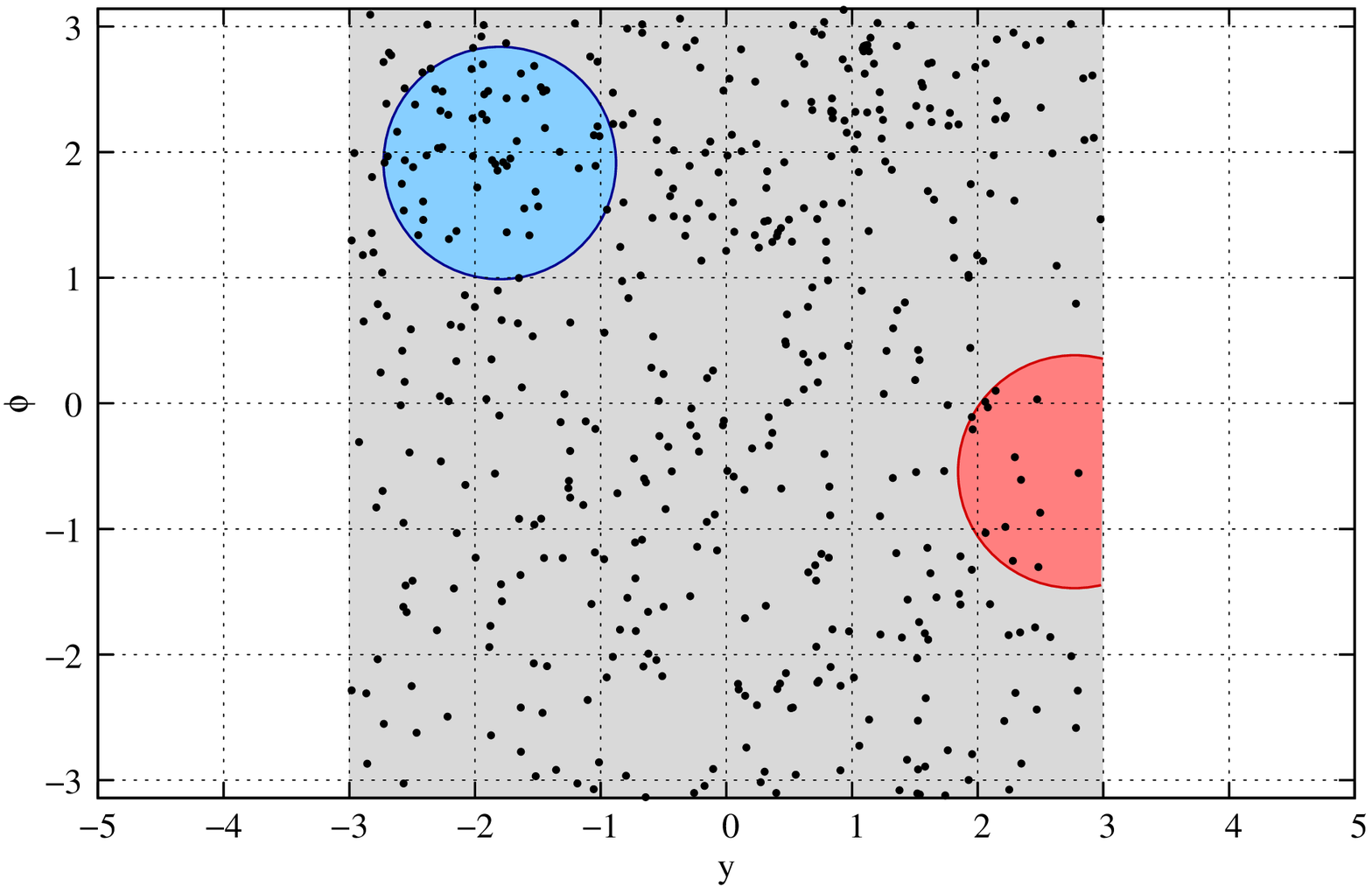}
  \caption{Two choices for ghost placement. The grey area in each plot
    indicates
    the region containing ghosts, while the dots are particles, which
    here are accepted up to $|y|<3$. 
    The circular regions indicate the areas that will be found for two
    particular jets.  
    In the left-hand case, with ghosts that extend well beyond the
    acceptance for particles, jet areas are unaffected by the particle
    acceptance; 
    in the right-hand case, with ghosts placed only up to the
    acceptance limit, jet areas are clipped at the edge of the
    acceptance.  }
  \label{fig:ghost-placement}
\end{figure}

The \ttt{ghost\_maxrap} variable defines the maximum rapidity up to
which ghosts are generated.
If one places ghosts well beyond the particle acceptance (at least $R$
beyond it), then jet areas also stretch beyond the acceptance, giving
a measure of the jet's full extent in rapidity and azimuth.
If ghosts are placed only up to the particle acceptance, then the jet
areas are clipped at that acceptance and correspond more closely
to a measure of the jet's susceptibility to contamination from
accepted soft particles.
This is relevant in particular for jets within a distance $R$ of the
particle acceptance boundary.
The two choices are illustrated in fig.~\ref{fig:ghost-placement}.
To define more complicated ghost acceptances it is possible to replace
\ttt{ghost\_maxrap} with a \ttt{Selector}, which must be purely
geometrical and have finite rapidity extent.

The \ttt{ghost\_area} parameter in the \ttt{GhostedAreaSpec}
constructor is the area associated with a single ghost. 
The number of ghosts is inversely proportional to the ghost area, and
so a smaller area leads to a longer CPU-time for clustering. However
small ghost areas give more accurate results. 
We have found the default ghost area given above to be adequate for
most applications.
Smaller ghost areas are beneficial mainly for high-precision
determinations of areas of jets with small $R$.

By default, one set of ghosts is generated for each event that is
clustered.
The small random fluctuations in ghost positions and $p_t$'s,
introduced to break clustering degeneracies, mean that for repeated
clustering of the same event a given hard jet's area will be different
after each clustering.
This is especially true for sparse events, where a jet's particle
content fails to accurately delineate the boundaries of the jet.
For the \texttt{active\_area} choice (and certain passive areas),
specifying $\ttt{repeat} > 1$ causes \fastjet to directly cluster the
same hard event with multiple ghost sets.
This results in a pseudo-Monte Carlo type evaluation of the jet areas.
A statistical uncertainty on the area is available for each jet,
through the \ttt{jet.area\_error()} call.
It is calculated as the standard deviation of areas obtained for that
jet, divided by $\sqrt{\ttt{repeat}-1}$.
While there are situations in which this facility is useful, for most
applications of jet areas it is sufficient to use
$\ttt{repeat}=1$.\footnote{%
  Several parameters are available to control the properties and
  randomness of the ghosts: each ghost's position differs from an
  exact grid vertex by a random amount distributed uniformly in the
  range $\pm \frac12 \texttt{grid\_scatter}$ times the grid spacing in
  both the rapidity and azimuth directions.
  Each ghost's $p_t$ is distributed randomly in the range $(1 \pm \frac12
  \texttt{pt\_scatter})$ times \texttt{mean\_ghost\_pt}.
  For nearly all applications, it makes sense to use the default
  values.
  Facilities are also available to set and retrieve the seeds for the
  random-number generator, notably through the
  \texttt{set\_random\_status(...)} and
  \texttt{get\_random\_status(...)} members of
  \texttt{GhostedAreaSpec}.  }

After initialisation, the parameters can be modified and retrieved
respectively with calls such as \ttt{set\_ghost\_area(...)} and
\ttt{ghost\_maxrap()} (similarly for the other
parameters\footnote{In versions of \fastjet prior to 3.0.1, the names
  \texttt{mean\_ghost\_kt} and \texttt{kt\_scatter} should be used
  rather than \texttt{mean\_ghost\_pt} and \texttt{pt\_scatter}. The
  former names will be maintained for the foreseeable future.}).
A textual description of the \ttt{GhostedAreaSpec} can be obtained, as
usual, with the \ttt{description()} member function.


\subsubsection{Voronoi Areas}
\label{sec:voronoi-areas}

The Voronoi areas of jets are evaluated by summing the corresponding
Voronoi areas of the jets' constituents. The latter are obtained 
by considering the intersection between
the Voronoi cell of each particle and a circle of radius $R$ centred 
on the particle's position in the rapidity-azimuth plane.

The jets' Voronoi areas can be obtained from 
\ttt{ClusterSequenceArea} by passing
the proper \ttt{VoronoiAreaSpec} specification to
\ttt{AreaDefinition}. Its constructors are
\begin{lstlisting}
  /// default constructor (effective_Rfact = 1)
  VoronoiAreaSpec() ;
  
  /// constructor that allows you to set effective_Rfact
  VoronoiAreaSpec(double effective_Rfact) ; 
\end{lstlisting}
The second constructor allows one to modify (by a multiplicative
factor \ttt{effective\_Rfact}) the radius of the circle which is
intersected with the Voronoi cells. With $\ttt{effective\_Rfact} = 1$,
for the $k_t$ algorithm, the Voronoi area is equivalent to the passive
area.
Information about the specification in use is returned by
\ttt{effective\_Rfact()} and \ttt{description()} member functions.

The Voronoi areas are calculated with the help of Fortune's ($N \ln
N$) Voronoi diagram generator for planar static point
sets~\cite{Fortune}.

One use for the Voronoi area is in background determination with the
$k_t$ algorithm (see below, section~\ref{sec:BackgroundEstimator}): with
the choice $\ttt{effective\_Rfact}\simeq0.9$ it provides an acceptable
approximation to the $k_t$ algorithm's active area and is often
significantly faster to compute than the active area.
Note that it is not currently possible to clip Voronoi areas with a
given particle acceptance. 
As a result, given particles up to $|y|<y_{\max}$, only jets with $|y|
\lesssim y_{\max} - R$ will have areas that reflect the jets'
sensitivity to accepted particle contamination.
It is only these jets that should then be used for background
determinations.

\subsection{\tt ClusterSequenceArea}

This is the class
\footnote{  
  \texttt{ClusterSequenceArea} is derived from
  \texttt{ClusterSequenceAreaBase} (itself derived from
  \texttt{ClusterSequence}) and makes use of one among
  \texttt{ClusterSequenceActiveAreaExplicitGhosts}, 
  \texttt{ClusterSequenceActiveArea}, 
  \texttt{ClusterSequencePassiveArea},
  \texttt{ClusterSequence1GhostPassiveArea} or 
  \texttt{ClusterSequenceVoronoiArea} (all of them in the \texttt{fastjet}
  namespace of course), according to the choice
  made with \texttt{AreaDefinition}. The user can also use these
  classes directly.
  \texttt{ClusterSequenceActiveAreaExplicitGhosts} is particularly
  useful in that it allows the user to specify their own set of ghost
  particles.
  This is exploited to provide area support in a number of the
  transformers of section~\ref{sec:transformers}.
}
used for producing a cluster sequence that also calculates jet areas.
Its constructor is
\begin{lstlisting}
  template<class L> ClusterSequenceArea(const std::vector<L> & input_particles, 
                                        const JetDefinition & jet_def,
	                                const AreaDefinition & area_def);
\end{lstlisting}
and the class includes the methods
\begin{lstlisting}
  /// Return a reference to the area definition
  virtual const AreaDefinition & area_def() const; 

  /// Returns an estimate of the area contained within the selector that is free of jets. 
  /// The selector needs to have a finite area and be applicable jet by jet.
  /// The function returns 0 if active_area_explicit_ghosts was used.
  virtual double empty_area(const Selector & selector) const;
\end{lstlisting}

As long as an instance of this class is in scope, a user can access
information about the area of its jets using the following methods of
\ttt{PseudoJet}:
\begin{lstlisting}
  /// Returns the scalar area associated with the given jet
  double area = jet.area();

  /// Returns the error (uncertainty) associated with the determination of the
  /// scalar area of the jet; gives zero when the repeat=1 and for passive/Voronoi areas
  double area_error = jet.area_error();

  /// Returns a PseudoJet whose 4-vector is defined by the following integral
  ///
  ///       $\int dy d\phi$ PseudoJet($y$,$\phi$,$p_t=1$) * $\Theta$("$y,\phi$ inside jet boundary")
  ///
  /// where PseudoJet($y$,$\phi$,$p_t=1$) is a 4-vector with the given rapidity ($y$),
  /// azimuth ($\phi$) and $p_t=1$, while $\Theta$("$y,\phi$ inside jet boundary") is 1 
  /// when $y,\phi$ define a direction inside the jet boundary and 0 otherwise.
  PseudoJet area_4vector = jet.area_4vector();

  /// When using active_area_explicit_ghosts, returns true for jets made 
  /// exclusively of ghosts and for ghost constituents.
  bool is_pure_ghost = jet.is_pure_ghost();
\end{lstlisting}

\section{Background estimation and subtraction}
\label{sec:BackgroundEstimator}

Events with hard jets are often accompanied by a more diffuse
``background'' of relatively soft particles, for example from the
underlying event (in $pp$ or PbPb collisions) or from pileup (in $pp$
collisions).
For many physical applications, it is useful to be able to
estimate characteristics of the background on an event-by-event basis,
for example the $p_t$ per unit area ($\rho$), or fluctuations from
point to point ($\sigma$).
One use of this information is to correct the hard jets for the soft
contamination, as discussed below in section~\ref{sec:subtractor}.

One of the issues in characterising the background is that it is
difficult to introduce a robust criterion to distinguish
``background'' jets from hard jets.
The main method that is available in \fastjet involves the
determination of the distribution of $p_t/A$ for the jets in a given
event (or region of the event) and then taking the median of the
distribution as an estimate of $\rho$, as proposed in~\cite{cs} and
studied in detail also in~\cite{Cacciari:2009dp,GridMedianLH}.
This is largely insensitive to the presence of a handful of hard jets, and
avoids any need for introducing a $p_t$ scale to distinguish hard and
background jets.

The original form of this method used the $k_t$ or Cambridge/Aachen
jet algorithms to find the jets.
These algorithms have the advantage that the resulting jets tend to
have reasonably uniform areas\footnote{Whereas anti-$k_t$ and SISCone
  suffer from jets with near zero areas or, for SISCone, sometimes
  huge, ``monster'' jets, biasing the $\rho$ determination. They are
  therefore not recommended.}
In the meantime a variant of the approach that has emerged is to
cluster the particles into rectangular grid cells in $y$ and $\phi$
and determine their median $p_t/A$.
This has the advantage of simplicity and much greater speed. 
One may worry that a hard jet will sometimes lie at a corner of
multiple grid cells, inducing larger biases in the median than with a
normal jet finder jets, however we have found this not
to be an issue in practice~\cite{GridMedianLH}.

\subsection{General Usage}\label{sec:bkg_general_usage}

\subsubsection{Background estimation}\label{sec:bkg_estim_usage}

The simplest workflow for background estimation is first, outside the
event loop, to create a background estimator.
For the jet-based method, one creates a
\ttt{fastjet::JetMedianBackgroundEstimator},
\begin{lstlisting}
  #include "fastjet/tools/JetMedianBackgroundEstimator.hh"
  // ...
  JetMedianBackgroundEstimator bge(const Selector & selector,
                                   const JetDefinition & jet_def,
                                   const AreaDefinition & area_def);
\end{lstlisting}
where the selector is used to indicate which jets are used for
background estimation (for simple use cases, one just specifies a
rapidity range, e.g.\ \ttt{SelectorAbsRapMax(4.5)} to use all jets
with $|y|<4.5$), together with a jet definition  and an area
definition. 
We have found that the $k_t$ or Cambridge/Aachen jet algorithms with $R
= 0.4 - 0.6$ generally provide adequate background estimates, with
the lower range of $R$ values to be preferred if the events are likely
to be busy~\cite{Cacciari:2009dp,GridMedianLH}.
An active area with explicit ghosts is generally
recommended.\footnote{With the $k_t$ algorithm one may also use a
  Voronoi area (\texttt{effective\_Rfact = 0.9} is recommended), which
  has the advantage of being deterministic and faster than ghosted
  areas. In this case however one must use a selector that is
  geometrical and selects only jets well within the range of event
  particles. 
  E.g. if particles are present up to $|y| = y_{\max}$ one should only
  use jets with $|y| \lesssim y_{\max} - R$.
  When using ghosts, the selector can instead go right up
  to the edge of the acceptance, if the ghosts also only go right up
  to the edge, as in the right-hand plot of
  fig.~\ref{fig:ghost-placement}.}

For the grid based method one creates a
\ttt{fastjet::GridMedianBackgroundEstimator},
\begin{lstlisting}
  #include "fastjet/tools/GridMedianBackgroundEstimator.hh"
  // ...
  GridMedianBackgroundEstimator bge(double max_rapidity,
                                    double requested_grid_spacing);
\end{lstlisting}
We have found grid spacings in the range $0.5-0.7$ to be
adequate~\cite{GridMedianLH}, with lower values preferred for events
that are likely to have high multiplicities.

Both of the above background estimators derive from a
\ttt{fastjet::BackgroundEstimatorBase} class and the remaining
functionality is common to both.
In particular, for each event, one tells the background estimator
about the event particles,
\begin{lstlisting}
  bge.set_particles(event_particles);
\end{lstlisting}
where \ttt{event\_particles} is a vector of \PJ, and then extracts the
background density and a measure of its fluctuations with the two
following calls 
\begin{lstlisting}
  // the median of ($p_t$/area) for grid cells, or for jets that pass the selection cut, 
  // making use also of information on empty area in the event (in the jets case)
  rho = bge.rho(); 

  // an estimate of the fluctuations in the $p_t$ density per unit $\smash{\sqrt{A}}$,
  // which is obtained from the 1-sigma half-width of the distribution of pt/A.
  // To be precise it is defined such that a fraction (1-0.6827)/2 of the jets
  // (including empty jets) have $p_t/A < \rho - \sigma \sqrt{\langle A \rangle}$
  sigma = bge.sigma(); 
\end{lstlisting}
Note that $\rho$ and $\sigma$ determinations count empty area within
the relevant region as consisting of jets of zero $p_t$.
Thus (roughly speaking), if more that half of the area covered by the
jets selector or grid rapidity range is empty, the median estimate for
$\rho$ will be zero, as expected and appropriate for quiet events.

\subsubsection{Background subtraction}\label{sec:subtractor}

A common use of an estimation of the background is to subtract its contamination 
from the
transverse momentum of hard jets, in the form
\begin{equation}
p_{t,jet}^{sub} = p_{t,jet}^{raw} - \rho A_{jet} 
\end{equation}
or its 4-vector version
\begin{equation}
p_{\mu,jet}^{sub} = p_{\mu,jet}^{raw} - \rho A_{\mu,jet} \, ,
\end{equation}
as first proposed in~\cite{cs}.

To this end, the \ttt{Subtractor} class is defined in
\ttt{include/tools/Subtractor.hh}. Its constructor takes a pointer to
a background estimator:
\begin{lstlisting}
  JetMedianBackgroundEstimator bge(....); // or a grid-based estimator
  Subtractor subtractor(&bge);
\end{lstlisting}
(it is also possible to construct the Subtractor with a fixed value
for $\rho$).
The subtractor can then be used as follows:
\begin{lstlisting}
  PseudoJet jet;
  vector<PseudoJet> jets;
  // ...
  PseudoJet subtracted_jet = subtractor(jet);
  vector<PseudoJet> subtracted_jets = subtractor(jets);
\end{lstlisting}
The subtractor normally returns \ttt{jet - bge.rho(jet)*jet.area\_4vector()}.
If \ttt{jet.perp() < bge.rho(jet)*jet.area\_4vector().perp()}, then
the subtractor instead returns a jet with zero 4-momentum (so that
\ttt{(subtracted\_jet==0)} returns \ttt{true}).
In both cases, the returned jet retains the user and structural
information of the original jet.

An example program is given in \ttt{example/07-subtraction.cc}.

Note that \ttt{Subtractor} derives from the \ttt{Transformer} class (see
section~\ref{sec:transformers}) and hence from 
\ttt{FunctionOfPseudoJet<PseudoJet>} (cf.\
appendix~\ref{app:function-of-pj}).

\subsection{Positional dependence of background}
\label{sec:BGE-positional}

The background density in $pp$ and heavy-ion collisions usually has
some non-negligible dependence on rapidity (and sometimes azimuth).
This dependence is not accounted for in the standard estimate of
$\rho$ based on all jets or grid cells from (say) $|y|<4.5$.
Two techniques are described below to help alleviate this problem.
In each case the properties of the background are to be obtained
through the methods (available for both
\ttt{JetMedianBackgroundEstimator} and
\ttt{GridMedianBackgroundEstimator})
\begin{lstlisting}
  double rho  (const PseudoJet & jet); // $p_t$ density per unit area $A$ near jet
  double sigma(const PseudoJet & jet); // fluctuations in the $p_t$ density near jet
\end{lstlisting}

\subsubsection{Local estimation}
\label{sec:local-bkgd-estimation}

The first technique, ``local estimation'', available for now only in
the case of the jet-based estimator, involves the use of a more local
range for the determination of $\rho$, with the help of a
\ttt{Selector} that is able to take a reference jet,
e.g. \ttt{SelectorStrip(}$\Delta y$\ttt{)}, a strip of half-width
$\Delta y$ (which might be of order $1$) centred on whichever jet is
set as its reference.
With this kind of selector, when the user calls either \ttt{rho(jet)}
or \ttt{sigma(jet)} a \ttt{selector.set\_reference(jet)} call is made
to centre the selector on the specified jet. Then only the jets in the
event that
pass the cut specified by this newly positioned \ttt{selector} are
used to estimate $\rho$ or $\sigma$.\footnote{If the selector does not
  take a reference jet, then these calls give identical results to the
  plain \texttt{rho()} and \texttt{sigma()} calls (unless a manual
  rapidity rescaling is also in effect, cf.\
  section~\ref{sec:rescaled-bkgd-estimation}).}
This method is adequate if the number of jets that pass the selector
is much larger than the number of hard jets in the range (otherwise
the median $p_t/A$ will be noticeably biased by the hard jets).
It therefore tends to be suitable for dijet events in $pp$ or PbPb
collisions, but may fare less well in event samples such as
hadronically decaying $t\bar t$ which may have many central hard jets.
One can attempt to remove some given number of hard jets before
carrying out the median estimation, e.g.\ with a \ttt{selector} such
as
\begin{lstlisting}
  selector = SelectorStrip($\Delta y$) * (!SelectorNHardest(2))
\end{lstlisting}
which removes the 2 hardest jets globally and then, of the remainder,
takes the ones within the strip.\footnote{If you use non-geometric
  selectors such as this in determining $\rho$, the area must have
  explicit ghosts in order to simplify the determination of the empty
  area. If it does not, an error will be thrown.}
This is however not always very effective, because one may not know
how many hard jets to remove.

\subsubsection{Rescaling method}
\label{sec:rescaled-bkgd-estimation}

A second technique to account for positional dependence of the
background is ``rescaling''.
First one parametrises the average shape of the rapidity
dependence from some number of pileup events.
Then for subsequent event-by-event background determinations, one
carries out a global $\rho$ determination and then applies the
previously determined average rescaling function to that global
determination to obtain an estimate for $\rho$ in the neighbourhood of
a specific jet.

The rescaling approach approach is available for both grid and
jet-based methods.
To encode the background shape, one defines an object such as 
\begin{lstlisting}
  // gives rescaling$(y) = 1.16 + 0\cdot y -0.023 \cdot y^2 + 0\cdot y^3 + 0.000041 \cdot y^4$
  fastjet::BackgroundRescalingYPolynomial rescaling(1.16, 0, -0.023, 0, 0.000041);
\end{lstlisting}
(for other shapes, e.g. parametrisation of elliptic flow in heavy ion
collisions, with both rapidity and azimuth dependence, derive a class
from \ttt{FunctionOfPseudoJet<double>} --- see appendix
\ref{app:function-of-pj}). Then one tells the background estimator
(whether jet or grid based) about the rescaling with the call
\begin{lstlisting}
  // tell JetMedianBackgroundEstimator or GridMedianBackgroundEstimator about the rescaling
  bge.set_rescaling_class(&rescaling);
\end{lstlisting}
Subsequent calls to \ttt{rho()} will return the median of the
distribution $p_t/A / \ttt{rescaling}(y)$ (rather than $p_t/A$).
Any calls to \ttt{rho(jet)} and \ttt{sigma(jet)} will include an
additional factor of \ttt{rescaling}$(y_\ttt{jet})$.
Note that any overall factor in the rescaling function cancels out for
\ttt{rho(jet)} and \ttt{sigma(jet)}, but not for calls to \ttt{rho()}
and \ttt{sigma()} (which are in any case less meaningful when a
rapidity dependence is being assumed for the background).

In ongoing studies \cite{GridMedianLH}, we have found that despite its
use of an average background shape, the rescaling method generally
performs comparably to local estimation in terms of its residual $p_t$
dispersion after subtraction.
Additionally, it has the advantage of reduced sensitivity to biases in
events with high multiplicities of hard jets.

\subsection{Other facilities}
\label{sec:BGE-other-facilities}

The \ttt{JetMedianBackgroundEstimator} has a number of enquiry
functions to access information used internally within the median
$\rho$ and $\sigma$ determination.
\begin{lstlisting}
  // Returns the mean area of the jets used to actually compute the background properties,
  // including empty area and jets (available also in grid-based estimator)
  double mean_area() const;
   
  // Returns the number of jets used to actually compute the background properties
  // (including empty jets)
  unsigned int n_jets_used() const;
   
  // Returns the estimate of the area (within the range defined by the selector) that
  // is not occupied by jets.
  double empty_area() const;
   
  // Returns the number of empty jets used when computing the background properties. 
  double n_empty_jets() const;
\end{lstlisting}
For area definitions with explicit ghosts the last two functions
return $0$.
For active areas without explicit ghosts the results are calculated
based on the observed number of internally recorded pure ghost jets
(and unclustered ghosts) that pass the selector; for Voronoi and
passive areas, they are calculated using the difference between the
total range area and the area of the jets contained in the range, with
the number of empty jets then being calculated based on the average
jet area for ghost jets ($0.55\pi R^2$~\cite{CSSAreas}).
All four function above return a result corresponding to the last call
to \ttt{rho} or \ttt{sigma} (as long as the particles, cluster sequence or
selector have not changed in the meantime).

\subsection{Alternative workflows}
To allow flexibility in the user's workflow, 
alternative constructors to \ttt{JetMedianBackgroundEstimator} are provided.
These can come in useful if, for example, the user wishes to carry out
multiple background estimations with the same particles but different
selectors, or wishes to take care of the jet clustering themselves,
e.g.\ because the results of that same jet clustering will be used in
multiple contexts and it is more efficient to perform it just once. These
constructors are:
\begin{lstlisting}
  // create an estimator that uses the inclusive jets from the supplied cluster sequence
  JetMedianBackgroundEstimator(const Selector & rho_range, 
                               const ClusterSequenceAreaBase & csa);
  // a default constructor that requires all information to be set later
  JetMedianBackgroundEstimator();
\end{lstlisting}
In the first case, the background estimator already has all the information it
needs. Instead, if the default constructor has been used, one can then employ
\begin{lstlisting}
  // (re)set the selector to be used for future calls to rho() etc.
  void set_selector(const Selector & rho_range_selector);
  // (re)set the cluster sequence to be used by future calls to rho() etc. 
  // (as with the cluster-sequence based constructor, its inclusive jets are used)
  void set_cluster_sequence(const ClusterSequenceAreaBase & csa);
\end{lstlisting}
to set the rest of the necessary information. If a list of jets is already
available, they can be submitted to the background estimator in place
of a cluster sequence:
\begin{lstlisting}
  // (re)set the jets to be used by future calls to rho() etc. 
  void set_jets(const std::vector<PseudoJet> & jets);
\end{lstlisting}
Note that the
jets passed via the \ttt{set\_jets()} call above must all originate from a common
\ttt{ClusterSequenceAreaBase} type class.

\section{Jet transformers (substructure, taggers, etc...)}
\label{sec:transformers}

Performing post-clustering actions on jets has in recent years become
quite widespread: for example, numerous techniques have been
introduced to tag boosted hadronically decaying objects, and various
methods also exist for suppressing the underlying event and pileup in
jets, beyond the subtraction approach discussed in
section~\ref{sec:BackgroundEstimator}.
\fastjet 3 provides a common interface for such
tools, intended to help simplify their usage and to guide authors of
new ones.
Below, we first discuss generic considerations about these tools, which
we call \ttt{fastjet::Transformer}s. 
We then describe some that have already been implemented. 
New user-defined transformers can be implemented as described in 
section~\ref{sec:transformerdetails}.

A transformer derived from \ttt{Transformer}, e.g. the
 class \ttt{MyTransformer}, will generally be used as follows:
\begin{lstlisting}
  MyTransformer transformer;
  PseudoJet transformed_jet = transformer(jet);
\end{lstlisting}
Often, transformers provide new structural information that is to be
associated with the returned result. 
For a given transformer, say \ttt{MyTransformer}, the new information
that is not already directly accessible from \ttt{PseudoJet} (like its
\ttt{constituents}, \ttt{pieces} or \ttt{area} when they are
relevant), can be accessed through
\begin{lstlisting}
  transformed_jet.structure_of<MyTransformer>()
\end{lstlisting}
which returns a reference to an object of type
\ttt{MyTransformer::StructureType}.
This is illustrated below on a case-by-case basis for each of the
transformers that we discuss.
Using the Boolean function
\ttt{transformed\_jet.has\_structure\_of<MyTransformer>()} it is possible to
check if \ttt{transformed\_jet} is compatible with the structure provided by
\ttt{MyTransformer}.

A number of the transformers that we discuss below are ``taggers'' for
boosted objects.
In some cases they will determine that a given jet does not satisfy
the tagging conditions (e.g., for a top tagger, because it seems not
to be a top jet).
We will adopt the convention that in such cases the result of the
transformer is a jet whose 4-momentum is zero, i.e.\ one that satisfies
\ttt{jet == 0}.
Such a jet may still have structural information however (e.g.\ to
indicate why the jet was not tagged).

\subsection{Noise-removal transformers}

In section \ref{sec:subtractor} we already saw one transformer for
noise removal, i.e.\ \ttt{Subtractor}. 
Others have emerged in the context of jet substructure studies and are
described here.

\subsubsection{Jet Filtering and Trimming using \texttt{Filter}}
\label{sec:filtering}

Filtering was first introduced in \cite{BDRS} to reduce the
sensitivity of a boosted Higgs-candidate jet's mass to the underlying
event.
Generally speaking, filtering clusters a jet's constituents with a
smaller-than-original jet radius $R_{\rm filt}$.
It then keeps just the $n_{\rm filt}$ hardest of the resulting
subjets, rejecting the others.
Trimming~\cite{trimming} is similar, but selects the subjets to be kept based on a
$p_t$ cut.
The use of filtering and trimming has been advocated in number of
contexts, beyond just the realm of boosted object reconstruction.

The \ttt{fastjet::Filter} class derives from \ttt{Transformer}, and
can be constructed 
using a \ttt{JetDefinition}, a \ttt{Selector} and (optionally) a
value for the background density,
\begin{lstlisting}
  #include "fastjet/tools/Filter.hh"
  // ...
  Filter filter(subjet_def, selector, rho);
\end{lstlisting}
This reclusters the jet's constituents with the jet definition
\ttt{subjet\_def}\footnote{
When the input jet was obtained with the Cambridge/Aachen
algorithm and the subjet definition also involves the Cambridge/Aachen
algorithm, the \ttt{Filter} uses the exclusive subjets of the input
jet to avoid having to recluster its constituents.
} 
and then
applies \ttt{selector} on the \ttt{inclusive\_jets} resulting from the
clustering to decide which of these (sub)jets have to be kept.
If \ttt{rho} is non-zero, each of the subjets is subtracted
(using the specified value for the background density) prior to the
selection of the kept subjets. Alternatively, the user can set a
\ttt{Subtractor} (see section~\ref{sec:subtractor}), e.g.
\begin{lstlisting}
  GridMedianBackgroundEstimator bge(...);
  Subtractor sub(&bge);
  filter.set_subtractor(sub);
\end{lstlisting}
When this is done, the subtraction operation is performed using the
\ttt{Subtractor}, independently of whether a value had been set for
\ttt{rho}.

If the jet definition to be used to recluster the jet's constituents is 
the Cambridge/Aachen algorithm, two additional constructors are available:
\begin{lstlisting}
  Filter(double Rfilt, Selector selector, double rho = 0.0);
  Filter(FunctionOfPseudoJet<double> * Rfilt_dyn, Selector selector, double rho = 0.0);
\end{lstlisting}
In the first one, only the radius parameter is specified instead of
the full subjet definition. 
In the second, one has to provide a
(pointer to) a class derived from \ttt{FunctionOfPseudoJet<double>}
which dynamically computes the filtering radius as a function of the
jet being filtered (as was originally used in \cite{BDRS} where
$R_{\rm filt}={\rm min}(0.3,R_{b\bar{b}/2})$, with $R_{b\bar b}$ the
distance between the parents of the jet). 

As an example, a simple filter, giving the subjets obtained clustering
with the Cambridge/Aachen algorithm with radius $R_{\rm filt}$ and
keeping the $n_{\rm filt}$ hardest subjets found, can be set up and applied using
\begin{lstlisting}
  Filter filter(Rfilt, SelectorNHardest(nfilt));
  PseudoJet filtered_jet = filter(jet);
\end{lstlisting}
The \ttt{pieces()} of the resulting
filtered/trimmed jet correspond to the subjets that were kept:
\begin{lstlisting}
  vector<PseudoJet> kept = filtered_jet.pieces();
\end{lstlisting} 
Additional structural information is available as follows:
\begin{lstlisting}
  // the subjets (on the scale Rfilt) not kept by the filtering
  vector<PseudoJet> rejected = filtered_jet.structure_of<Filter>().rejected();
\end{lstlisting}

Trimming, which keeps the subjets with a $p_t$ larger than a fixed
fraction of the input jet, can be obtained defining
\begin{lstlisting}
  Filter trimmer(Rfilt, SelectorPtFractionMin(pt_fraction_min));
\end{lstlisting}
and then applying \ttt{trimmer} similarly to \ttt{filter} above.

Note that the jet being filtered must have constituents. Furthermore,
if \ttt{rho} is non-zero or if a \ttt{Subtractor} is set, the input
jet must come from a cluster sequence with area support and explicit
ghosts. If any of these requirements fail, an exception is thrown.
In cases where the filter/trimmer has been defined with just a jet
radius, the reclustering of the jet is performed with the same
recombination scheme as was used in producing the original jet
(assuming it can be uniquely determined).

\subsubsection{Jet pruning}
\label{sec:pruning}

Pruning was introduced in \cite{Ellis:2009su}. 
It works by reclustering a jet's constituents with some given
sequential recombination algorithm, but vetoing soft and large-angle
recombinations between pseudojets $i$ and $j$, specifically when the
two following conditions are met
\begin{enumerate}
\item the geometric distance between $i$ and $j$ is larger than a
  parameter \ttt{Rcut}, with \ttt{Rcut} = \ttt{Rcut\_factor}$\times
  2m/p_t$, where $m$ and $p_t$ are the mass and transverse momentum of
  the original jet being pruned;
\item one of $p_t^i, p_t^j $ is $<$~\ttt{zcut}$\times p_t^{i+j}$.
\end{enumerate}
When the veto condition occurs, the softer of $i$ and $j$ is
discarded, while the harder one continues to participate in the
clustering.

Pruning bears similarity to filtering in that it reduces the
contamination of soft noise in a jet while aiming to retain hard
perturbative radiation within the jet.
However, because by default the parameters for the noise removal
depend on the original mass of the jet, the type of radiation that is
discarded depends significantly on the initial jet structure.
As a result pruning, in its default form, is better thought of as a
noise-removing boosted-object tagger (to be used in conjunction with a
pruned-jet mass cut) rather than a generic noise-removal procedure.

The \ttt{fastjet::Pruner} class, derived from \ttt{Transformer}, can be used as
follows, using a \ttt{JetAlgorithm} and two \ttt{double} parameters:
\begin{lstlisting}
  #include "fastjet/tools/Pruner.hh"
  // ...
  Pruner pruner(jet_algorithm, zcut, Rcut_factor);
  // ...
  PseudoJet pruned_jet = pruner(jet);
\end{lstlisting}
The \ttt{pruned\_jet} will have a valid associated cluster sequence, so that one
can, for instance, ask for its constituents with
\ttt{pruned\_jet.constituents()}.
In addition, the subjets that have been rejected by the pruning algorithm (i.e.
have been `pruned away') can be obtained with
\begin{lstlisting}
  vector<PseudoJet> rejected_subjets = pruned_jet.structure_of<Pruner>().rejected();
\end{lstlisting}
and each of these subjets will also have a valid associated clustering sequence.

When using the constructor given above, the jet radius used by the pruning clustering 
sequence is set internally to the functional equivalent of infinity. Alternatively, 
a pruner transformer can be constructed with a \ttt{JetDefinition} instead of just a
\ttt{JetAlgorithm}:
\begin{lstlisting}
  JetDefinition pruner_jetdef(jet_algorithm, Rpruner);
  Pruner pruner(pruner_jetdef, zcut, Rcut_factor);
\end{lstlisting}
In this situation, the jet definition \ttt{pruner\_jetdef} should normally have a radius
\ttt{Rpruner}
large enough to ensure that
all the constituents of the jet being pruned are reclustered into a single jet. 
If this is not the case, pruning is applied to the entire reclustering
and it is the hardest resulting pruned jet that is returned; the
others can be retrieved using
\begin{lstlisting}
  vector<PseudoJet> extra_jets = pruned_jet.structure_of<Pruner>().extra_jets();
\end{lstlisting}

Finally, note that a third constructor for \ttt{Pruner} exists, that allows one
to construct the pruner using functions that dynamically compute \ttt{zcut} and
\ttt{Rcut} for the jet being pruned:
\begin{lstlisting}
  Pruner (const JetDefinition &jet_def, 
          FunctionOfPseudoJet< double > *zcut_dyn, 
	  FunctionOfPseudoJet< double > *Rcut_dyn);
\end{lstlisting}

\subsection{Boosted-object taggers}
\label{sec:taggers}

A number of the taggers developed to distinguish 2- or 3-pronged
decays of massive objects from plain QCD jets (see the review
\cite{Abdesselam:2010pt}) naturally fall into the category of
transformers.
Typically they search for one or more hard branchings within the jet
and then return the part of the jet that has been identified as
associated with those hard branchings.
They share the convention that if they were not able to identify
suitable substructure, they return a \ttt{jet} with zero momentum,
i.e.\ one that has the property \ttt{jet == 0}.

At the time of writing, we provide only a small set of taggers.
These include one main two-body tagger, the
\ttt{fastjet::MassDropTagger} introduced in \cite{BDRS} and one main
boosted top tagger, \ttt{fastjet::JHTopTagger} from
\cite{Kaplan:2008ie} (\ttt{JHTopTagger} derives from the
\ttt{fastjet::TopTaggerBase} class, expressly included to provide a
common framework for all top taggers capable of also returning a $W$).
In addition, to help provide a more complete set of examples of coding
methods to which users may refer when writing their own taggers, we
have also included the
\ttt{fastjet::CASubJetTagger} introduced in~\cite{Butterworth:2009qa},
which illustrates the use of a \ttt{WrappedStructure} (cf.\
appendix~\ref{sec:transformerdetails}) and the
rest-frame \ttt{fastjet::RestFrameNSubjettinessTagger}
from Ref.~\cite{nsubtagger}, which makes use of facilities to boost a
cluster sequence.

We refer the reader to the original papers for a more extensive
description of the physics use of these taggers.

More taggers may be provided in the future, either through native
implementations or, potentially, through a ``contrib'' type area.
Users are invited to contact the \fastjet authors for further
information in this regard.

\subsubsection{The mass-drop tagger}

Introduced in \cite{BDRS} for the purpose of identifying a boosted Higgs
decaying into a $b\bar b$ pair, this is a general 2-pronged tagger. It starts with a fat jet obtained
with a Cambridge/Aachen algorithm (originally, $R=1.2$ was suggested
for boosted Higgs tagging). Tagging then proceeds as follows:
\begin{enumerate}
\item the last step of the clustering is undone: $j \to j_1,j_2$, with
  $m_{j_1} > m_{j_2}$;
\item if there is a significant mass drop, $\mu \equiv m_{j_1}/m_j <
  \mu_{\rm cut}$, and the splitting is sufficiently symmetric,
  $y\equiv {\rm min}(p_{tj_1}^2, p_{tj_2}^2)\Delta R_{j_1j_2}^2/m_j^2
  > y_{\rm cut}$, then $j$ is the resulting heavy particle candidate
  with $j_1$ and $j_2$ its subjets;
\item otherwise, redefine $j$ to be equal to $j_1$ and go back to step
  1.
\end{enumerate}
The tagger can be constructed with
\begin{lstlisting}
  #include "fastjet/tools/MassDropTagger.hh"
  // ...
  MassDropTagger mdtagger(double $\mu_\text{cut}$, double $y_\text{cut}$);
\end{lstlisting}
and applied using
\begin{lstlisting}
  PseudoJet tagged_jet = mdtagger(jet);
\end{lstlisting}

This tagger will run with any jet that comes from a \CS. A warning
will be issued if the \CS is not based on the C/A algorithm.
If the \ttt{JetDefinition} used in the \CS involved a non-default
recombiner, that same recombiner will be used when joining the final
two prongs to form the boosted particle candidate.

For a jet that is returned by the tagger and has the property that
\ttt{tagged\_jet != 0}, two enquiry functions can be used to return
the actual value of $\mu$ and $y$ for the clustering that corresponds
to the tagged structure:
\begin{lstlisting}
  tagged_jet.structure_of<MassDropTagger>.mu();
  tagged_jet.structure_of<MassDropTagger>.y(); 
\end{lstlisting}

Note that in \cite{BDRS} the mass-drop element of the tagging was
followed by a filtering stage using $\min(0.3, R_{jj}/2)$ as the
reclustering radius and selecting the three hardest subjects. That can
be achieved with
\begin{lstlisting}
  vector<PseudoJet> tagged_pieces = tagged_jet.pieces();
  double Rfilt = min(0.3, 0.5 * pieces[0].delta_R(pieces[1]));
  PseudoJet filtered_tagged_jet = Filter(Rfilt, SelectorNHardest(3))(tagged_jet);
\end{lstlisting}
(It is also possible to use the \ttt{Rfilt\_dyn} option to the filter
discussed in section~\ref{sec:filtering}).

\subsubsection{The Johns-Hopkins top tagger}

The Johns Hopkins top tagger~\cite{Kaplan:2008ie} is a 3-pronged tagger
specifically designed to identify top quarks.
It recursively breaks a jet into pieces, finding up to 3 or 4 subjets
and then looking for a $W$ candidate among them.
The parameters used to identify the relevant subjets include a
momentum fraction cut and a minimal separation in Manhattan distance
($|\Delta y| + |\Delta \phi|$) between subjets obtained from a
declustering.

The tagger will run with any jet that comes from a \CS, however
to conform with the original formulation of~\cite{Kaplan:2008ie}, the
\CS should be based on the C/A algorithm. A warning will be issued if
this is not the case.
If the \ttt{JetDefinition} used in the \CS involves a non-default
recombiner, that same recombiner will be used when joining the final
two prongs to form the boosted particle candidate.
The tagger can be used as follows:
\begin{lstlisting}
  #include "fastjet/tools/JHTopTagger.hh"
  // ...
  double delta_p = 0.10; // subjets must carry at least this fraction of original jet's $p_t$
  double delta_r = 0.19; // subjets must be separated by at least this Manhattan distance
  double cos_theta_W_max = 0.7; // the maximal allowed value of the W helicity angle
  JHTopTagger top_tagger(delta_p, delta_r, cos_theta_W_max);
  // indicate the acceptable range of top, W masses 
  top_tagger.set_top_selector(SelectorMassRange(150,200));
  top_tagger.set_W_selector  (SelectorMassRange( 65, 95));
  // now try and tag a jet
  PseudoJet top_candidate = top_tagger(jet); // jet should come from a C/A clustering
  if (top_candidate != 0) { // successful tagging
    double top_mass = top_candidate.m();
    double W_mass   = top_candidate.structure_of<JHTopTagger>().W().m();
  }
\end{lstlisting}
Other information available through the
\ttt{structure\_of<JHTopTagger>()} call includes: \ttt{W1()} and
\ttt{W2()}, the harder and softer of the two $W$ subjets;
\ttt{non\_W()}, the part of the top that has not been identified with
a $W$ (i.e.\ the candidate for the $b$); and \ttt{cos\_theta\_W()}.
The \ttt{top\_candidate.pieces()} call will return 2 pieces, where the
first is the $W$ candidate (identical to
\ttt{structure\_of<JHTopTagger>().W()}), while the second is the
remainder of the top jet (i.e.\ \ttt{non\_W}).

Note the above calls to \ttt{set\_top\_selector()} and
\ttt{set\_W\_selector()}. If these calls are not made, then the tagger
places no cuts on the top or $W$ candidate masses and it is then the
user's responsibility to verify that they are in a suitable range.

Note further that  \ttt{JHTopTagger} does not derive directly from 
\ttt{Transformer}, but from the
\ttt{fastjet::TopTaggerBase} class instead. This class (which itself derives
from \ttt{Transformer}) has been included to provide a proposed common
interface for all the top taggers. In particular, \ttt{TopTaggerBase} provides
(via the associated structure)
\begin{lstlisting}
  top_candidate.structure_of<TopTaggerBase>().W()
  top_candidate.structure_of<TopTaggerBase>().non_W()
\end{lstlisting}
and standardises the fact that the resulting top candidate is a \ttt{PseudoJet}
made of these two pieces.

The benefits of the base class for top taggers will of course be more
evident once more than a single top tagger has been implemented.

\subsubsection{The Cambridge/Aachen subjet tagger}

The Cambridge/Aachen subjet
tagger~\cite{Butterworth:2009qa}, originally implemented in a
3-pronged context, is really a generic 2-body tagger, which can also be
used in a nested fashion to obtained multi-pronged tagging.
It can be obtained through the include
\begin{lstlisting}
  #include "fastjet/tools/CASubjetTagger.hh"
\end{lstlisting}
As it is less widely used than the taggers mentioned above, we refer
the user to the online doxygen documentation for further details.

\subsubsection{The rest-frame $N$-subjettiness tagger}

The rest-frame $N$-subjettiness
tagger~\cite{nsubtagger}, meant to identify a highly boosted colour
singlet particle decaying to $2$ partons, can be obtained through the include
\begin{lstlisting}
  #include "fastjet/tools/RestFrameNSubjettinessTagger.hh"
\end{lstlisting}
As it is less widely used than the taggers mentioned above, we refer
the user to the online doxygen documentation for further details.

\section{Compilation notes}

Compilation and installation make use of the standard
\begin{verbatim}
  % ./configure
  % make
  % make check
  % make install
\end{verbatim}
procedure. Explanations of available options are given in the
\ttt{INSTALL} file in the top directory, and a list can also be obtained running
\ttt{./configure --help}.

In order to access the \ttt{NlnN} strategy for the $k_t$ algorithm,
the \fastjet library needs to be compiled with support for the
Computational Geometry Algorithms Library \ttt{CGAL} \cite{CGAL}. This
same strategy gives $N\ln N$ performance for Cambridge/Aachen and
$N^{3/2}$ performance for anti-$k_t$ (whose sequence for jet
clustering triggers a worst-case scenario for the underlying
computational geometry methods.)
CGAL can be enabled with the \verb|--enable-cgal| at the
\ttt{configure} stage.
\ttt{CGAL} may be obtained in source form from
\url{http://www.cgal.org/} and is also available in binary form for
many common Linux distributions.
For CGAL versions 3.4 and higher, the user can specify
\verb|--with-cgaldir=...| if the CGAL files are not installed in a
standard location.\footnote{For events with near degeneracies in their
  Delaunay triangulation, issues have been found with versions 3.7 and
  3.8 of CGAL. We recommend the use of earlier or later versions.}

The \ttt{NlnNCam} strategy does not require CGAL, since it is based on
a considerably simpler computational-geometry structure~\cite{Chan}.

\section*{Acknowledgements}

Many people have provided bug reports, suggestions for development and
in some cases explicit code for plugin algorithms. We would in
particular like to thank
%
%
Vanya Belyaev,
Andy Buckley,
Timothy Chan,
Pierre-Antoine Delsart,
Olivier Devillers,
Robert Harris,
Joey Huston,
Sue Ann Koay,
Andreas Oehler,
Sal Rappoccio,
Juan Rojo,
Sebastian Sapeta,
Mike Seymour,
Jessie Shelton,
Lars Sonnenschein,
Hartmut Stadie,
Chris Vermilion,
Markus Wobisch.

Since its inception, this project has been supported in part by grants
ANR-05-JCJC-0046-01, ANR-09-BLAN-0060 and ANR-10-CEXC-009-01 from the
French Agence Nationale de la Recherche, PITN-GA-2010-264564 from the
European Commission and DE-AC02-98CH10886 from the U.S.\ Department of
Energy.

We would also like to thank the numerous institutes that have hosted
us for shorter or longer stays while \fastjet was being developed,
including the GGI in Florence, KITP at Santa Barbara, Rutgers
University and Brookhaven National Laboratory.
\appendix

\section{Clustering strategies and performance}
\label{app:strategies}

The constructor for a \ttt{JetDefinition} can take a strategy argument
(cf.\ section~\ref{sec:JetDefinition}), which selects the algorithmic
``strategy'' to use while clustering.
It is an \ttt{enum} of type \ttt{Strategy} with relevant
values listed in table~\ref{tab:Strategies}.
\begin{table}
  \begin{center}
    \begin{tabular}{ll}\toprule
      \ttt{N2Plain} & a plain $N^2$ algorithm (fastest for $N
      \lesssim 30$)\\
      \ttt{N2Tiled} & a tiled $N^2$ algorithm (fastest for $30 \lesssim
      N \lesssim 400$)\\ 
      \ttt{N2MinHeapTiled} & a tiled $N^2$ algorithm with a heap for
      tracking the minimum of \\& $d_{ij}$ (fastest for $400 \lesssim
      N \lesssim 15000$)\\ 
      \ttt{NlnN} & the Voronoi-based $N\ln N$ algorithm (fastest for $
      N \gtrsim 15000$)\\ 
      \ttt{NlnNCam} & based on Chan's $N\ln N$ closest pairs
      algorithm (fastest for \\&$ 
      N \gtrsim 6000$), suitable only for the Cambridge jet algorithm\\ 
      \ttt{Best} & automatic selection of the best of these based on
      $N$ and $R$\\\bottomrule
    \end{tabular}
    \caption{The more interesting of the various algorithmic
      strategies for clustering. Other strategies are given
      \ttt{JetDefinition.hh} --- note however that strategies not
      listed in the above table may disappear in future releases.
      For jet algorithms with spherical distance measures (those whose
      name starts with ``\ttt{ee\_}''), only the \ttt{N2Plain} strategy is
      available.  }
    \label{tab:Strategies}
  \end{center}
\end{table}
Nearly all strategies are based on the factorisation of energy and
geometrical distance components of the $d_{ij}$
measure~\cite{fastjet}. In particular they involve the dynamic
maintenance of a nearest-neighbour graph for the geometrical
distances.  They apply equally well to any of the internally
implemented hadron-collider jet algorithms.
The one exception is \ttt{NlnNCam}, which is based on a computational
geometry algorithm for dynamic maintenance of closest pairs
\cite{Chan} (rather than the more involved nearest neighbour graph),
and is suitable only for the Cambridge algorithm, whose distance
measure is purely geometrical.

The \ttt{N2Plain} strategy uses a ``nearest-neighbour heuristic''
\cite{Anderberg} approach to maintaining the geometrical
nearest-neighbour graph; \ttt{N2Tiled} tiles the $y-\phi$ cylinder
to limit the set of points over which nearest-neighbours are searched
for,%
\footnote{Tiling is a textbook approach in computational geometry,
  where it is often referred to as bucketing. It has been used also in
  certain cone jet algorithms, notably at trigger level and in
  \cite{Sonnenschein}.} %
and \ttt{N2MinHeapTiled} differs only in that it uses an $N\ln N$
(rather than $N^2$) data structure for maintaining in order the subset
of the $d_{ij}$ that involves nearest neighbours.
The \ttt{NlnN} strategy uses CGAL's Delaunay Triangulation
\cite{CGAL} for the maintenance of the nearest-neighbour graph.
Note that $N \ln N$ performance of is an \emph{expected} result, and
it holds in practice for the $k_t$ and Cambridge algorithms, while for
anti-$k_t$ and generalised-$k_t$ with $p<0$, hub-and-spoke
(or bicycle-wheel) type configurations emerge dynamically during the
clustering and these break the conditions needed for the expected
result to hold (this however has a significant impact only for $N
\gtrsim 10^5$).
A further comment about the $N \ln N$ strategy is that it currently
has the limitation that it cannot be used in events with perfectly
collinear particles.
This is related to the fact that the underlying computation geometry
structures cannot cleanly accommodate multiple particles in the same
location, because of the degeneracies that are induced. 
A workaround for this problem may be provided on request.

If \ttt{strategy} is omitted then the \ttt{Best} option is set.  
Note that the $N$ ranges quoted in table~\ref{tab:Strategies} for
which a given strategy is optimal hold for $R=1$; the general $R$
dependence can be significant and non-trivial.
While some attempt has been made to account for the $R$-dependence in
the choice of the strategy with the ``\ttt{Best}'' option, there may
exist specific regions of $N$ and $R$ in which a manual choice of
strategy can give faster execution.
Furthermore the \ttt{NlnNCam} strategy's timings may depend
strongly on the size of the cache.
Finally for a given $N$ and $R$, the optimal strategy may also depend
on the event structure.

\begin{figure}[t]
  \centering
  \includegraphics[width=0.7\textwidth]{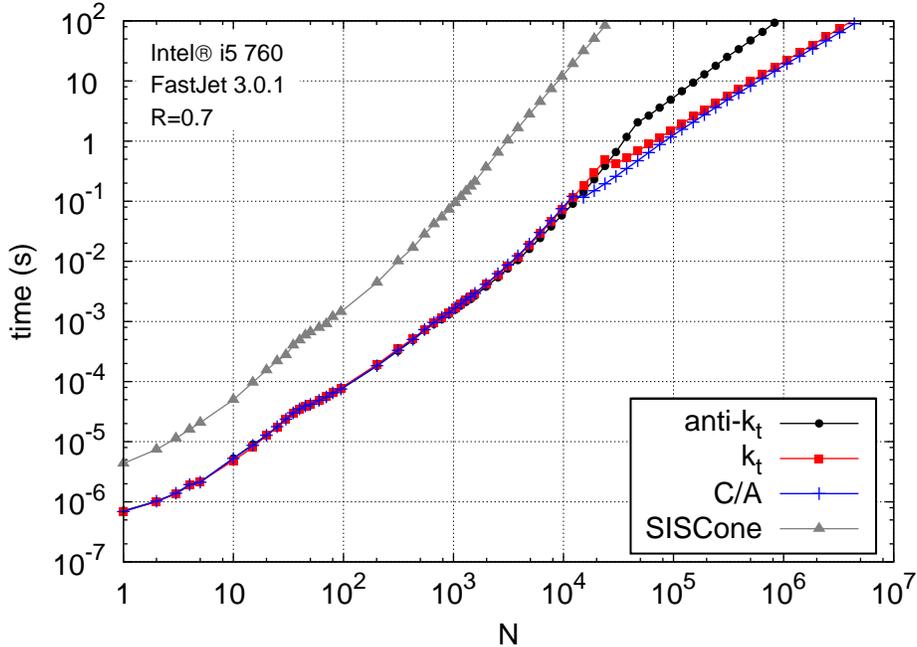}
  \caption{Time required to perform the clustering of $N$ particles in
    \fastjet 3.0.1 with the \ttt{Best} strategy. The anti-$k_t$,
    $k_t$, and Cambridge/Aachen (C/A) native algorithms are shown,
    together with the SISCone plugin.  All use $R=0.7$.
    Shown for an Intel i5 760 processor with 8~MB of
    cache.
    For small $N$, $N$ was varied by taking a single hard dijet event
    generated with Pythia~6~\cite{Sjostrand:2006za} and extracting the
    $N$ hardest particles. 
    Large $N$ values were obtained by taking a single hard dijet event
    and adding simulated minimum-bias events to it.  
    The results include the time to extract the inclusive jets with
    $p_t > 5\GeV$ and sort them into decreasing $p_t$.  }
\label{fig:timings}
\end{figure}

Illustrative timings for the \ttt{Best} strategy are shown as a
function of $N$ in figure~\ref{fig:timings} for the anti-$k_t$, $k_t$
and the Cambridge/Aachen algorithms. Results for the SISCone plugin
are given for comparison purposes.
Kinks in the timings of the native algorithms are visible at the $N$
values where there is a switch from one strategy to another.
There can be imperfections in this choice, e.g. as seen for the $k_t$
algorithm near $N=20\,000$. 
While their impact is generally modest, depending on event structure
there can be cases where a manual choice of strategy can have
significant benefits.

We note that there are a few places where there remains scope for
timing improvements. 
In particular at low $N$ the overheads related to copying and sorting
of a vector of \ttt{PseudoJet} objects are a substantial fraction of the
total time, and could be reduced.
Additionally, for the Cambridge/Aachen algorithm at moderate to large
$N$, the use of multiple grid sizes could bring an $\order{1}$
benefit;
for the anti-$k_t$ algorithm one can envisage $\order{1}$ improvements
at moderate to large $N$ when $N$ is dominated by ghost particles,
making use of the fact that for the anti-$k_t$ algorithm one may
neglect the ghosts' self clustering in the determination of hard jets'
areas.
Should users have applications where such improvements would be
critical, they are encouraged to contact the \fastjet authors.

\section{User Info in PseudoJets}
\label{app:user-info}

One method for associating extra user information with a
\ttt{PseudoJet} is via its user index
(section~\ref{sec:PseudoJet}). This is adequate for encoding simple
information. such as an input particle's barcode in a HepMC
event.
However, it can quickly show its limitations; for example, when
simulating pileup one might have several HepMC events and it is then
useful for each particle to additionally store information about which
HepMC event it comes from.

A second method for supplementing a \PJ with extra user information is for
the user to derive a class from \ttt{PseudoJet::UserInfoBase} and
associate the \ttt{PseudoJet} with a pointer to an instance of that
class:
\begin{lstlisting}
  void set_user_info(UserInfoBase * user_info);
  const UserInfoBase* user_info_ptr() const;
\end{lstlisting}
The function \ttt{set\_user\_info(...)} transfers ownership of the
pointer to the \ttt{PseudoJet}. 
This is achieved internally with the help of a shared pointer. Copies
of the \ttt{PseudoJet} then point to the same \ttt{user\_info}.
When the \ttt{PseudoJet} and all its copies go out of scope, the
\ttt{user\_info} is automatically deleted.
Since nearly all practical uses of \ttt{user\_info} require it
to be cast to the relevant derived class of  \ttt{UserInfoBase}, we also
provide the following member function for convenience:
\begin{lstlisting}
  template<class L> const L & user_info() const;
\end{lstlisting}
which explicitly performs the cast of the extra info to type \ttt{L}.
If the cast fails, or the user info has not been set, an error
will be thrown.\footnote{%
  For clustering with explicit ghosts, even if the particles being
  clustered have user information, the ghosts will not. 
  The user should take care therefore not to ask for user information
  about the ghosts, e.g.\ with the help of the \texttt{PseudoJet::is\_pure\_ghost()}
  or \texttt{PseudoJet::has\_user\_info<L>()} calls.
  The \texttt{SelectorIsPureGhost()} can also be used for this purpose.
}

The user may wonder why we have used shared pointers internally (i.e.\
have ownership transferred to the \ttt{PseudoJet}) rather than normal
pointers.
An example use case where the difference is important is if, for
example, one wishes to write a \ttt{Recombiner} that sets the
\ttt{user\_info} in the recombined \PJ.
Since this is likely to be new information, the \ttt{Recombiner} will
have to allocate some memory for it. 
With a normal pointer, there is then no easy way to clean up that
memory when the \PJ is no longer relevant (e.g.\ because the \CS that
contains it has gone out of scope). 
In contrast, with a shared pointer the memory is handled automatically.\footnote{
  The user may also wonder why we didn't simply write a templated
  version of \PJ in order to contain extra information.
  The answer here is that to introduce a templated \PJ would imply
  that every other class in \fastjet should then also be templated. 
}

The shared pointer type in \fastjet is a template class called
\ttt{SharedPtr}, available through 
\begin{lstlisting}
  #include "fastjet/SharedPtr.hh"
\end{lstlisting}
It behaves almost identically to the \ttt{C++0x} \ttt{shared\_ptr}.\footnote{
Internally it has been designed somewhat differently, in order
to limit the memory footprint of the \PJ that contains it. One
consequence of this is that dynamic casts of \ttt{SharedPtr}'s are not
supported.}
The end-user should not usually need to manipulate the
\ttt{SharedPtr}, though the \ttt{SharedPtr} to \ttt{user\_info} is
accessible through \PJ's \ttt{user\_info\_shared\_ptr()} member.

An example of the usage might be the following. First you define a
class \ttt{MyInfo}, derived from \ttt{PseudoJet::UserInfo},
\begin{lstlisting}
  class MyInfo: public PseudoJet::UserInfoBase {
     MyInfo(int id) : _pdg_id(id);
     int pdg_id() const {return _pdg_id;}
     int _pdg_id;
  };
\end{lstlisting}
Then you might set the info as follows
\begin{lstlisting}
  PseudoJet particle(...);
  particle.set_user_info(new MyInfo(its_pdg_id));
\end{lstlisting}
and later access the PDG id through the function
\begin{lstlisting}
  particle.user_info<MyInfo>().pdg_id();
\end{lstlisting}
More advanced examples can be provided on request, including code that
helps handle particle classes from third party tools such as
Pythia~8~\cite{Sjostrand:2007gs}.

\section{Structural information for various kinds of \texttt{PseudoJet}}
\label{app:structure_table}

\begin{table}[t]\centering
\begin{tabular}{lccccccc}
\toprule
  && particle & jet & jet (no CS) & constituent & \ttt{join(}$j_1,j_2$\ttt{)} & \ttt{join(}$p_1,p_2$\ttt{)} \\
\midrule
\ttt{has\_associated\_cs()}
                     && false   & true    &  true   & true    & false    & false   \\
\ttt{associated\_cs()}     
                     && NULL    &  CS     &  NULL   &  CS     & NULL     & NULL    \\
\midrule                       
\ttt{has\_valid\_cs()}
                     && false   & true    &  false  & true    & false    & false   \\
\ttt{validated\_cs()}&& \throws &  CS     & \throws &  CS     & \throws  & \throws \\
\midrule                       
\ttt{has\_constituents()}
                     && false   & true    & true    & true    & true     & true    \\
\ttt{constituents()} && \throws & from CS & \throws & itself  & recurse  & pieces  \\
\midrule                       
\ttt{has\_pieces()}  && false   & true    & \throws & false   & true     & true    \\ 
\ttt{pieces()}       && \throws & parents & \throws & empty   & pieces   & pieces  \\ 
\midrule                       
\ttt{has\_parents(...)}
                     && \throws & from CS & \throws & from CS & \throws  & \throws \\ 
\ttt{has\_child(...)}&& \throws & from CS & \throws & from CS & \throws  & \throws \\ 
\ttt{contains(...)}  && \throws & from CS & \throws & from CS & \throws  & \throws \\ 
\bottomrule
\end{tabular}
\caption{summary of the behaviour obtained when requesting
  structural information from different kinds of \ttt{PseudoJet}. A
  particle (also $p_1,p_2$) is a \ttt{PseudoJet}
  constructed by the user, without structural information; a ``jet''
  (also $j_1,j_2$) is the output from a
  \ttt{ClusterSequence}; ``from CS'' means that the information is
  obtained from the associated \ttt{ClusterSequence}. 
  A ``jet (no CS)'' is one whose \ttt{ClusterSequence} has gone out of
  scope.
  All other entries should be self-explanatory.}\label{tab:structure}
\end{table}

Starting with \fastjet version 3.0, a \ttt{PseudoJet} can access
information about its structure, for example its constituents if it came
from a \ClusterSequence, or its pieces if it was the result of a
\ttt{join(...)} operation.
In this appendix, we summarise what the various structural access
methods will return for different types of \ttt{PseudoJet}s: input
particles, jets resulting from a clustering, etc.
Table \ref{tab:structure} provides the information for the most
commonly-used methods. 

Additionally, all the methods that access information related to the
clustering (\ttt{has\_partner()}, \ttt{is\_inside()},
\ttt{has\_exclusive\_subjets()}, \ttt{exclusive\_subjets()},
\ttt{n\_exclusive\_subjets()}, \ttt{exclusive\_subdmerge()}, and
\ttt{exclusive\_subdmerge\_max}) require the presence of an associated
cluster sequence and throw an error if none is available (except for
\ttt{has\_exclusive\_subjets()} which just returns \ttt{false}).

For area-related calls, \ttt{has\_area()} will be \ttt{false} unless
the jet is obtained from a \ttt{ClusterSequenceAreaBase} or is a
composite jet made from such jets. All other area calls
(\ttt{validated\_csab()}, \ttt{area()}, \ttt{area\_error()},
\ttt{area\_4vector()}, \ttt{is\_pure\_ghost()}) will return the
information from the \ttt{ClusterSequenceAreaBase}, or from the pieces
in case of a composite jet. An error will be thrown if the jet does
not have area information.

\paragraph{Internal storage of structural information.}
The means by which information about a jet's structure is stored is
generally transparent to the user.
The main exception that arises is when the user wishes to create jets
with a new kind of structure, for example when writing boosted-object
taggers.
Here, we simply outline the approach adopted. For concrete usage
examples one can consult section~\ref{sec:transformers} and appendix \ref{sec:transformerdetails},
where we discuss transformers and taggers.

To be able to efficiently access structural information, each \PJ has
a shared pointer to a class of type \ttt{fastjet::PseudoJetStructureBase}.
For plain {\PJ}s the pointer is null.
For {\PJ}s obtained from a \CS the pointer is to a class
\ttt{fastjet::ClusterSequenceStructure}, which derives from
\ttt{PseudoJetStructureBase}.
For {\PJ}s obtained from a \ttt{join(...)} operation, the pointer is
to a class \ttt{fastjet::CompositeJetStructure}, again derived from
\ttt{PseudoJetStructureBase}.
It is these classes that are responsible for answering structural
queries about the jet, such as returning its constituents, or
indicating whether it \ttt{has\_pieces()}.
Several calls are available for direct access to the internal structure
storage, among them
\begin{lstlisting}
  const PseudoJetStructureBase* structure_ptr() const;
  PseudoJetStructureBase*       structure_non_const_ptr();
  template<typename StructureType> const StructureType & structure() const;
  template<typename TaggerType> const TaggerType::StructureType & structure_of() const;
\end{lstlisting}
where the first two return simply the structure pointer, while the
last two cast the pointer to the desired derived structure type.

\section{Functions of a \texttt{PseudoJet}}
\label{app:function-of-pj}

A concept that is new to \fastjet 3 is that of a
\ttt{fastjet::FunctionOfPseudoJet}.
Functions of \ttt{PseudoJet}s arise in many contexts: many
boosted-object taggers take a jet and return a modified version of a
jet; background subtraction does the same; so does a simple Lorentz
boost.
Other functions return a floating-point number associated
with the jet: for example jet shapes, but also the rescaling functions
used to provide local background estimates in
section~\ref{sec:BGE-positional}.

To help provide a uniform interface for functions of a \PseudoJet,
\fastjet has the following template base class:
\begin{lstlisting}
  // a generic function of a PseudoJet
  template<typename TOut> class FunctionOfPseudoJet{
    // the action of the function (this _has_ to be overloaded in derived classes)
    virtual TOut result(const PseudoJet &pj) const = 0;
  };
\end{lstlisting}
Derived classes should implement the \ttt{result(...)} function. 
In addition it is good practice to overload the \ttt{description()}
member,
\begin{lstlisting}
  virtual std::string description() const {return "";}
\end{lstlisting}

Usage of a \ttt{FunctionOfPseudoJet} is simplest through the
\ttt{operator(...)} member functions
\begin{lstlisting}
  TOut operator()(const PseudoJet & pj) const;
  vector<TOut> operator()(const vector<PseudoJet> & pjs) const;
\end{lstlisting}
which just call \ttt{result(...)} either on the single jet, or
separately on each of the elements of the vector of
{\PseudoJet}s.\footnote{Having \ttt{result(...)} and
  \ttt{operator(...)} doing the same thing may seem redundant,
  however, it allows one to redefine only \ttt{result} in derived
  classes.  If we had had a virtual \ttt{operator(...)} instead, both
  the \ttt{PseudoJet} and \ttt{vector<PseudoJet>} versions would have
  had to be overloaded.}.

The \ttt{FunctionOfPseudoJet} framework makes it straightforward to
pass functions of \ttt{PseudoJet}s as arguments. This is, e.g.,
used for the background rescalings in section~\ref{sec:BGE-positional},
which are just derived from \ttt{FunctionOfPseudoJet<double>}.
It is also used for the \ttt{Transformer}s of
section~\ref{sec:transformers}, which all derive from
\ttt{FunctionOfPseudoJet<PseudoJet>}.
The use of a class for these purposes, rather than a pointer to a
function, provides the advantage that the class can be initialised
with additional arguments.

\section{User-defined extensions of \fastjet}

\subsection{External Recombination Schemes}
\label{sec:recombiner}

A user who wishes to introduce a new recombination scheme may
do so by writing a class derived from \ttt{JetDefinition::Recombiner}:
\begin{lstlisting}
  class JetDefinition::Recombiner {
  public:
    /// return a textual description of the recombination scheme implemented here
    virtual std::string description() const = 0;
    
    /// recombine pa and pb and put result into pab
    virtual void recombine(const PseudoJet & pa, const PseudoJet & pb, 
                           PseudoJet & pab) const = 0;

    /// routine called to preprocess each input jet (to make all input
    /// jets compatible with the scheme requirements (e.g. massless).
    virtual void preprocess(PseudoJet & p) const {};
    
    /// a destructor to be replaced if necessary in derived classes...
    virtual ~Recombiner() {};
  };
\end{lstlisting}
A jet definition can then be constructed by providing a pointer to an
object derived from \ttt{JetDefinition::Recombiner} instead of the
\ttt{RecombinationScheme} index:
\begin{lstlisting}
  JetDefinition(JetAlgorithm jet_algorithm, 
                double R, 
                const JetDefinition::Recombiner * recombiner,
                Strategy strategy = Best);
\end{lstlisting}
The derived class \ttt{JetDefinition::DefaultRecombiner} is what is
used internally to implement the various recombination schemes if an
external \ttt{Recombiner} is not provided. It provides a useful
example of how to implement a new \ttt{Recombiner} class.

The recombiner can also be set with a \ttt{set\_recombiner(...)} call.
If the recombiner has been created with a \ttt{new} statement and the
user does not wish to manage the deletion of the corresponding memory
when the \ttt{JetDefinition} (and any copies) using the recombiner
goes out of scope, then the user may wish to call the
\ttt{delete\_recombiner\_when\_unused()} function, which tells the
\ttt{JetDefinition} to acquire ownership of the pointer to the
recombiner and delete it when it is no longer needed.


\subsection{Implementation of a plugin jet algorithm}
\label{sec:new-plugin}

The base class from which plugins derive has the following structure:
\begin{lstlisting}
  class JetDefinition::Plugin{
  public:
    /// returns a textual description of the jet-definition implemented in this plugin
    virtual std::string description() const = 0;
  
    /// given a ClusterSequence that has been filled up with initial particles, 
    /// the following function should fill up the rest of the ClusterSequence, 
    /// using the following member functions of ClusterSequence:
    ///   - plugin_do_ij_recombination(...)
    ///   - plugin_do_iB_recombination(...)
    virtual void run_clustering(ClusterSequence &) const = 0;
  
    /// a destructor to be replaced if necessary in derived classes...
    virtual ~Plugin() {};

    //------- ignore what follows for simple usage! ---------
    /// returns true if passive areas can be efficiently determined by
    /// (a) setting the ghost_separation scale (see below)
    /// (b) clustering with many ghosts with $p_t$ $\ll$ ghost_separation_scale
    /// (c) counting how many ghosts end up in a given jet
    virtual bool supports_ghosted_passive_areas() const {return false;}

    /// sets the ghost separation scale for passive area determinations
    /// in future runs (NB: const, so should set internal mutable var)
    virtual void set_ghost_separation_scale(double scale) const;
    virtual double ghost_separation_scale() const;

  };
\end{lstlisting}
Any plugin class must define the \ttt{description} and
\ttt{run\_clustering} member functions. The former just returns a
textual description of the jet algorithm and its options (e.g.\ radius,
etc.), while the latter does the hard work of running the user's own
jet algorithm and transferring the information to the
\ttt{ClusterSequence} class. This is best illustrated with an example:
\begin{lstlisting}
using namespace fastjet;

void CDFMidPointPlugin::run_clustering(ClusterSequence & clust_seq) {
  
  // when run_clustering is called, the clust_seq.jets() has already been
  // filled with the initial particles
  const vector<PseudoJet> & initial_particles = clust_seq.jets();

  // it is up to the user to do their own clustering on these initial particles
  // ...
\end{lstlisting}
Once the plugin has run its own clustering it must transfer the
information back to the \ttt{clust\_seq}. This is done by recording
mergings between pairs of particles or between a particle and the
beam. The new momenta are stored in the \ttt{clust\_seq.jets()}
vector, after the initial particles. Note though that the plugin is
not allowed to modify \ttt{clust\_seq.jets()} itself. Instead it must
tell \ttt{clust\_seq} what recombinations have occurred, via the
following (\ttt{ClusterSequence} member) functions
\begin{lstlisting}
  /// record the fact that there has been a recombination between jets()[jet_i] 
  /// and jets()[jet_j], with the specified dij, and return the index (newjet_k) 
  /// allocated to the new jet. The recombined PseudoJet is determined by 
  /// applying the JetDefinition's recombiner to the two input jets.
  /// (By default E-scheme recombination, i.e. a 4-vector sum)
  void plugin_record_ij_recombination(int jet_i, int jet_j, double dij, int & newjet_k);

  /// as for the simpler variant of plugin_record_ij_recombination, except 
  /// that the new jet is attributed the momentum and user information of newjet
  void plugin_record_ij_recombination(int jet_i, int jet_j, double dij, 
				      const PseudoJet & newjet, int & newjet_k);

  /// record the fact that there has been a recombination between jets()[jet_i] 
  /// and the "beam", with the specified diB; this jet will then be returned to
  /// the user when they request inclusive_jets() from the cluster sequence.
  void plugin_record_iB_recombination(int jet_i, double diB);
\end{lstlisting}
The \ttt{dij} recombination functions return the index
\ttt{newjet\_k} of the newly formed pseudojet. The plugin may need to
keep track of this index in order to specify subsequent
recombinations.

Certain (cone) jet algorithms do not perform pairwise clustering ---
in these cases the plugin must invent a fictitious series of pairwise
recombinations that leads to the same final jets. Such jet algorithms
may also produce extra information that cannot be encoded in this way
(for example a list of stable cones), but to which one may still want
access. For this purpose, during \verb|run_clustering(...)|, the
plugin may call the \verb|ClusterSequence| member function:
\begin{lstlisting}
  inline void plugin_associate_extras(std::auto_ptr<ClusterSequence::Extras> extras);
\end{lstlisting}
where \verb|ClusterSequence::Extras| is an abstract base class, which
the plugin should derive from so as to provide the relevant information:
\begin{lstlisting}
  class ClusterSequence::Extras {
  public:
    virtual ~Extras() {}
    virtual std::string description() const;
  };
\end{lstlisting}
A method of \verb|ClusterSequence| then provides the user with access
to the extra information:
\begin{lstlisting}
  /// returns a pointer to the extras object (may be null) const
  ClusterSequence::Extras * extras() const;
\end{lstlisting}
The user should carry out a dynamic cast so as to convert the extras
back to the specific plugin extras class, as illustrated for
SISCone in section~\ref{sec:siscone-plugin}.

\subsubsection{Building new sequential recombination algorithms}
\label{sec:new-seq-rec}

To enable users to more easily build plugins for new sequential
recombination algorithms, \fastjet also provides a class \verb|NNH|,
which provides users with access to an implementation of the
nearest-neighbour heuristic for establishing and maintaining
information about the closest pair of objects in a dynamic set of
objects (see \cite{EppsteinHierarchical} for an introduction to this
and other generic algorithms).
In good cases (C/A-like) this allows one to construct clustering that runs in
$N^2$ time, though its worst case can be as bad as $N^3$ (e.g.\ anti-$k_t$).
It is a templated class and the template argument should be a class
that stores the minimal information for each jet so as to be able to
calculate interjet distances.
It underlies the implementations of the Jade and $\ee$ Cambridge
plugins.
The interested user should consult those codes for more information,
as well as the header for the \verb|NNH| class.

\subsection{Implementing new selectors}
\label{sec:new-selectors}

Technically a \ttt{Selector} contains a shared pointer to a
\ttt{SelectorWorker}.
Classes derived from \ttt{SelectorWorker} actually do the work. 
So, for example, the call to the function \ttt{SelectorAbsRapMax(2.5)}
first causes a new instance of the internal \ttt{SW\_AbsRapMax} class
to be constructed with the information that the limit on $|y|$
is 2.5 (\ttt{SW\_AbsRapMax} derives from \ttt{SelectorWorker}).
Then a \ttt{Selector} is constructed with a pointer to the
\ttt{SW\_AbsRapMax} object, and it is this \ttt{Selector} that is
returned to the user:
\begin{lstlisting}
  Selector SelectorAbsRapMax(double absrapmax) {
    return Selector(new SW_AbsRapMax(absrapmax));
  }
\end{lstlisting}
Since \ttt{Selector} is really nothing more than a shared pointer to
the \ttt{SW\_AbsRapMax} object, it is a lightweight object. 
The fact that it's a shared pointer also means that it looks after
the memory management issues associated with the \ttt{SW\_AbsRapMax}
object. 

If a user wishes to implement a new selector, they should write a
class derived from \ttt{SelectorWorker}.
The base is defined with sensible defaults, so for simple usage, only
two \ttt{SelectorWorker} functions need to be overloaded:
\begin{lstlisting}
  /// returns true if a given object passes the selection criterion.
  pass(const PseudoJet & jet) const = 0;

  /// returns a description of the worker
  virtual std::string description() const {return "missing description";}
\end{lstlisting}
For information on how to implement more advanced workers (for example
workers that do not apply jet-by-jet, or that take a reference), users
may wish to examine the extensive in-code documentation of
\ttt{SelectorWorker}, the implementation of the existing workers
and/or consult the authors.
A point to be aware of in the case of constructors that take a
reference is the need to implement the \ttt{SelectorWorker::copy()}
function.

\subsection{User-defined transformers}
\label{sec:transformerdetails}

All transformers are derived from  the \ttt{Transformer} base class, 
declared in the \ttt{fastjet/tools/Transformer.hh} header:
\begin{lstlisting}
  class Transformer : public FunctionOfPseudoJet<PseudoJet> {
  public:
    // the result of the Transformer acting on the PseudoJet.
    // this has to be overloaded in derived classes
    virtual PseudoJet result(const PseudoJet & original) const = 0;
  
    // should be overloaded to return a description of the Transformer
    virtual std::string description() const = 0;
  
    // information about the associated structure type
    typedef PseudoJetStructureBase StructureType;

    // destructor is virtual so that it can be safely overloaded
    virtual ~Transformer(){}
  };
\end{lstlisting}
Relative to the \ttt{FunctionOfPseudoJet<PseudoJet>} (cf.\
appendix~\ref{app:function-of-pj}) from which it derives, the
\ttt{Transformer}'s main additional feature is that the jets resulting
from the transformation are generally expected to have standard
structural information, e.g.\ constituents, and will often have
supplemental structural information, which the \ttt{StructureType}
\ttt{typedef} helps access.
As for a \ttt{FunctionOfPseudoJet<PseudoJet>}, the action of a
\ttt{Transformer} is to be implemented in the \ttt{result(...)} member
function,
though typically it will be used through the \ttt{operator()}
function, as discussed in appendix~\ref{app:function-of-pj}.

To help understand how to create user-defined transformers, it is
perhaps easiest to consider the example of a filtering/trimming class.
The simplest form of such a class is the following:%
\footnote{The actual \texttt{Filter} class is somewhat more elaborate
  than this, since it also handles areas, pileup subtraction and avoids
  reclustering when the jet and subjet definitions are C/A based.}
\begin{lstlisting}
  /// a simple class to carry out filtering and/or trimming
  class SimpleFilter: public Transformer {
  public:
    SimpleFilter(const JetDefinition & subjet_def, const Selector & selector) :
                                 _subjet_def(subjet_def), _selector(selector) {}
  
    virtual std::string description() const {
      return "Filter that finds subjets with " + _subjet_def.description()
             + ", using a (" + _selector.description() + ") selector" ;}
     
    virtual PseudoJet result(const PseudoJet & jet) const;
   
    // CompositeJetStructure is the structural type associated with the 
    // join operation that we use shall use to create the returned jet below
    typedef CompositeJetStructure StructureType;
   
  private:
    JetDefinition _subjet_def;
    Selector      _selector;
  };
\end{lstlisting}
The function that does the work in this class is \ttt{result(...)}:
\begin{lstlisting}
  PseudoJet SimpleFilter::result(const PseudoJet & jet) const {
    // get the subjets
    ClusterSequence * cs = new ClusterSequence(jet.constituents(), _subjet_def);
    vector<PseudoJet> subjets = cs->inclusive_jets();
     
    // signal that the cluster sequence should delete itself when
    // there are no longer any of its (sub)jets in scope anywhere
    cs->delete_self_when_unused();
     
    // get the selected subjets 
    vector<PseudoJet> selected_subjets = _selector(subjets);
    // join them using the same recombiner as was used in the subjet_def
    PseudoJet joined = join(selected_subjets, *_subjet_def.recombiner());
    return joined;
  }
\end{lstlisting}
This provides almost all the basic functionality that might be needed
from a filter, including access to the \ttt{pieces()} of the filtered
jet since it is formed with the \ttt{join(...)} function.
The one part that is potentially missing is that the user does not
have any way of accessing information about the subjets that were not
kept by the filter.
This requires adding to the structural information that underlies the
returned jet.
The \ttt{join(...)}  function creates a structure of type
\ttt{CompositeJetStructure}. There is also a templated version,
\ttt{join<ClassDerivedFromCompositeJetStructure>(...)}, which allows
the user to choose the structure created by the \ttt{join} function.
In this case we therefore create 
\begin{lstlisting}
  #include "fastjet/CompositeJetStructure.hh"
  class SimpleFilterStructure: public CompositeJetStructure { 
  public:
    // the form of constructor expected by the join<...> function
    SimpleFilterStructure(const vector<PseudoJet> & pieces, 
                          const Recombiner *recombiner = 0) :
                                     CompositeJetStructure(pieces, recombiner) {}
    // provide access to the rejected subjets from the filtering
    const vector<PseudoJet> & rejected() const {return _rejected;}
  private: 
    vector<PseudoJet> _rejected; 
    friend class SimpleFilter;
  };
\end{lstlisting}
and then replace the last few lines of the
\ttt{SimpleFilter::result(...)} function with
\begin{lstlisting}
  // get the selected and rejected subjets
  vector<PseudoJet> selected_subjets, rejected_subjets;
  _selector.sift(subjets, selected_subjets, rejected_subjets);

  // join the selected ones, now with a user-chosen structure
  PseudoJet joined = join<SimpleFilterStructure>(selected_subjets, *_subjet_def.recombiner());

  // and then set the structure's additional elements
  SimpleFilterStructure * structure = 
        	    static_cast<SimpleFilterStructure *>(joined.structure_non_const_ptr());
  structure->_rejected = rejected_subjets;
  return joined;
\end{lstlisting}
Finally, with the replacement of the \ttt{typedef} in the
\ttt{SimpleFilter} class with
\begin{lstlisting}
  typedef SimpleFilterStructure StructureType;
\end{lstlisting}
then on a jet returned by the \ttt{SimpleFilter} one can simply call
\begin{lstlisting}
  filtered_jet.structure_of<SimpleFilter>().rejected();
\end{lstlisting}
as with the fully fledged \ttt{Filter} of section~\ref{sec:filtering}.

A second way of extending the structural information of an existing
jet is to ``wrap'' it. This can be done with the help of the
\ttt{WrappedStructure} class.
\begin{lstlisting}
  #include "fastjet/WrappedStructure.hh"
  /// a class to wrap and extend existing jet structures with information about 
  /// "rejected" pieces
  class SimpleFilterWrappedStructure: public WrappedStructure {
  public:
    SimpleFilterWrappedStructure(const SharedPtr<PseudoJetStructureBase> & to_be_wrapped,
				 const vector<PseudoJet> & rejected_pieces) :
	       WrappedStructure(to_be_wrapped), _rejected(rejected_pieces) {}
  
    const vector<PseudoJet> & rejected() const {return _rejected;}
  private:
    vector<PseudoJet> _rejected;
  };
\end{lstlisting}
The \ttt{WrappedStructure}'s constructor takes a \ttt{SharedPtr} to an
existing structure and simply redirects all standard structural
queries to that existing structure. A class derived from it can then
reimplement some of the standard queries, or implement non-standard
ones, as done above with the \ttt{rejected()} call.
To use the wrapped class one might proceed as in the following lines:
\begin{lstlisting}
  // create a jet with some existing structure
  PseudoJet joined = join(selected_subjets, *_subjet_def.recombiner());
  // create a new structure that wraps the existing one and supplements it with new info
  SharedPtr<PseudoJetStructureBase> structure(new
     SimpleFilterWrappedStructure(joined.structure_shared_ptr(), rejected_subjets));
  // assign the new structure to the original jet
  joined.set_structure_shared_ptr(structure);  
\end{lstlisting}
The \ttt{SharedPtr}s ensure that memory allocated for the structural
information is released when no jet remains that refers to it.
For the above piece of code to be used in the \ttt{SimpleFilter} it
would then suffice to include a
\begin{lstlisting}
  typedef SimpleFilterWrappedStructure StructureType;
\end{lstlisting}
line in the \ttt{SimpleFilter} class definition.

In choosing between the templated \ttt{join<...>} and
\ttt{WrappedStructure} approaches to providing advanced structural
information, two elements are worth considering: on one hand, the
\ttt{WrappedStructure} can be used to extend arbitrary structural
information; on the other, while \ttt{join<...>} is more limited in
its scope, it involves fewer pointer indirections when accessing
structural information and so may be marginally more efficient.


\section{Error handling}
\label{sec:error-handling}

\fastjet provides warning and error messages through the classes
\ttt{fastjet::LimitedWarning} and \ttt{fastjet::Error} respectively.
A user does not normally need to interact with them,
however, they do provide some customisation facilities, especially to
redirect and summarise their output.

Each different kind of warning is written out a maximum number of
times (the current default is 5) before its output is suppressed. The
program is allowed to continue. 
At the end of the run (or at any other stage) it is possible to obtain
a summary of all warnings encountered, both explicit or suppressed,
through the following static member function of the LimitedWarning
class:
\begin{lstlisting}
  #include "fastjet/LimitedWarning.hh"
  // ...
  cout << LimitedWarning::summary() << endl;
\end{lstlisting}
The throwing of an \ttt{Error} aborts the program. One can use
\begin{lstlisting}
  /// controls whether the error message (and the backtrace, if its printing is enabled) 
  /// is printed out or not
  static void Error::set_print_errors(bool print_errors);

  /// controls whether the backtrace is printed out with the error message or not.
  /// The default is "false".
  static void Error::set_print_backtrace(bool enabled);
\end{lstlisting}
to control whether an error message is printed (default = \ttt{true})
and whether a full backtrace is also given (default = \ttt{false}).
Switching off the printing of error messages can be useful, for
example, if the user expects to repeatedly catch \fastjet errors.
The \ttt{message()} member function can then be used to access the
specific error message.

The output of both \ttt{LimitedWarning} and \ttt{Error}, which by default
goes to \ttt{std::cerr}, can be redirected to a file using their 
\ttt{set\_default\_stream(std::ostream * ostr)} functions. For instance,
\begin{lstlisting}
  #include "fastjet/LimitedWarning.hh"
  #include "fastjet/Error.hh"
  #include <iostream>
  #include <fstream>
  // ...
  ostream * myerr = new ofstream("warnings-and-errors.txt");
  LimitedWarning::set_default_stream(myerr);
  Error::set_default_stream(myerr);
  Error::set_print_backtrace(true);
  // ...
  cout << LimitedWarning::summary() << endl;
 
\end{lstlisting}
will send the output of both classes to the file
\ttt{warnings-and-errors.txt} (as well as provide the backtrace of
errors).  
Note that the output of \ttt{LimitedWarning::summary()} will only be
present if the program did not abort earlier due to an error.

With a suitable design of the output stream, the output redirection
facility can also be used by the user to record additional information
when an error or warning occurs, for example the event number.
One only \ttt{stream << string} type operation is performed for each
warning or error, so as to help with formatting in such cases.


\section{Evolution of FastJet across versions}
\label{sec:fastjet-history}

\subsection{History}
Version 1 of \fastjet provided the first fast implementation of the
longitudinally invariant $k_t$ clustering~\cite{ktexcl,ktincl}, based
on the factorisation of momentum and geometry in that algorithm's
distance measure~\cite{fastjet}.

Version 2.0 brought
the implementation of the inclusive Cambridge/Aachen algorithm
\cite{CamOrig,CamWobisch} and of jet areas and background
estimation~\cite{cs,CSSAreas}; other changes include a new
interface,\footnote{The old one was retained through v2} and new
algorithmic strategies that could provide a factor of two improvement
in speed for events whose number $N$ of particles was $\sim 10^4$.
Choices of recombination schemes and plugins for external jet
algorithms were new features of version 2.1.  
The initial set of plugins included SISCone~\cite{SISCone}, the CDF
midpoint~\cite{RunII-jet-physics} and JetClu~\cite{Abe:1991ui} cones
and PxCone~\cite{PxCone,Seymour:2006vv}.
The plugins helped provide a uniform interface to a range of different
jet algorithms and made it possible to overlay \fastjet features such as areas
onto the external jet algorithms.
Version 2.2 never made it beyond the beta-release stage, but
introduced a number of the features that eventually were released in 2.3.
The final 2.3 release included the anti-$k_t$ algorithm~\cite{antikt},
a broader set of area measures, improved access to background
estimation, means to navigate the ClusterSequence and a new build system
(GNU autotools).
Version 2.4 included the new version 2.0 of SISCone (including the
spherical variant), as well as
plugins to the \Dzero Run II cone, the ATLAS cone, the CMS cone,
TrackJet and a range of $e^+e^-$ algorithms, and also further tools to
help investigate jet substructure.
It also added a wrapper to \fastjet 
allowing one to run SISCone and some of the sequential recombination
algorithms from Fortran programs.

A major practical change in version 3.0 was that \ttt{PseudoJet}
acquired knowledge 
(where relevant) about its underlying ClusterSequence, allowing one to
write {\em e.g.}  \ttt{jet.constituents()}
It also became possible to associate extra information with a
\ttt{PseudoJet} beyond just a user index.
It brought the first of a series of \fastjet tools to help with
advanced jet analyses, namely the \ttt{Selector} class, filters,
  pruners, taggers and new background estimation classes.
Version~3 also added the D0-Run I cone~\cite{Abbott:1997fc} plugin and
support for native jet algorithms to be run with $R>\pi/2$.

\subsection{Deprecated and removed features}
\label{sec:deprecated}

While we generally aim to maintain backwards compatibility for
software written with old versions of \fastjet, there are occasions
where old interfaces or functionality no longer meet the standards that
are demanded of a program that is increasingly widely used.
Table~\ref{tab:deprecated} lists the cases where such considerations
have led us to deprecate and/or remove functionality.

\begin{table}
  \centering
  \begin{tabular}{lccl}\toprule
    Feature, class or include file
                    & Dep. & Rem.  & Suggested replacement\\\midrule
    FjClusterSequence.hh 
                    & 2.0   & 3.0  & fastjet/ClusterSequence.hh\\
    FjPseudoJet.hh  & 2.0   & 3.0  & fastjet/PseudoJet.hh\\\midrule
    CS::set\_jet\_finder(...)    & 2.1 & 3.0 & pass a JetDefinition to constructor\\
    CS::set\_jet\_algorithm(...) & 2.1 & 3.0 & pass a JetDefinition to constructor\\
    CS::CS(particles, R, ...)    & 2.1 & 3.0 & CS::CS(particles, jet\_def)\\
    JD(jet\_alg, R, strategy)    & 2.1 & -   & JD(jet\_alg, R, recomb\_scheme, strategy)\\
    \midrule
    JetFinder       & 2.3    & -   & JetAlgorithm \\
    SISConePlugin.hh  
                    & 2.3    & 3.0  & fastjet/SISConePlugin.hh (idem.\ other plugins)\\
    ActiveAreaSpec  & 2.3   &  --  & AreaDefinition \& GhostedAreaSpec\\
    ClusterSequenceWithArea 
                    & 2.3   &  --  & ClusterSequenceArea\\\midrule
    default $f=0.5$ in some cone plugins
                    & --    &  2.4 & include $f$ explicitly in constructor\\
    default $R=1$ in JetDefinition
                    & --    &  2.4 & include $R$ explicitly in constructor\\
    \midrule 
    RangeDefinition & 3.0   &  --  & Selector(s)        \\
    CircularRange   & 3.0   &  --  & SelectorCircle     \\
    CSAB::median\_pt\_per\_unit\_area(...)
                   & 3.0    & --   & BackgroundEstimator\\
    CSAB::parabolic\_pt\_per\_unit\_area(...)
                   & 3.0    & --   & BackgroundEstimator (cf.\ section \ref{sec:BGE-positional})\\
    GAS::set\_fj2\_placement(...)
                   & 3.0    & --   & use new default ghost placement instead\\
    \bottomrule 
  \end{tabular}
  \caption{Summary of interfaces and features of earlier versions that have been
    deprecated and/or removed. For brevity we have used the following
    abbreviations: Dep. = version since which a feature has been
    deprecated, Rem. = version where removed, CS
    = ClusterSequence, JD = JetDefinition, CSAB = ClusterSequenceAreaBase, GAS =
    GhostedAreaSpec.
    \label{tab:deprecated}
  } 
\end{table}

\subsection{Backwards compatibility of background estimation facilities}
\label{sec:BGE-backwards}

The \ttt{JetMedianBackgroundEstimator} and
\ttt{GridMedianBackgroundEstimator} classes are new to \fastjet 3.
In \fastjet versions 2.3 and 2.4, the background estimation tools were
instead integrated into the \ttt{ClusterSequenceAreaBase} class.
Rather than using selectors to specify the jets used in the background
estimation, they used the \ttt{RangeDefinition} class.
For the purpose of backwards compatibility, these facilities will
remain present in all 3.0.x versions.
Note that \ttt{ClusterSequenceAreaBase} now actually uses a selector
in its background estimation interface, and that a
\ttt{RangeDefinition} is automatically converted to a selector.

An explicit argument in $\rho$-determination calls in \fastjet 2.4
concerned the choice between the use of scalar areas and the
transverse component of the 4-vector area in the denominator of
$p_t/A$.
The transverse component gives the more accurate $\rho$ determination
and that is now the default in \ttt{JetMedianBackgroundEstimator}. 
The behaviour can be changed with a member function call of the form 
\begin{lstlisting}
  set_use_area_4vector(false);
\end{lstlisting}
Finally, the calculation of $\sigma$ in \fastjet 2.x incorrectly
handled the limit of a small number of jets. This is now fixed in \fastjet 3, but 
a call to
\ttt{set\_provide\_fj2\_sigma(true)} causes \ttt{JetMedianBackgroundEstimator}
to reproduce that behaviour.

\fastjet 2.x also placed the ghosts differently, resulting in different
event-by-event rho estimates, and possibly a small systematic offset
(scaling as the square-root of the ghost area) when ghosts and
particles both covered identical (small) regions.
This offset is no longer present with the \fastjet 3 ghost placement.
If the old behaviour is needed, a call to a specific
\ttt{GhostedAreaSpec}'s \ttt{set\_fj2\_placement(true)} function
causes ghosts to placed as in the 2.x series.



\newpage

\begin{thebibliography}{99}

\bibitem{StermanWeinberg}
  G.~Sterman and S.~Weinberg,
  ``Jets From Quantum Chromodynamics,''
  Phys.\ Rev.\ Lett.\  {\bf 39} (1977) 1436.

\bibitem{Moretti:1998qx}
  S.~Moretti, L.~Lonnblad and T.~Sjostrand,
   ``New and old jet clustering algorithms for electron positron events,''
  JHEP {\bf 9808} (1998) 001
  [arXiv:hep-ph/9804296].

\bibitem{RunII-jet-physics}
  G.~C.~Blazey {\it et al.},
  hep-ex/0005012.

\bibitem{Ellis:2007ib}
  S.~D.~Ellis, J.~Huston, K.~Hatakeyama, P.~Loch and M.~Tonnesmann,
  ``Jets in Hadron-Hadron Collisions,''
  Prog.\ Part.\ Nucl.\ Phys.\  {\bf 60} (2008) 484
  [arXiv:0712.2447 [hep-ph]].

\bibitem{Salam:2009jx}
  G.~P.~Salam,
  Eur.\ Phys.\ J.\  {\bf C67 } (2010)  637-686
  [arXiv:0906.1833 [hep-ph]].


\bibitem{Ali:2010tw}
  A.~Ali, G.~Kramer,
  Eur.\ Phys.\ J.\  {\bf H36 } (2011)  245-326.
  [arXiv:1012.2288 [hep-ph]].

\bibitem{GPLv2} 
  \url{http://www.gnu.org/licenses/gpl-2.0.html}

\bibitem{ktexcl}
  S.~Catani, Y.~L.~Dokshitzer, M.~H.~Seymour and B.~R.~Webber,
  Nucl.\ Phys.\ B {\bf 406}  (1993)  187.

\bibitem{ktincl}
  S.~D.~Ellis and D.~E.~Soper,
  Phys.\ Rev.\ D {\bf 48} (1993) 3160 
  [hep-ph/9305266]. 

\bibitem{fastjet}
  M.~Cacciari and G.~P.~Salam,
  Phys.\ Lett.\ B {\bf 641} (2006) 57
  [hep-ph/0512210].

\bibitem{KtClus}
  M. Seymour,
  \url{http://hepwww.rl.ac.uk/theory/seymour/ktclus/}.


\bibitem{KtJet}
  \url{http://hepforge.cedar.ac.uk/ktjet/};
  J.~M.~Butterworth, J.~P.~Couchman, B.~E.~Cox and B.~M.~Waugh,
  Comput.\ Phys.\ Commun.\  {\bf 153}, 85 (2003)
  [hep-ph/0210022].

\bibitem{CGAL}
A.~Fabri {\it et al.},
Softw.~Pract.~Exper.~ {\bf 30} (2000) 1167;
J.-D.~Boissonnat {\it et al.},
Comp.~Geom.~{\bf 22} (2001) 5; \url{http://www.cgal.org/}
  
\bibitem{antikt}
  M.~Cacciari, G.~P.~Salam and G.~Soyez,
  JHEP {\bf 0804} (2008) 063
  [arXiv:0802.1189 [hep-ph]].

\bibitem{Abdesselam:2010pt}
  A.~Abdesselam, E.~B.~Kuutmann, U.~Bitenc, G.~Brooijmans, J.~Butterworth, P.~Bruckman de Renstrom, D.~Buarque Franzosi, R.~Buckingham {\it et al.},
  Eur.\ Phys.\ J.\  {\bf C71 } (2011)  1661.
  [arXiv:1012.5412 [hep-ph]].

\bibitem{SpartyJet} P.A.~Delsart, K.~Geerlings, J.~Huston,
  B.~Martin and C.~Vermilion, SpartyJet,
  \url{http://projects.hepforge.org/spartyjet}

\bibitem{CSSAreas} 
  M.~Cacciari, G.~P.~Salam and G.~Soyez,
  JHEP {\bf 0804} (2008) 005,
  [arXiv:0802.1188 [hep-ph]].

\bibitem{cs}
  M.~Cacciari and G.~P.~Salam,
  Phys.\ Lett.\  B {\bf 659} (2008) 119
  [arXiv:0707.1378 [hep-ph]].

\bibitem{Cacciari:2010te}
  M.~Cacciari, J.~Rojo, G.~P.~Salam, G.~Soyez,
  Eur.\ Phys.\ J.\  {\bf C71 } (2011)  1539.
  [arXiv:1010.1759 [hep-ph]].

\bibitem{Buttar:2008jx}
  C.~Buttar {\it et al.},
  arXiv:0803.0678 [hep-ph].

\bibitem{Ellis:2009su}
  S.~D.~Ellis, C.~K.~Vermilion, J.~R.~Walsh,
  Phys.\ Rev.\  {\bf D80 } (2009)  051501.
  [arXiv:0903.5081 [hep-ph]].


\bibitem{CamOrig}
  Y.~L.~Dokshitzer, G.~D.~Leder, S.~Moretti and B.~R.~Webber,
  JHEP {\bf 9708}, 001 (1997)
  [hep-ph/9707323];

\bibitem{CamWobisch}
  M.~Wobisch and T.~Wengler,
   ``Hadronization corrections to jet cross sections in deep-inelastic
  arXiv:hep-ph/9907280;
  M.~Wobisch,
   ``Measurement and QCD analysis of jet cross sections in deep-inelastic
DESY-THESIS-2000-049.

\bibitem{eekt}
  S.~Catani, Y.~L.~Dokshitzer, M.~Olsson, G.~Turnock and B.~R.~Webber,
  Phys.\ Lett.\ B {\bf 269}, 432 (1991);

\bibitem{Lonnblad:1992qd}
  L.~Lonnblad,
  Z.\ Phys.\  {\bf C58 } (1993)  471-478.


\bibitem{SISCone}
  G.P.~Salam and G.~Soyez,
  JHEP {\bf 0705} 086 (2007),
  [arXiv:0704.0292 [hep-ph]]; 
standalone code available from \url{http://projects.hepforge.org/siscone}.


\bibitem{EHT}
  S.~D.~Ellis, J.~Huston and M.~Tonnesmann,
in {\it Proc. of the APS/DPF/DPB Summer Study on the Future of
  Particle Physics (Snowmass 2001) } ed. N.~Graf, p. P513
  [hep-ph/0111434].

\bibitem{TeV4LHC}
  TeV4LHC QCD Working Group {\it et al.},
  hep-ph/0610012.

\bibitem{Weinzierl:2011jx}
  S.~Weinzierl,
  arXiv:1108.1934 [hep-ph].


\bibitem{CDFCones} The CDF code has been taken from \\
  \url{http://www.pa.msu.edu/~huston/Les_Houches_2005/JetClu+Midpoint-StandAlone.tgz}\,.

\bibitem{Abe:1991ui}
  F.~Abe {\it et al.}  [CDF Collaboration],
  ``The Topology of three jet events in $\bar{p}p$ collisions at $\sqrt{s} =
  1.8$ TeV,''
  Phys.\ Rev.\ D {\bf 45} (1992) 1448.

\bibitem{Abbott:1997fc}
  B.~Abbott {\it et al.} [D0 Collaboration],
  FERMILAB-PUB-97-242-E.

\bibitem{arXiv:1110.3771}
  V.~M.~Abazov {\it et al.} [D0 Collaboration],
  arXiv:1110.3771 [hep-ex].

\bibitem{Seymour:2006vv}
  M.~H.~Seymour and C.~Tevlin,
  JHEP {\bf 0611} (2006) 052
  [arXiv:hep-ph/0609100].

\bibitem{PxCone} L.~A.~del~Pozo and M.~H.~Seymour, unpublished.

\bibitem{Affolder:2001xt}
  T.~Affolder {\it et al.} [ CDF Collaboration ],
  Phys.\ Rev.\  {\bf D65 } (2002)  092002.

\bibitem{Bartel:1986ua}
  W.~Bartel {\it et al.}  [JADE Collaboration],
  Z.\ Phys.\ C {\bf 33} (1986) 23; 

\bibitem{Bethke:1988zc}
  S.~Bethke {\it et al.}  [JADE Collaboration],
  Phys.\ Lett.\ B {\bf 213} (1988) 235.

\bibitem{SpheriSISCone} G.~P.~Salam and G.~Soyez, April 2009,
  unpublished. 

\bibitem{Fortune}
  S.~Fortune,
  Algorithmica {\bf 2} (1987) 1.

\bibitem{Cacciari:2009dp}
  M.~Cacciari, G.~P.~Salam, S.~Sapeta,
  JHEP {\bf 1004 } (2010)  065.
  [arXiv:0912.4926 [hep-ph]].

\bibitem{GridMedianLH} M.~Cacciari, G.~P.~Salam and G.~Soyez, contribution in
  preparation to proceedings of ``Workshop on TeV Colliders'', Les
  Houches, June 2011.


\bibitem{BDRS}
  J.~M.~Butterworth, A.~R.~Davison, M.~Rubin and G.~P.~Salam,
  Phys.\ Rev.\ Lett.\  {\bf 100} (2008) 242001
  [arXiv:0802.2470 [hep-ph]].

\bibitem{trimming}
  D.~Krohn, J.~Thaler and L.~T.~Wang,
  JHEP {\bf 1002} (2010) 084
  [arXiv:0912.1342 [hep-ph]].

\bibitem{Kaplan:2008ie}
  D.~E.~Kaplan, K.~Rehermann, M.~D.~Schwartz, B.~Tweedie,
  Phys.\ Rev.\ Lett.\  {\bf 101 } (2008)  142001
  [arXiv:0806.0848 [hep-ph]].

\bibitem{Butterworth:2009qa}
  J.~M.~Butterworth, J.~R.~Ellis, A.~R.~Raklev, G.~P.~Salam,
  Phys.\ Rev.\ Lett.\  {\bf 103 } (2009)  241803.
  [arXiv:0906.0728 [hep-ph]].

\bibitem{nsubtagger}
  J.~H.~Kim,
  Phys.\ Rev.\  D {\bf 83} (2011) 011502
  [arXiv:1011.1493 [hep-ph]].


\bibitem{Chan}
  T.~M.~Chan,
  ``Closest-point problems simplified on the RAM,''
  in Proc.\ 13th ACM-SIAM Symposium on Discrete Algorithms (SODA),
  p.~472 (2002).


\bibitem{Anderberg}
  M.~R.~Anderberg, 
  Cluster Analysis for Applications,
  (Number 19 in Probability and Mathematical Statistics, Academic
  Press, New York, 1973).


\bibitem{Sonnenschein}
  L.~Sonnenschein, Ph.D. Thesis, RWTH Aachen 2001; \\
  \url{http://cmsdoc.cern.ch/documents/01/doc2001_025.ps.Z}

\bibitem{Sjostrand:2006za}
  T.~Sjostrand, S.~Mrenna and P.~Skands,
  ``{\tt Pythia} 6.4 physics and manual,''
  JHEP {\bf 0605} (2006) 026,
  [arXiv:hep-ph/0603175].

\bibitem{Sjostrand:2007gs}
  T.~Sjostrand, S.~Mrenna, P.~Z.~Skands,
  Comput.\ Phys.\ Commun.\  {\bf 178 } (2008)  852-867.
  [arXiv:0710.3820 [hep-ph]].

\bibitem{EppsteinHierarchical}
  D.~Eppstein
  J. Experimental Algorithmics {\bf 5} (2000) 1-23 [cs.DS/9912014].



\end{thebibliography}
\end{document}